\documentclass[prd,nofootinbib,floats,aps,twocolumn,tightenlines,superscriptaddress]{revtex4-2}
\usepackage{natbib}
\usepackage{graphicx}
\usepackage{mathtools}
\usepackage[normalem]{ulem}
\usepackage{hyperref}
\hypersetup{
    colorlinks=true,
    linkcolor=blue,
    filecolor=magenta,      
    urlcolor=cyan,
}
\usepackage{commath}
\usepackage{bm}
\usepackage{amsfonts,amsmath,amssymb,amsthm}
\usepackage[usenames,dvipsnames]{xcolor}
\usepackage{dsfont}
\usepackage{enumitem}
\usepackage{color}
\usepackage{tikz}
\usepackage{physics}
\usepackage{verbatim}

\newcommand{\ii}{\mathrm{i}}
\newcommand{\proj}[2]{\left| {#1} \right\rangle\!\left\langle {#2} \right|}

\renewcommand*\d[2][]{%
	\mathrm{d}%
	\ifx\relax#1\relax\else
	\rule{-0.02em}{1.5ex}^{#1}\rule{0.08em}{0ex}\!
	\fi
	#2\,
}

\newcommand{\phih}{\hat \phi}

\newcommand{\muh}{\hat{\mu}}

\newcommand{\crea}{\hat{a}^{\dagger}}
\newcommand{\anni}{\hat{a}}

\newcommand{\rhoh}{\hat{\rho}}

\newcommand\restr[2]{{% we make the whole thing an ordinary symbol
  \left.\kern-\nulldelimiterspace % automatically resize the bar with \right
  #1 % the function
  \vphantom{\normal|} % pretend it's a little taller at normal size
  \right|_{#2} % this is the delimiter
  }}
\newcommand{\diff}{\mathrm{d}}
\newcommand{\spatial}[1]{\bm{#1}}
\newcommand{\sptime}[1]{\mathsf{#1}}
\newcommand{\R}[1]{\mathbb{R}^{#1}}
\newcommand{\beq}{\begin{equation}}
\newcommand{\eeq}{\end{equation}}
\newtheorem*{claim*}{Claim}

\begin{document}

\title{A detector-based measurement theory for quantum field theory}

\author{Jos\'{e} Polo-G\'{o}mez}
\email{jpologomez@uwaterloo.ca}
\affiliation{Department of Applied Mathematics, University of Waterloo, Waterloo, Ontario N2L 3G1, Canada}
\affiliation{Institute for Quantum Computing, University of Waterloo, Waterloo, Ontario N2L 3G1, Canada}

\author{Luis J. Garay}
\email{luisj.garay@ucm.es}
\affiliation{Departamento de F\'{i}sica Te\'{o}rica and IPARCOS, Universidad Complutense de Madrid, 28040 Madrid, Spain}
\affiliation{Instituto de Estructura de la Materia (IEM-CSIC), Serrano 121, 28006 Madrid, Spain}

\author{Eduardo Mart\'{i}n-Mart\'{i}nez}
\email{emartinmartinez@uwaterloo.ca}
\affiliation{Department of Applied Mathematics, University of Waterloo, Waterloo, Ontario N2L 3G1, Canada}
\affiliation{Institute for Quantum Computing, University of Waterloo, Waterloo, Ontario N2L 3G1, Canada}
\affiliation{Perimeter Institute for Theoretical Physics, Waterloo, Ontario N2L 2Y5, Canada}

\begin{abstract}

We propose a measurement theory for quantum fields based on measurements made with localized non-relativistic systems that couple covariantly to quantum fields (like the Unruh-DeWitt detector). Concretely, we analyze the positive operator-valued measure (POVM) induced on the field when an idealized measurement is carried out on the detector after it coupled to the field. Using an information-theoretic approach, we provide a relativistic analogue to the quantum mechanical L\"uders update rule to update the field state following the measurement on the detector. We argue that this proposal has all the desirable characteristics of a proper measurement theory. In particular it does not suffer from the ``impossible measurements'' problem pointed out by Rafael Sorkin in the 90s which shows that idealized measurements cannot be used in quantum field theory.

\end{abstract}
	
\maketitle

\section{Introduction}\label{section: introduction}

In any physical theory, it is necessary to describe the mechanism that allows us to gather information about the physical systems we are modelling, that is, it is necessary to describe measurements. In classical theories, the description of measurements is frequently not explicit, often hidden under the assumption that we can neglect the effect of measurements on the state of the system of interest. However, in quantum mechanics, describing measurement processes has been problematic and a subject of discussion since its very inception (see e.g.,~\cite{Wheeler1983}). Nevertheless, from an operational point of view the problem can be bypassed in the context of non-relativistic quantum mechanics by employing \mbox{L\"uders rule}, also known as the projection postulate~\cite{Luders1951}. This rule prescribes how to update the state of the system after the measurement in a way consistent with the measurement outcome, through projection-valued measures (PVMs). This model of measurement is called \textit{projective measurement} or \textit{idealized measurement}. 

However, projective measurements are not suitable to describe measurements in quantum field theory, since they are not compatible with relativistic causality, and therefore they are not consistent with the very foundational framework of quantum field theory (QFT). Specifically, there are no local projectors of finite rank in QFT~\cite{ReehSchlieder,Schlieder1968,Hellwig1970formal,Redhead1995}. Any finite rank projector in QFT, such as a projector onto some single particle wave-packet state, is inherently non-local, and so any attempt to generalize the projection postulate with such a projector leads to spurious faster-than-light signalling~\cite{Sorkin1993,Lin2014,Benincasa2014,Borsten2021,Jubb2021}. It should be clear from the beginning that when we talk about the causality issues of the projection postulate, we are indeed referring to superluminal causation, and not the non-locality that arises from entanglement, which can be present even between non-relativistic systems~\cite{EPR}. The latter is perfectly compatible with causality as long as it does not enable signalling---it just tells us that quantum theories are non-local in nature, and correlations can be present between quantum systems that are spacelike separated even in QFT~\cite{Summers1985,Summers1987}.  

The impossibility of naively generalizing the projection postulate to QFT has been addressed mainly in three different ways. 

First, one could consider localized ideal measurements\footnote{For a more thorough analysis where general CPTP maps (not necessarily ideal measurements) on quantum fields are characterized according to its causal behaviour, see~\cite{Jubb2021}.} (in the form of infinite rank projectors) and try to modify the projection postulate in a covariant way, as in Hellwig and Kraus's proposal~\cite{Hellwig1970formal,Hellwig1970operations,Hellwig1969}. This prescription however suffers from the same faster-than-light signalling that Sorkin pointed out in~\cite{Sorkin1993}, as discussed in~\cite{Borsten2021}. 

Second, another way consists of formulating a measurement theory completely within quantum field theory, such as Fewster and Verch's framework~\cite{FewsterVerch,Fewster2019covariant,Bostelmann2021}. %\sout{Such a measurement scheme is consistent with QFT by definition and is therefore completely safe from any causality issues} 
By giving a covariant update rule, they obtain a measurement scheme consistent with QFT and therefore completely safe from any causality issues. In this framework, however, being entirely within QFT, the localized measurement probes are also quantum fields, and we are still left with the problem of how to access the information of that second ancillary field~\cite{DanBruno2021}. This is because low energy measurement apparatuses (like atoms, photodetectors, photomultipliers, human retinas, etc.) are not well described by a free field theory, and the treatment of bound states in QFT is still very much an open problem~\cite{WeinbergQFT}. 

Finally, the third option relies on coupling so-called \textit{particle detectors}---localized non-relativistic quantum systems---to quantum fields, such as the Unruh-DeWitt (UDW) model~\cite{Unruh,DeWitt}. Although pointlike detector models are fully compatible with relativistic causality~\cite{Edu2015,TalesBruno2020,TalesBruno2021,Pipo2021}, their singular nature leads, in certain contexts, to divergences~\cite{Schlicht2004}. However, those divergences are not present for detectors that are adiabatically switched on~\cite{Satz2007}, or that are spatially smeared~\cite{Louko2006,Langlois}. In the latter case, even though the unitary evolution is perfectly compatible with causality and does not allow faster-than-light signalling with a second detector~\cite{Edu2015,Pipo2021}, the use of a (non-pointlike) spatial smearing along with the non-relativistic approximation can indeed enable some degree of faster-than-light signalling between two detectors with the help of a third ancillary one in between~\cite{Pipo2021}. However, unlike in the case of projective measurements, in the smeared setups that show any degree of superluminal signalling, its impact is bounded by the smearing lengthscale of the ancillary detector (which by approximating it to be non-relativistic we already neglected in our frame of reference, to start with) and moreover does not show up at leading orders in perturbation theory~\cite{Pipo2021}. These results, together with the fact that detector models can realistically represent the way fields are measured experimentally~\cite{Edu2013,Rodriguez2018,Lopp2020}, make this option especially appealing for modelling measurements in QFT. %\luisprop{However, unlike in the case of projective measurements, in this scenario the spacelike distance between the superluminal receiver and the future light cone of the emitter is bounded by the size of the smearing of the ancillary detector, as illustrated in Figure~\ref{fig: superminal signalling with detectors}. By approximating this ancillary detector to be non-relativistic, we are willfully neglecting its size in our frame of reference. Moreover, this kind of faster-than-light signalling does not show up at leading orders in perturbation theory~\cite{Pipo2021}. These results, together with the fact that detector models can realistically represent the way fields are measured experimentally~\cite{Edu2013,Rodriguez2018,Lopp2020}, make this option especially appealing for modelling measurements in QFT in the authors' opinion.} 

In the particle detector approach, however, not every step of the process is already well understood. We still need to describe the mechanism through which we go from a field state and a detector that are originally decoupled and uncorrelated, to a measurement outcome that an experimentalist can put in a plot or write on a notepad. After the experiment is performed and some classical information has been obtained about the field, how does one take into account this information for the description of future experiments involving the field? 

This question is particularly relevant for the field of relativistic quantum information. Indeed, there are several landmark protocols and experimental proposals in the context of quantum information (e.g., the quantum Zeno effect~\cite{Misra1977,Patil2015}, the delayed choice quantum eraser experiment~\cite{Scully1982,Scully1991,Kwiat1992,Kim2000,Ma2013}, or the Wigner's friend experiment~\cite{Wignerchapter1961,Frauchiger2018}, among others) in which the ability to perform measurements and using the information of the outcomes to update the state is essential for their implementation. To be able to formulate these scenarios in relativistic contexts, it is necessary to have a well-understood measurement theory that works in the context of quantum field theory and that connects to experimentally measurable quantities.

In this paper we aim to formulate consistently a measurement theory for QFT using detectors as measuring tools. First, in Section~\ref{section: setup} we present our working model. In Section~\ref{section: the updated state of the field} we describe the measurement process (including field-detector interaction and idealized measurement of the detector) and obtain the field state update according to the measurement outcome. In Section~\ref{section: causal behaviour} we analyze this update in order to determine whether this kind of measurement abides by relativistic causality. In Section~\ref{section: the update rule} we present a context-dependent update rule consistent with QFT. In Section~\ref{section: update of n-point functions} we explicitly formulate it in terms of $n$-point functions and in Section~\ref{section: generalization to the presence of entangled third parties} we analyze the most general initial scenario. Section~\ref{section: discussion} is devoted to discussing how the framework presented in this manuscript constitutes a valid measurement theory for QFT. Finally we present our conclusions in Section~\ref{section: conclusions}.

\section{Setup}\label{section: setup}

In this work we consider a spatially smeared Unruh-DeWitt model~\cite{Unruh,DeWitt} for a detector coupled to a real scalar field in a (1+$d$)-dimensional flat spacetime. This is a simplified model that is both covariant~\cite{TalesBruno2020,TalesBruno2021} and yet captures the phenomenology of light-matter interaction neglecting angular momentum exchange but without any further common quantum optics approximations---such as the rotating-wave or single (or few) mode approximation~\cite{Edu2013,Pozas2016,Rodriguez2018,Lopp2020}. For our purposes, let us consider that the detector is inertial and at rest in the frame of coordinates $(t,\bm{x})$ so that its  proper time coincides with the coordinate time $t$. Then, in the interaction picture, the interaction Hamiltonian is~\cite{Langlois} 
\beq\label{UDW Hamiltonian}
\hat{H}_{I}(t)=\lambda \chi(t)\muh(t) \int\diff^{d}\spatial{x}\,F(\spatial{x})\phih(t,\spatial{x}) \;.
\eeq
In this equation, the scalar field $\hat\phi$ can be expanded in terms of plane-wave solutions in the quantization frame $(t,\bm x)$ as
\beq
\phih(t,\bm{x})\!=\!\!\int\!\frac{\diff^{d}\bm{k}}{\sqrt{(2\pi)^{d}2\omega_{\bm{k}}}}\left( \anni_{\bm{k}}e^{-\ii(\omega_{\bm{k}}t-\bm{k}\cdot\bm{x})}+\crea_{\bm{k}}e^{\ii(\omega_{\bm{k}}t-\bm{k}\cdot\bm{x})} \right),
\eeq
where $\hat a^\dagger_{\bm k}$ and $\hat a_{\bm k}$ are canonical creation and annihilation operators satisfying the commutation relations $[\hat{a}^{\phantom{\dagger}}_{\bm{k}},\hat{a}_{\bm{k'}}^{\dagger}]=\delta(\bm{k}-\bm{k}')$. The internal degree of freedom (monopole moment) of the detector, which we choose to have two levels (ground $\ket{g}$ and excited $\ket{e}$) with an energy gap $\Omega$ between them is given by 
\beq
\muh(t)=\proj{g}{e} e^{-\ii\Omega t}+\proj{e}{g} e^{\ii\Omega t} \;.
\eeq
$\lambda$ is the coupling strength and $\chi(t)$ is the switching function controlling the time dependence of the coupling. The interaction is on only for times in the support of $\chi(t)$. For simplicity we assume this to be a finite interval $[t_\textrm{on},t_\textrm{off}]$ (i.e. $\chi(t)$ is compactly supported). $F(\bm{x})$ is the spatial smearing function that models the localization of the detector, and therefore the support of the product of $\chi$ and $F$,
\beq\label{interaction region}
\mathcal{D}\coloneqq \operatorname{supp}\{\chi(t)\cdot F(\bm{x})\} \;,
\eeq
is what we will call \textit{interaction region} or, slightly abusing nomenclature, \textit{detector}. Its \textit{causal future/past} $\mathcal{J}^\pm(\mathcal{D})$ is the union of the future/past light cones of all its points and their interiors. In Minkowski spacetime\footnote{The metric in a coordinate system associated with inertial observers is $\eta_{\mu\nu}=\text{diag}(-1,1,\hdots,1)$, $\{x^0,\dots,x^{d}\}$ are the coordinates of the event $\mathsf{x}$, and   $\mathsf{x}\cdot\mathsf{y}\coloneqq \eta_{\mu\nu}x^\mu y^\nu$.} $\mathcal{M}$,
\begin{equation}\label{causal future of the interaction region}
\mathcal{J}^{\pm}(\mathcal{D})\!:=\!\{\,\mathsf{y}\! \in\! \mathcal{M}\!:\!\exists \:\mathsf{x}\in\!\mathcal{D}\:,\: \mathsf{x}\cdot\mathsf{y} \leq 0 \:,\: \pm(y^{0}-x^{0})\geq 0\} \;.
\end{equation}
The \textit{causal support} of $\mathcal{D}$ is the union of both its causal past and its causal future,
\begin{equation}
\mathcal{J}(\mathcal{D}):=\mathcal{J}^{+}(\mathcal{D})\cup\mathcal{J}^{-}(\mathcal{D}) \;.
\end{equation}

\section{The updated state of the field}\label{section: the updated state of the field}

In this section we will compute what is the update on the field state after an observable of the detector is measured by an experimenter through an idealized measurement. In other words, we will compute what is the POVM that is applied to update the field state if the detector is updated by a projective measurement. Although we will consider the most general case, a physical example of this kind of situation could be an experiment in the lab where we check  whether the detector clicks (gets excited) or not (stays in the ground state) after interacting with an excited state of the electromagnetic field. 

It is reasonable to consider that the detector and the field are  initially uncorrelated. That is, in the absence of third parties\footnote{We will generalize to cases where the field is entangled with third parties in Section~\ref{section: generalization to the presence of entangled third parties}.}, the state of the system before the interaction is $\rhoh=\rhoh_{\textrm{d}} \otimes \rhoh_{\phi}$. At time \mbox{$t=T$}, the experimenter performs a rank-one projective measurement \mbox{$P=\ket{s}\!\bra{s}$} of an arbitrary observable of the detector. One can think of the idealized measurement as performing a measurement of an observable of the detector, and of $\proj{s}{s}$ as being the eigenprojector associated to the obtained measurement outcome. For simplicity we assume that the measurement takes place after the interaction between the field and the detector has been switched off, that is, $T\geq t_\textrm{off}$. Then, after the projective measurement, the updated state of the joint system is~\cite{Nielsen2010}
\begin{equation}
\rhoh^{P}=\frac{(P \otimes \mathds{1})\hat{U}\rhoh \hat{U}^{\dagger}(P\otimes \mathds{1})}{\textrm{tr}\big[(P\otimes \mathds{1})\hat{U}\rhoh \hat{U}^{\dagger}\big]} \;, 
\end{equation}
where the unitary evolution operator $\hat{U}$ is given by 
\beq\label{unitary}
\hat{U}=\mathcal{T}\exp(-\ii\int_{-\infty}^{\infty}\!\!\!\!\diff t'\, \hat{H}_I(t')) \;.
\eeq
The assumption that $\chi(t)=0$ for all $t\geq t_{\textrm{off}}$ allows us to safely extend the integration range to infinity. From now on, we will use the integral sign without specifying limits whenever the integral is carried out in the whole domain of the integrand. We thus have, for the updated state of the field,
\beq
\rhoh^{P}_{\phi}= \tr_{\text{d}}(\rhoh^{P}) \propto \textrm{tr}_{\text{d}}\big[(P\otimes\mathds{1})\hat{U}\rhoh \hat{U}^{\dagger}(P\otimes \mathds{1})\big] \;,
\eeq
where $\tr_{\text{d}}(\cdot)$ stands for the partial trace over the Hilbert space of the detector. We note that the matrix elements of $\rhoh^{P}_{\phi}$ satisfy that  
\beq
\bra{\varphi_{1}}\rhoh^{P}_{\phi} \ket{\varphi_{2}} \propto \bra{s,\varphi_1}\hat{U}\rhoh \hat{U}^{\dagger} \ket{s,\varphi_2} \;.
\eeq
where $\ket{s,\varphi_i}\equiv\ket{s}\otimes\ket{\varphi_i}$ and $\ket{\varphi_i}$ is a vector in the field Hilbert space. From now on, we will use the following notation: if $\hat{\mathcal{O}}$ is an operator acting on the detector-field Hilbert space and $\ket{\psi_1},\ket{\psi_2} \in \mathcal{H}_\textrm{d}$ are detector states, then we will understand $\bra{\psi_1}\hat{\mathcal{O}}\ket{\psi_2}$ to be the field operator that satisfies
\begin{equation}\label{notation for sandwiched operators}
\bra{\varphi_1}\bra{\psi_1}\hat{\mathcal{O}}\ket{\psi_2}\ket{\varphi_2}=\bra{\psi_1,\varphi_1}\hat{\mathcal{O}}\ket{\psi_2,\varphi_2}
\end{equation}
for any field states $\ket{\varphi_1},\ket{\varphi_2}\in\mathcal{H}_\phi$. Finally, let us assume that the initial state of the detector is pure\footnote{This simplification can be easily dropped, and the result straightforwardly generalized, in the same way that happens with POVMs in non-relativistic quantum mechanics~\cite{Nielsen2010}.}, $\rhoh_\textrm{d}=\proj{\psi}{\psi}$. Thus, using the convention in~\eqref{notation for sandwiched operators},
\begin{equation}
\bra{s,\varphi_1}\hat{U}\rhoh\hat{U}^\dagger\ket{s,\varphi_2}=\bra{s}\hat{U}\ket{\psi}\rhoh_\phi \bra{\psi}\hat{U}^\dagger\ket{s} \;.
\end{equation}
We can therefore write the updated state of the field for the projection over the state $\ket{s}$ of the detector as
\beq\label{updated state field}
\rhoh^{s,\psi}_{\phi}=\frac{\hat{M}_{s,\psi}\rhoh_{\phi}\hat{M}_{s,\psi}^{\dagger}}{\textrm{tr}_{\phi}\big(\rhoh_{\phi}\hat{E}_{s,\psi}\big)} \;,
\eeq
where
\beq\label{M operator}
\hat{M}_{s,\psi}=\bra{s}\hat{U}\ket{\psi}
\eeq
is an operator acting on the field Hilbert space, and the POVM elements~\cite{Nielsen2010} are
\beq
\hat{E}_{s,\psi}=\hat{M}_{s,\psi}^{\dagger}\hat{M}_{s,\psi} \;.
\eeq
For our system, we can get a tractable expression for the $\hat{M}_{s,\psi}$ operators proceeding perturbatively in $\lambda$. First, the unitary $\hat{U}$ in \eqref{unitary} can be written as
\beq
\hat{U}=\mathds{1}+\sum_{n=1}^{\infty}\lambda^n \hat{U}^{(n)} \;.
\eeq
For the first two orders, substituting~\eqref{UDW Hamiltonian}, we have
\beq
\hat{U}^{(1)}=-\ii \int\diff t \,\diff^{d}\spatial{x} \, \chi(t) F(\spatial{x}) \muh(t) \phih(t,\spatial{x})
\eeq
and
\begin{align}
\hat{U}^{(2)}=&-\int\diff t \,\diff t' \, \diff^{d}\spatial{x} \, \diff^{d}\spatial{x}'\, \theta(t-t')\chi(t)\chi(t') \\
& \times \, F(\spatial{x})F(\spatial{x}') \muh(t)\muh(t')\phih(t,\spatial{x})\phih(t',\spatial{x}') \;. \nonumber
\end{align}
As a result, we can apply the same expansion to $\hat{M}_{s,\psi}$,
\beq
\hat{M}_{s,\psi}=\hat{M}_{s,\psi}^{(0)}+\lambda\hat{M}_{s,\psi}^{(1)}+\lambda^2\hat{M}_{s,\psi}^{(2)}+\hdots \;,
\eeq
where we are denoting \mbox{$\hat{M}_{s,\psi}^{(n)}=\bra{s}\hat{U}^{(n)}\ket{\psi}$}. In particular,
\begin{align}
&\hat{M}_{s,\psi}^{(0)}=\braket{s}{\psi}\mathds{1}_{\phi}\;,\label{M operator order 0}\\
&\hat{M}_{s,\psi}^{(1)}=-\ii\int \diff t \,\diff^{d}\spatial{x} \,\chi(t) F(\spatial{x}) \bra{s}\muh(t)\ket{\psi}\phih(t,\spatial{x})\;,\label{M operator order 1}\\
&\hat{M}_{s,\psi}^{(2)}=-\int \diff t\,\diff t'\, \diff^{d}\spatial{x}\,\diff^{d}\spatial{x}'\,\theta(t-t')\chi(t)\chi(t') \nonumber \\
&\qquad \times\, F(\spatial{x})F(\spatial{x}') \bra{s}\muh(t)\muh(t')\ket{\psi}\phih(t,\spatial{x})\phih(t',\spatial{x}') \;.\label{M operator order 2}
\end{align}

\section{Causal behaviour}\label{section: causal behaviour}

Once the form of the POVM that updates the state of the field~\eqref{updated state field} has been obtained, we want to analyze whether this update respects relativistic causality. In this section we will study whether the measurement defined in the previous section influences the field state outside of the causal future of the measurement.

Concretely, in order to understand the causal behaviour of the update of the field state that arises from performing a projective measurement \textit{on the detector}, we need to compare the updated state of the field $\rhoh_{\phi}^\textrm{u}$ (post-measurement) and the initial state of the field $\rhoh_{\phi}$ (pre-measurement) and see that there is no influence on the field state outside the causal future of the interaction region. Since the state of the field is fully characterized by its \mbox{$n$-point} functions, the analysis can be reduced to studying how the $n$-point functions change after the measurement process (consisting of \mbox{(i) interaction} with the detector, and---after switching off the interaction---\mbox{(ii) idealized} measurement of the detector and corresponding POVM update on the field) in the region that is spacelike separated from the detector. %\sout{Note that if the state is Gaussian the analysis can be further reduced to the study of one-point and two-point functions. However, we do not need to make this assumption here, the arguments will follow for $n$-point functions in general.} 

Regarding the updated field state $\rhoh_\phi^\textrm{u}$, note that the update given by Eqs.~\eqref{updated state field} and~\eqref{M operator} corresponds to a \textit{selective measurement}~\cite{Luders1951,Hellwig1970formal}---the measurement is performed and its outcome is checked, updating the state of the field accordingly. However, if an observer is spacelike separated from the detector, then they might know that the measurement was prearranged to be performed, but they cannot know the outcome of such measurement since information cannot be transmitted to them. Thus, from that observer's perspective, the update of the state has to be the one associated with a \textit{non-selective measurement}~\cite{Luders1951,Hellwig1970formal}---the state of the field is updated taking into account that the measurement has been carried out, but without knowing its outcome. This measurement model respects causality if the spacelike separated observer cannot tell with local operations whether the measurement was carried out or not, i.e. if the non-selective update does not impact the outcome of local operations performed outside the causal support of the measurement.

A non-selective measurement has to be understood as having made the projective measurement on the detector when the outcome is not made concrete. Therefore, to update the state we apply a convex mixture of all the projectors over all the possible proper subspaces associated with every potential outcome of the measurement, weighted by its associated probabilities given by Born's rule (see again \cite{Luders1951,Hellwig1970formal}).

Since we are considering a two-level Unruh-DeWitt detector, the most general non-selective projective measurement can be described by two complementary rank-one projections, $\ket{s}\!\bra{s}$ and $\ket{\bar{s}}\!\bra{\bar{s}}$, such that
\beq\label{s and r are a basis}
\mathds{1}_{\textrm{d}}-\ket{s}\!\bra{s}=\ket{\bar{s}}\!\bra{\bar{s}},
\eeq
where $\ket{s}$ and $\ket{\bar{s}}$ are two orthonormal vectors in the detector's Hilbert space, $\mathcal{H}_{\textrm{d}}$. The state of the field updated by a non-selective measurement can then be written as the mixture of the updates for each projection $\rhoh_{\phi}^{s,\psi}$ and $\rhoh_{\phi}^{\bar{s},\psi}$ given by \eqref{updated state field}, weighted by their respective probabilities, $\langle\hat{E}_{s,\psi}\rangle_{\rhoh_{\phi}}$ and $\langle\hat{E}_{\bar{s},\psi}\rangle_{\rhoh_{\phi}}$,
\begin{align}\label{non-selective}
\rhoh_{\phi}^\textrm{u}&=\langle\hat{E}_{s,\psi}\rangle_{\rhoh_{\phi}}\rhoh_{\phi}^{s,\psi}+\langle\hat{E}_{\bar{s},\psi}\rangle_{\rhoh_{\phi}}\rhoh_{\phi}^{\bar{s},\psi}\\
&=\hat{M}_{s,\psi}\rhoh_{\phi}\hat{M}_{s,\psi}^{\dagger}+\hat{M}_{\bar{s},\psi}\rhoh_{\phi}\hat{M}_{\bar{s},\psi}^{\dagger} \;. \nonumber
\end{align}
By~\eqref{M operator}, recalling the notation described in~\eqref{notation for sandwiched operators}, from~\eqref{non-selective},
\begin{align}\label{non-selective updated state as a trace}
\rhoh_{\phi}^{\textsc{NS}}&=\bra{s}\hat{U}\ket{\psi}\rhoh_\phi \bra{\psi}\hat{U}^\dagger\ket{s}+\bra{\bar{s}}\hat{U}\ket{\psi}\rhoh_\phi \bra{\psi}\hat{U}^\dagger\ket{\bar{s}} \nonumber \\
&=\textrm{tr}_{\textrm{d}}\big[\hat{U}(\ket{\psi}\!\bra{\psi}\otimes \rhoh_{\phi})\hat{U}^{\dagger}\big] \;,
\end{align}
where we have used~\eqref{s and r are a basis} to reduce the sum to a trace over the detector Hilbert space. 

Let $\hat{A}$ be a field observable. Then we get that its expectation value for the non-selective update of the field state is
\begin{align}\label{A rho phi}
&\langle \hat{A} \rangle_{\rhoh_{\phi}^{\textsc{NS}}}=\textrm{tr}_{\phi}\big[\textrm{tr}_{\textrm{d}}[\hat{U}(\ket{\psi}\!\bra{\psi}\otimes \rhoh_{\phi})\hat{U}^{\dagger}]\hat{A}\big] \\
&=\textrm{tr}\big[\hat{U}(\ket{\psi}\!\bra{\psi}\otimes \rhoh_{\phi})\hat{U}^{\dagger}\hat{A}\big]=\textrm{tr}\big[(\ket{\psi}\!\bra{\psi}\otimes \rhoh_{\phi})\hat{U}^{\dagger}\hat{A}\hat{U}\big] \;, \nonumber
\end{align}
where \mbox{$\langle \hat{\mathcal{O}}\rangle_{\rhoh}=\textrm{tr}\big( \rhoh \hat{\mathcal{O}} \big)$} as usual. We have used the cyclic property of trace and we have denoted the detector-field operator $\mathds{1}_\textrm{d}\otimes\hat{A}$ simply as $\hat{A}$, omitting the identity of the detector. Now, taking into account the form of the UDW interaction Hamiltonian~\eqref{UDW Hamiltonian} and the unitary evolution operator~\eqref{unitary}, if $\hat{A}$ is a field observable supported outside the causal support of the interaction region, then microcausality ensures that
\begin{equation}
[\hat{A},\phih(t,\bm{x})]=0 
\end{equation}
for every $(t,\bm{x}) \in \mathcal{D}$, and therefore
\begin{equation}
[\hat{A},\hat{U}]=0 \;.
\end{equation}
This means that Eq.~\eqref{A rho phi} yields
\begin{align}\label{non-selective and initial are the same when spacelike}
&\langle \hat{A} \rangle_{\rhoh_\phi^\textsc{NS}}=\textrm{tr}_\phi\big(\rhoh_\phi^\textsc{NS}\hat{A}\big)=\textrm{tr}\big[ (\proj{\psi}{\psi}\otimes\rhoh_\phi)\hat{U}^\dagger \hat{A} \hat{U} \big] \\
&=\textrm{tr}\big[ (\proj{\psi}{\psi}\otimes\rhoh_\phi)\hat{U}^\dagger \hat{U} \hat{A} \big]=\textrm{tr}_\phi\big(\rhoh_\phi\hat{A}\big)=\langle \hat{A} \rangle_{\rhoh_\phi} \;. \nonumber
\end{align}
We conclude that the non-selective POVM does not affect the expectation value of any local observable outside the causal influence region of the detector. Of particular importance is the case when we take \mbox{$\hat{A}=\phih(t_1,\bm{x}_1)\cdots\phih(t_n,\bm{x}_n)$}, with all $(t_1,\bm{x}_1),\hdots,(t_n,\bm{x}_n)$ outside the causal support of the interaction region, i.e. spacelike separated from the interaction region. Then Eq.~\eqref{non-selective and initial are the same when spacelike} allows us to conclude that the corresponding $n$-point functions do not change under the non-selective update.

The effect of the rank-one projective measurement performed \textit{on the detector} is thus restricted to the causal future of the interaction region between the detector and the field. In particular, it is bounded in every spacelike hypersurface of the Minkowski spacetime. This feature contrasts with the effect (studied by Sorkin in~\cite{Sorkin1993}) of a finite-rank projective measurement performed \textit{on the field}, which affects the whole future half of the spacetime determined by the spacelike hypersurface in which the measurement is considered to be performed~\cite{Sorkin1993}. Therefore, if we are in the regimes where causality is respected by the coupling between detector and field (e.g., pointlike detectors in any scenario or spatially smeared detectors in the scales identified in~\cite{Pipo2021}), the projective measurement performed on the detector is safe from any causality issues\footnote{For the sake of brevity, from now on we will not restate the conditions under which particle detector models behave causally and will just state that the particle detector models are causal. The facts that should be acknowledged throughout this manuscript are that a pointlike detector is fully causal, that its quantum dynamics is non-singular (when switched on carefully) and that the causality violations (if any) of the model come through (well-controlled) smearing scales. For a more careful recapitulation of these conditions, see Section~\ref{section: introduction}.} of the kind exposed in ``Impossible measurements on quantum fields''~\cite{Sorkin1993}. The existence of physically motivated regimes that set the limits of validity of the particle detector model distinguishes this approach from the performance of infinite-rank projective measurements on the field, where faster-than-light signalling is allowed in general~\cite{Borsten2021}. Moreover, Eq.~\eqref{A rho phi} also shows that, for the non-selective update, expectation values of arbitrary observables only depend on the joint state of the field and the detector after the interaction, and not on the measurement performed on the detector. Indeed, the non-selective measurement eliminates the entanglement that the detector and the field acquired through the interaction, but it does not change the partial state of the field, as Eq.~\eqref{non-selective updated state as a trace} explicitly shows. This is of course a physically reasonable feature of the update: we have specified that the measurement is performed after the interaction is switched off, but it could be some arbitrary amount of time after this. The physical change of the field state due to the measurement is due only to the physical coupling between the detector and the field, and not to the fact that we decide to do a projective measurement on the detector after this interaction. This is because the projective measurement acts only on the detector once the interaction has been switched off, and it does not provide additional information since being non-selective the outcome is not known. This important interpretational point will be revisited when we consider the update rule for selective measurements, where the state of the field has to be updated consistently with the concrete outcome of the measurement.

%would be just another form of faster-than-light signalling, which Eqs.~\eqref{non-selective updated state as a trace} and \eqref{A rho phi} rule out nevertheless.

\section{The update rule}\label{section: the update rule}

In the previous section we have shown that the process of measuring a quantum field through locally coupling an Unruh-DeWitt detector and then carrying out an idealized measurement on the detector---which corresponds to a field state update with the appropriate (non-selective) POVM---does not introduce causality violations. We are now in a position to build an update rule for the field when we assume that the experimenter knows the concrete outcome of the idealized measurement carried out on the detector, which is akin to considering what is the field state update induced by a selective measurement on the detector after the detector finished interacting with the field.  

\subsection{Issues of a non-contextual update}

Prescribing an update rule for selective measurements in a way that is compatible with the relativistic nature of QFT requires more care than in regular quantum mechanics. An update rule for selective measurements based on particle detectors should:
\begin{itemize}
\item[(1)]{Include the knowledge of the measurement outcome in the description of the field and implement the compatibility between measurements that are sequentially applied to the field, in the spirit of L\"uders rule~\cite{Luders1951}.}
\item[(2)]{Be compatible with relativity.}
\end{itemize}
To guarantee that condition (1) is fulfilled, it is necessary to use the update of the state of the field given by~\eqref{updated state field}. However, in Appendix~\ref{appendix: causal behaviour of the selective update} we show that this update cannot be applied outside the causal future of the detector in a way consistent with relativity. Hence, we see that any non-contextual update (i.e. an update where one gives the density operator a global nature and its change affects all observers regardless of whether they are in causal contact with the detector or not) cannot satisfy conditions (1) and (2) simultaneously. To bypass this difficulty, a first attempt that one could try is to prescribe that the selective update given by~\eqref{updated state field} should only be used in the causal future of the measurement. This prescription, however, is ill-defined since the density operator does not naturally depend on points of the spacetime manifold. In particular, this kind of prescription does not provide a way to calculate arbitrary $n$-point functions, since by naively looking at the formula $w_n(\mathsf{x},\mathsf{x}',\hdots)=\textrm{tr}_{\phi}\big(\hat\sigma_{\phi} \hat \phi(\mathsf{x})\hat \phi(\mathsf{x'})\cdots\big)$---where $\hat\sigma_\phi$ is an arbitrary field state---it is not clear what density matrix $\hat{\sigma}_\phi$ we should use when considering points $\mathsf{x},\mathsf{x}'$ in regions of spacetime with different updates. 

We conclude that a non-contextual update that includes the information obtained from a selective measurement performed on the detector is at odds with relativistic causality. Instead, in order to satisfy conditions (1) and (2) we must partially give up on the physical significance of density operators \mbox{$\rhoh_{\phi}$ and $\rhoh_{\phi}^\textrm{u}$} as representatives of observer-independent field states and simply treat them as states of information about the field (much like it is done in quantum informational approaches to the measurement problem in quantum mechanics~\cite{Wheelerchapter1977,Wheelerchapter1996,Zehchapter2004,Spekkens2007,Leifer2014,Brukner2015}). This is precisely what the next subsection focuses on.

%, and in particular to its global character, by introducing a dependence on the observer.

\subsection{A contextual update rule}

As we just concluded, to properly formulate an update rule that is respectful of relativity we need to consider the field density operators to be observer-dependent. In particular, we propose that the update depends on the context of the observer, i.e. the information available to them  according to their position in spacetime. Moreover, because it depends on the observer, once they receive information about a measurement the update only takes place inside their causal future. It is perhaps interesting to remark here the distinction between \textit{the measurement}, that is performed by the experimenter, and \textit{the update}, that is performed by each observer according to the information they have about the field. It is in this sense that we say that the update is observer-dependent. As such, when we write that a certain observer updates their field state, we mean that they are updating their information about the field and changing the field density operator that describes the field state for them, without acting upon the field in any way whatsoever. This operational approach can be summarized as follows:
\begin{enumerate}
    \item{After an experimenter provided with a detector performs a projective measurement on the detector, an observer that becomes aware that the measurement has been performed can either have information about its outcome or not. If they do, they apply the selective update ~\eqref{updated state field}; if they do not, they apply the non-selective update~\eqref{non-selective}. Both updates take place in the causal future of the observer.}
    \item{If an observer is spacelike separated from the interaction region, at most they can be aware of the measurement being performed, but they do not have access to the outcome of the measurement. Their update, if anything, should be non-selective, and we have already seen that the non-selective update does not have any effect on the outcome of spacelike separated operations. Hence, the spacelike separated observer does not have to take into account at all that a measurement has been performed. As it is desirable in a relativistic measurement theory, spacelike operations do not affect each other.}
    \item To update the $n$-point functions we need to take into account where the information of the measurement is accessible. As such, the $n$-point functions will only be non-trivially updated (selectively or non-selectively) if any point of their $n$ arguments is in the causal future of the measurement region. This will be addressed in-depth in Section~\ref{section: update of n-point functions}.
 %   \item{Every non-local observable can only be considered in the intersection of the causal futures of the sets where it has support, since that is the only region of spacetime where information from all the sets involved is accessible. In particular, this provides a meaningful way of calculating arbitrary $n$-point functions, as will be addressed more in depth in Section~\ref{section: update of n-point functions}.}
\end{enumerate}

This update rule respects causality by fiat, and its consistency for spacelike separated observers is guaranteed by the fact that the non-selective update is causal. However, since it only updates the state in the causal future of the detector, one could legitimately wonder if the measurement prescription takes into account the correlations present in spacelike separated regions of the field that are well-known to exist~\cite{ReehSchlieder,Redhead1995,Summers1985,Summers1987}. Condition 3 tells us how to proceed in order to ensure that this is the case. Consider two experimenters, Alba and Blanca, each provided with their own detector. The initial state of Alba's detector is \mbox{$\rhoh_{\textsc{a}}=\ket{\xi}\!\bra{\xi}$}, while Blanca's is \mbox{$\rhoh_{\textsc{b}}=\ket{\zeta}\!\bra{\zeta}$}. While being spacelike separated, they perform measurements, i.e. 1)~they couple their detectors to the field and 2)~after switching off the interaction they perform a projective measurement on the detectors and update their field states selectively with the information obtained in their local measurements. Alba gets a result associated to state $\ket{a}$ of her detector, while Blanca gets another associated to $\ket{b}$. Their corresponding updates are
\begin{equation}
    \rhoh_{\phi}^{\textsc{a}}=\frac{\hat{M}_{a,\xi}\rhoh_{\phi}\hat{M}_{a,\xi}^{\dagger}}{\textrm{tr}_{\phi}\big(\rhoh_{\phi}\hat{E}_{a,\xi}\big)}  \quad \textrm{and}\quad \rhoh_{\phi}^{\textsc{b}}=\frac{\hat{M}_{b,\zeta}\rhoh_{\phi}\hat{M}_{b,\zeta}^{\dagger}}{\textrm{tr}_{\phi}\big(\rhoh_{\phi}\hat{E}_{b,\zeta}\big)} \;.
\end{equation}
In the future, they eventually meet and inform each other of their results. Their final updates based on the exchanged information are as follows: for Alba,
\begin{equation}\label{rho AlbaBlanca}
    \rhoh_{\phi}^{\textsc{ab}}=\frac{\hat{M}_{b,\zeta}\rhoh_{\phi}^{\textsc{a}}\hat{M}_{b,\zeta}^{\dagger}}{\textrm{tr}_{\phi}\big(\rhoh_{\phi}^{\textsc{a}}\hat{E}_{b,\zeta}\big)}=\frac{\hat{M}_{b,\zeta}\hat{M}_{a,\xi}\rhoh_{\phi}\hat{M}_{a,\xi}^{\dagger}\hat{M}_{b,\zeta}^{\dagger}}{\textrm{tr}_{\phi}\big( \rhoh_{\phi}\hat{E}_{a,\xi} \big)\textrm{tr}_{\phi}\big(\rhoh_{\phi}^{\textsc{a}}\hat{E}_{b,\zeta}\big)} \;,
\end{equation}
while for Blanca
\begin{equation}\label{rho BlancaAlba}
    \rhoh_{\phi}^{\textsc{ba}}=\frac{\hat{M}_{a,\xi}\rhoh_{\phi}^{\textsc{b}}\hat{M}_{a,\xi}^{\dagger}}{\textrm{tr}_{\phi}\big(\rhoh_{\phi}^{\textsc{b}}\hat{E}_{a,\xi}\big)}=\frac{\hat{M}_{a,\xi}\hat{M}_{b,\zeta}\rhoh_{\phi}\hat{M}_{b,\zeta}^{\dagger}\hat{M}_{a,\xi}^{\dagger}}{\textrm{tr}_{\phi}\big( \rhoh_{\phi}\hat{E}_{b,\zeta} \big)\textrm{tr}_{\phi}\big(\rhoh_{\phi}^{\textsc{b}}\hat{E}_{a,\xi}\big)} \;.
\end{equation}
%Their updates have to be consistent, since once they meet they have the same information, that is,
%\begin{equation}\label{consistency of updates}
%    \rhoh_{\phi}^{AB}=\rhoh_{\phi}^{BA} \;.
%\end{equation}
Now, taking into account the form of the $\hat{M}$ operators~\eqref{M operator}, in terms of the unitary~\eqref{unitary} and therefore the Hamiltonian~\eqref{UDW Hamiltonian}, it is straightforward to prove that if Alba's and Blanca's measurements are carried out in spacelike separated regions, then 
\begin{equation}\label{Ms commute!!}
    [\hat{M}_{a,\xi},\hat{M}_{b,\zeta}]=[\hat{M}_{a,\xi},\hat{M}_{b,\zeta}^{\dagger}]=0 \;.
\end{equation}
This means that the numerators in \eqref{rho AlbaBlanca} and \eqref{rho BlancaAlba} are the same. Since the denominators are normalization factors, we first conclude that the updates are consistent. Once they meet they have the same information, and indeed it holds that
\begin{equation}\label{consistency of updates}
    \rhoh_{\phi}^{\textsc{ab}}=\rhoh_{\phi}^{\textsc{ba}} \;.
\end{equation}
Moreover, by \eqref{Ms commute!!}
\begin{equation}
    \textrm{tr}_{\phi}\big(\hat{M}_{b,\zeta}\hat{M}_{a,\xi}\rhoh_{\phi}\hat{M}_{a,\xi}^{\dagger}\hat{M}_{b,\zeta}^{\dagger}\big)=\textrm{tr}_{\phi}\big( \rhoh_{\phi}\hat{E}_{a,\xi}\hat{E}_{b,\zeta} \big)
\end{equation}
so that we can write
\begin{equation}\label{correlations are there}
    \textrm{tr}_{\phi}\big( \rhoh_{\phi}^{\textsc{a}}\hat{E}_{b,\xi} \big)=\frac{\textrm{tr}_{\phi}\big( \rhoh_{\phi}\hat{E}_{a,\xi}\hat{E}_{b,\zeta} \big)}{\textrm{tr}_{\phi}\big( \rhoh_{\phi}\hat{E}_{a,\xi} \big)} \neq \textrm{tr}_\phi\big( \rhoh_\phi \hat{E}_{b,\xi} \big) \;.
\end{equation}
For a POVM, the probability of getting an outcome $r$ from a generic state $\hat{\sigma}_\phi$ is the trace $\textrm{tr}_{\phi}(\hat{\sigma}_\phi \hat{E}_{r})$, where $\hat{E}_{r}$ is the POVM element associated to the outcome $r$~\cite{Nielsen2010}. This means that \eqref{correlations are there} displays the correlations between the measurements due to the initial correlations in the field state. Indeed, Eq.~\eqref{correlations are there} can be read in terms of probabilities as
\begin{align}
    \textrm{Prob}(& \textrm{Blanca gets }b \;|\; \textrm{Alba gets }a ) \nonumber \\
    &\phantom{==}=\frac{\textrm{Prob}(\textrm{Alba gets $a$ and Blanca gets $b$})}{\textrm{Prob}(\textrm{Alba gets }a)} \\
    &\phantom{==}\neq \textrm{Prob}(\textrm{Blanca gets }b) \;, \nonumber 
\end{align}
where the first equality is the formula for conditional probability, and in particular shows that Alba's and Blanca's outcomes are not independent.

We conclude that the proposed update rule respects causality, is consistent for spacelike separated measurements (and trivially for timelike separated measurements), and accurately accounts for spacelike correlations. Therefore it is a suitable contextual rule for updating the state of the field after measuring with detectors in QFT.

It is remarkable that the proposed update rule, for a particular observer, is somewhat similar in its structure to that proposed by Hellwig and Kraus~\cite{Hellwig1970formal}. The formalism proposed in our work, however, relies on particle detectors instead of on local projections, establishing a direct connection with experiments~\cite{Edu2013,Rodriguez2018,Lopp2020,Pozas2016}. %Moreover, the state of the field represented by the density operator is not in the center of this does noit does not assign the central role in the description of evolution, by prescribing that it be updated independently by each observer according to the information available to them, thus making the update rule explicitly causality-abiding.

It should be noted that by giving up on density operators as global descriptions of the field state we are displacing the focus from the Hilbert space description to another based on what experimenters measure and the correlations between the possible measurements. This is precisely the approach adopted in algebraic quantum field theory (see e.g.~\cite{Haag1996,Hollands2015,AdvancesAQFT,Fewster2019AQFT}), where the \textit{algebraic state} is interpreted to be the complex linear form that associates to each observable\footnote{As an element of the direct limit of the net of local algebras~\cite{Fewster2019AQFT}.} its expectation value. Indeed, the contextual update rule described above should be given just in terms of updated $n$-point functions, as we will show in the next section.

\section{Update of n-point functions}\label{section: update of n-point functions}

In a free quantum field theory, the state of the field  can be described in two interchangeable ways: either by a density operator in some particular Hilbert space representation or, equivalently, by the set of the field $n$-point functions. However, in Section~\ref{section: the update rule}, we have argued that there are serious difficulties to apply a selective update to a field density operator because of the incompatibility with a context-independent description. Fortunately, the formalism of $n$-point functions is still adequate for describing the contextual update rule proposed in the previous section. In the present section, we will formulate the update rule proposed in the previous section explicitly in terms of the $n$-point functions that fully characterize the state of the field. The $n$-point functions directly appear in the most common expressions for the response of particle detectors (see e.g.~\cite{Takagi1986,Louko2006,Satz2007} among others), so having an update rule for all the $n$-point functions not only fully characterizes the updated state but is also of practical interest for any calculations involving particle detectors. 

Notice that for the update after the measurement to be given \textit{just} as an update of $n$-point functions, we need to initially assume that in our particular experiment the only relevant systems are the field and the detector. If the field is entangled with a third party in the past of the detector, we will assume for now that this third party will not be addressed in this measurement experiment, leaving the more complicated case for future sections\footnote{This initial assumption can indeed be relaxed: treated with some care, the update rule for $n$-point functions which we are about to formulate can also be used in arbitrarily general scenarios. The reason is that the scheme given in Section~\ref{section: the update rule} applies to arbitrary states $\rhoh_{\phi}$ that may be extended to include third-party systems in addition to the field. We will show how this more general scenario can be straightforwardly dealt with in Section~\ref{section: generalization to the presence of entangled third parties}. }.

For this section, this simplifying constraint will allow us to ``forget'' about the causal past of the measurement and define the update only in the region of spacetime outside of it. %We thus shall not consider $n$-point functions involving points in the causal past of the detector. 
In the spirit of the discussion in the previous section, we will distinguish whether the measurement performed on the detector is non-selective or selective. 

\subsection{Non-selective update}\label{subsection: non-selective update}

The non-selective update is straightforward to implement from the state update in Section~\ref{section: causal behaviour}. Since, as we showed, non-selective updates do not affect the state in regions spacelike separated from the measurement, there is no need to prescribe different updates whether the arguments are in the causal future of the detector or in the spacelike separated region. Hence, the updated $n$-point function is
\begin{align}\label{non-selective updated n-point function non-perturbative}
w_{n}^{\textsc{NS}}(\mathsf{x}_1,\hdots,\mathsf{x}_n)&=\textrm{tr}_{\phi}\big( \rhoh_{\phi}^\textrm{u}\phih(\mathsf{x}_1)\cdots\phih(\mathsf{x}_n) \big) \nonumber \\ 
&=\langle \hat{M}_{s,\psi}^{\dagger}\phih(\mathsf{x}_1)\cdots\phih(\mathsf{x}_n)\hat{M}_{s,\psi}\rangle_{\rhoh_{\phi}}\\
&\phantom{==}+\langle \hat{M}_{\bar{s},\psi}^{\dagger}\phih(\mathsf{x}_1)\cdots\phih(\mathsf{x}_n)\hat{M}_{\bar{s},\psi}\rangle_{\rhoh_{\phi}} \nonumber
\end{align}
for every $\mathsf{x}_1,\hdots,\mathsf{x}_n \in \mathcal{M}$ outside the causal past of the interaction region. This update can be given explicitly in terms of the $n$-point functions of the initial state of the field. In particular, for the one-point function and to first order in $\lambda$,
\begin{align}\label{non-selective updated one-point function}
&w_{1}^{\textsc{NS}}(t_1,\bm{x}_1)=w_{1}(t_1,\bm{x}_1) \nonumber\\
&\;+2\lambda\int\diff t\,\diff^{d}{\bm{x}}\,\chi(t)F(\bm{x})\bra{\psi}\hat{\mu}(t)\ket{\psi}\\
&\phantom{=========}\times\Im(w_{2}(t_1,\bm{x}_1,t,\bm{x})) \nonumber\\
&\;+O(\lambda^2) \nonumber
\end{align}
where $w_{n}$ is the $n$-point function of the initial state of the field $\rhoh_{\phi}$. Analogously, for the two-point function, 
\begin{align}\label{non-selective updated two-point function}
&w_{2}^{\textsc{NS}}(t_1,\bm{x}_1,t_2,\bm{x}_2)=w_{2}(t_1,\bm{x}_1,t_2,\bm{x}_2) \nonumber\\
&\;+\ii\lambda\int\diff t\,\diff^{d}{\bm{x}}\,\chi(t)F(\bm{x})\bra{\psi}\hat{\mu}(t)\ket{\psi}\\
&\;\times \big( w_3(t,\bm{x},t_1,\bm{x}_1,t_2,\bm{x}_2)-w_3(t_1,\bm{x}_1,t_2,\bm{x}_2,t,\bm{x}) \big) \nonumber\\[2mm]
&\;+O(\lambda^2) \;. \nonumber
\end{align}
And in general,
\begin{align}\label{non-selective updated n-point function}
&w_{n}^{\textsc{NS}}(t_1,\bm{x}_1,\hdots,t_n,\bm{x}_n)=w_{n}(t_1,\bm{x}_1,\hdots,t_n,\bm{x}_n) \nonumber \\
&\;+\ii\lambda\int\diff t\,\diff^{d}{\bm{x}}\,\chi(t)F(\bm{x})\bra{\psi}\hat{\mu}(t)\ket{\psi} \nonumber\\
&\;\times \big( w_{n+1}(t,\bm{x},t_1,\bm{x}_1,\hdots,t_n,\bm{x}_n)\\[2mm]
&\phantom{=======}-w_{n+1}(t_1,\bm{x}_1,\hdots,t_n,\bm{x}_n,t,\bm{x}) \big) \nonumber \\[2mm]
&\;+O(\lambda^2) \;. \nonumber
\end{align}
The details of these calculations can be seen in Appendix~\ref{appendix: update rules for n-point functions}. The second order terms in $\lambda$ for the previous perturbative expressions are also displayed in Eq.~\eqref{non-selective updated n-function up to second order in lambda} of Appendix~\ref{appendix: update rules for n-point functions}. It is worth remarking that microcausality ensures that Eqs.~\eqref{non-selective updated one-point function}, \eqref{non-selective updated two-point function} and \eqref{non-selective updated n-point function} reduce to the unchanged $n$-point function whenever their arguments are outside the causal future of the detector, since in that case \mbox{$[\phih(t,\bm{x}),\phih(t_j,\bm{x}_j)]=0$} for every $j\in\{1,\hdots,n\}$ and therefore
\begin{align}
&w_{n+1}(t,\bm{x},t_1,\bm{x}_1,\hdots,t_n,\bm{x}_n) \nonumber\\
&\phantom{===}=\textrm{tr}_\phi\big(\rhoh_\phi \phih(t,\bm{x})\phih(t_1,\bm{x}_1)\cdots\phih(t_n,\bm{x}_n)\big)  \\
&\phantom{===}=\textrm{tr}_\phi\big(\rhoh_\phi \phih(t_1,\bm{x}_1)\cdots\phih(t_n,\bm{x}_n)\phih(t,\bm{x})\big) \nonumber\\
&\phantom{===}=w_{n+1}(t_1,\bm{x}_1,\hdots,t_n,\bm{x}_n,t,\bm{x}) \;.\nonumber
\end{align}

\subsection{Selective update}\label{subsection: selective update}

For selective measurements, we will first present the update for the one-point function. Second, we will consider the two-point function, which involves a few subtleties that deserve attention. And finally, with the one-point and two-point functions as landmarks, we will generalize the update scheme to $n$-point functions. As before, the details of the calculations (as well as results at higher orders in the coupling strength) can be found in Appendix~\ref{appendix: update rules for n-point functions}. 

\subsubsection{One-point function}\label{subsubsection: one-point function}

As shown in Appendix~\ref{appendix: causal behaviour of the selective update}, when dealing with selective measurements the update cannot be applied outside the causal future of the detector. Moreover, it should be noticed that in full rigour, and unlike for non-selective measurements, the selective update does depend on the region of spacetime in which the projective measurement on the detector is performed, whose future we shall denote $\mathcal{P}$. Therefore we have to distinguish three cases depending on the argument \mbox{$\mathsf{x}_1\in\mathcal{M}$} of the one-point function: 
\begin{itemize}
\item[(a)] If $\mathsf{x}_1\in\mathcal{P}$, then we should consider the state of the field to be updated by the selective rule \eqref{updated state field}.
\item[(b)] If $\mathsf{x}_1\in\mathcal{J}^{+}(\mathcal{D})\setminus\mathcal{P}$, with $\mathcal{J}^+(\mathcal{D})$ being the causal future of the interaction region\footnote{Note that, since the projective measurement on the detector is performed in the causal future of the interaction region, \mbox{$\mathcal{P}\subset\mathcal{J}^+(\mathcal{D})$}.} as defined in~\eqref{causal future of the interaction region}, then we only have to take into account the interaction, which as shown in~\eqref{non-selective updated state as a trace} yields the same update as the non-selective rule~\eqref{non-selective}.
\item[(c)] If $\mathsf{x}_1$ is spacelike separated from $\mathcal{D}$ (i.e. it is outside the causal support of the interaction region, \mbox{$\mathsf{x}_1\notin\mathcal{J}(\mathcal{D})$}), then we should use the initial state of the field, or equivalently the non-selective update.
\end{itemize}
However, we saw in Subsection~\ref{subsection: non-selective update} that the non-selective update can be safely applied to spacelike separated regions. Therefore, we can consider cases (b) and (c) jointly when prescribing the update rule. All together, 
\begin{equation}\label{selective updated one-point function def}
w_{1}^{\textsc{S}}(\mathsf{x}_1)= 
\begin{dcases*}
\frac{\langle \hat{M}_{s,\psi}^{\dagger}\phih(\mathsf{x}_1)\hat{M}_{s,\psi} \rangle_{\rhoh_{\phi}}}{\langle \hat{E}_{s,\psi}\rangle_{\rhoh_{\phi}}}  & if $\mathsf{x}_1\in\mathcal{P}$, \\[1mm]
w_{1}^{\textsc{NS}}(\mathsf{x}_1)  & otherwise.\\
\end{dcases*} 
\end{equation}
Note that all expectation values are calculated for the initial state of the field, $\rhoh_{\phi}$. For the case in which $\mathsf{x}_1\in\mathcal{P}$, we have used~\eqref{updated state field} and the cyclic property of trace. Therefore, the update can be given in terms of the $n$-point functions of the initial state of the field. In particular, if $\braket{s}{\psi}\neq 0$, to first order in $\lambda$,
\begin{align}\label{selective update one-point function perturbative}
&w_{1}^{\textsc{S}}(t_{1},\bm{x}_{1})= w_{1}(t_{1},\bm{x}_{1})+\frac{2\lambda}{|\!\braket{s}{\psi}\!|^2}\int\diff t\,\diff^{d} \bm{x}\,\chi(t) \nonumber\\
&\;\times F(\bm{x})\,\textrm{Im}\big[\braket{\psi}{s}\bra{s}\hat{\mu}(t)\ket{\psi} \big( w_{2}(t_1,\bm{x}_1,t,\bm{x}) \\[1mm]
&\;\;\;\;\;-w_1(t_1,\bm{x}_1)w_{1}(t,\bm{x})\big) \big]+O(\lambda^2) \nonumber
\end{align}
whenever $(t_1,\bm{x}_1)\in\mathcal{P}$. The more cumbersome case in which $\braket{s}{\psi}=0$ is displayed in Eq.~\eqref{selective update n-point function orthogonal} of Appendix~\ref{appendix: update rules for n-point functions}, along with the case $\braket{s}{\psi}\neq 0$ up to order 2 in $\lambda$, that can be seen in Eq.~\eqref{second order of selective update n-point functions non-orthogonal}. 

\subsubsection{Two-point function}\label{subsubsection: two-point function}

For prescribing the update of the two-point function we also need to distinguish different cases. Following the same spirit of the prescription of the one-point function, when both arguments $\mathsf{x}_1,\mathsf{x}_2 \in \mathcal{P}$ are in the causal future of the projective measurement, we consider the field state to be updated by~\eqref{updated state field}, while if both $\mathsf{x}_1,\mathsf{x}_2$ are outside $\mathcal{P}$, the information of the measurement cannot propagate to those points and therefore we should use the non-selective update of the field state to calculate the expectation value. However, what should we do when we have a mixed situation (e.g. if $\mathsf{x}_1 \in \mathcal{P}$ and $\mathsf{x}_2\notin\mathcal{P}$)? %To figure it out, we can look at the circumstances in which this kind of argument for the two-point function becomes relevant. 
First, note that the two-point function is a non-local object that is only relevant in non-local experiments (for example, coordinating several labs around the world, or an interaction that is extended in space). However, it is only pertinent to ask about the result of a non-local experiment if we assume that the information obtained by the different measurements can be combined in a ``processing'' region\footnote{Notice the similarity with the notion of  processing region introduced in~\cite{Ruep2021}.} that intersects the causal futures of all the experiments. It is reasonable then that when the two-point function has mixed arguments inside and outside $\mathcal{P}$, the information about the outcome of the selective measurement is accessible to the processing region, as it has to have a non-zero intersection with $\mathcal{P}$. Hence, as long as one of the points of the two-point function is inside $\mathcal{P}$ we must use the selective update of the field state.

This is consistent with treating the field state as a state of information about the field. To update the field in accordance with the outcome of a measurement we need to look at where in spacetime the information obtained in the measurement can be accessed. Conversely, if an observer never accesses the causal future of a region in spacetime, it does not make sense for them to ask about the correlations between the field in that region and the region they have access to\footnote{For example, if two observers never get to communicate, directly or indirectly, it lacks physical meaning that they can ask any question that involves the correlations between their operations.}. All this considered, we shall prescribe the selective update for the two-point function as
\begin{equation}\label{selective updated two-point function def}
w_{2}^{\textsc{S}}(\mathsf{x}_1,\mathsf{x}_2)\!=\! 
\begin{dcases*}
\frac{\langle \hat{M}_{s,\psi}^{\dagger}\phih(\mathsf{x}_1)\phih(\mathsf{x}_2)\hat{M}_{s,\psi} \rangle_{\rhoh_{\phi}}}{\langle \hat{E}_{s,\psi}\rangle_{\rhoh_{\phi}}}  & if $\mathsf{x}_1$ or $\mathsf{x}_2\in\mathcal{P}$, \\
w_{2}^{\textsc{NS}}(\mathsf{x}_1,\mathsf{x}_2)  & otherwise.\\
\end{dcases*} 
\end{equation} 
%w_{2}^{\textsc{NS}}(\mathsf{x}_1,\mathsf{x}_2)  & $\mathsf{x}_1 \notin \mathcal{P}$, $\mathsf{x}_2 \notin \mathcal{P}$\\
Again, the update can be given in terms of the $n$-point functions of the initial state of the field. In particular, if $\braket{s}{\psi}\neq 0$, to first order in $\lambda$,
\begin{align}\label{selective update two-point function perturbative}
&w_{2}^{\textsc{S}}(t_1,\bm{x}_1,t_2,\bm{x}_2)=w_{2}(t_1,\bm{x}_1,t_2,\bm{x}_2) \nonumber\\
& +\frac{\lambda}{|\!\braket{s}{\psi}\!|^2}\int\diff t\,\diff^{d} \bm{x} \,\chi(t)F(\bm{x}) \nonumber \\
&\times\Big(  \ii\braket{s}{\psi}\!\bra{\psi}\hat{\mu}(t)\ket{s} w_3(t,\bm{x},t_1,\bm{x}_1,t_2,\bm{x}_2)\\[0.5mm]
&-\ii\braket{\psi}{s}\!\bra{s}\hat{\mu}(t)\ket{\psi}w_3(t_1,\bm{x}_1,t_2,\bm{x}_2,t,\bm{x}) \nonumber\\
&-2\Im(\braket{\psi}{s}\!\bra{s}\hat{\mu}(t)\ket{\psi})w_{2}(t_1,\bm{x}_1,t_2,\bm{x}_2)w_1(t,\bm{x}) \Big) \nonumber\\
&+O(\lambda^2) \nonumber 
\end{align}
whenever $(t_1,\bm{x}_1) \in \mathcal{P}$ or $(t_2,\bm{x}_2) \in \mathcal{P}$ (or both). As before, the case in which $\braket{s}{\psi}=0$ and the second order term of the previous expression are left to be displayed in Appendix~\ref{appendix: update rules for n-point functions}.

\subsubsection{n-point function}\label{subsubsection: n-point function}

The arguments given to justify the prescription for the two-point function immediately generalize to arbitrary \mbox{$n$-point} functions, for which the selective update is 
\begin{equation}\label{selective updated n-point function def 1}
w_{n}^{\textsc{S}}(\mathsf{x}_1,\hdots,\mathsf{x}_n)=w_{n}^{\textsc{NS}}(\mathsf{x}_1,\hdots,\mathsf{x}_n)  
\end{equation}
if all $\mathsf{x}_1,\hdots,\mathsf{x}_n$ are outside $\mathcal{P}$, and
\begin{equation}\label{selective updated n-point function def 2}
w_{n}^{\textsc{S}}(\mathsf{x}_1,\hdots,\mathsf{x}_n)= \frac{\langle \hat{M}_{s,\psi}^{\dagger}\phih(\mathsf{x}_1)\cdots\phih(\mathsf{x}_n)\hat{M}_{s,\psi} \rangle_{\rhoh_{\phi}}}{\langle \hat{E}_{s,\psi}\rangle_{\rhoh_{\phi}}} 
\end{equation}
otherwise. Once again, the update can be given in terms of the $n$-point functions of the initial state of the field, and in particular, if $\braket{s}{\psi}\neq 0$, to first order in $\lambda$,
\begin{align}\label{selective update n-point function perturbative}
&w_{n}^{\textsc{S}}(t_1,\bm{x}_1,\hdots,t_n,\bm{x}_n)=w_{n}(t_1,\bm{x}_1,\hdots,t_n,\bm{x}_n)\\
& +\frac{\lambda}{|\!\braket{s}{\psi}\!|^2}\int\diff t\,\diff^{d} \bm{x} \,\chi(t)F(\bm{x}) \nonumber\\
&\times\Big(  \ii\braket{s}{\psi}\!\bra{\psi}\hat{\mu}(t)\ket{s} w_{n+1}(t,\bm{x},t_1,\bm{x}_1,\hdots,t_n,\bm{x}_n) \nonumber\\[0.5mm]
&-\ii\braket{\psi}{s}\!\bra{s}\hat{\mu}(t)\ket{\psi}w_{n+1}(t_1,\bm{x}_1,\hdots,t_n,\bm{x}_n,t,\bm{x}) \nonumber\\
&-2\Im(\braket{\psi}{s}\!\bra{s}\hat{\mu}(t)\ket{\psi})\,w_{n}(t_1,\bm{x}_1,\hdots,t_n,\bm{x}_n)\,w_1(t,\bm{x}) \Big) \nonumber\\
&+O(\lambda^2) \; \nonumber
\end{align}
whenever $(t_i,\bm{x}_i) \in \mathcal{P}$ for some \mbox{$i\in\{1,\hdots,n\}$}. Just as before, the second order terms and the more tedious case in which $\braket{s}{\psi}=0$ can be found in Appendix~\ref{appendix: update rules for n-point functions}.

\section{Generalization to the presence of entangled third parties}\label{section: generalization to the presence of entangled third parties}

In the previous sections the analysis was performed considering that the initial entanglement of the field with systems other than the detector is not addressable and hence irrelevant for the scenarios considered. However, the update rule given in sections~\ref{section: the updated state of the field} and~\ref{section: the update rule} is not restricted to these situations. Generalizing beyond these situations is rather straightforward and conceptually identical to the prescription given in previous sections. For completeness, we will show here how the prescribed update rule can be generalized to the case in which there are other physical systems apart from the field and the detector that are relevant for the experiments under analysis. %This generalization, as will become clear, is conceptually identical to the previous prescriptions. It is nevertheless worth developing it explicitly for the sake of completeness.}%In this section we will clarify that the applicability of the update rule given in sections~\ref{section: the updated state of the field} and~\ref{section: the update rule} is not restricted to situations in which 

First, note that in Section~\ref{section: the updated state of the field} we considered the initial state of the system to be $\rhoh=\rhoh_\textrm{d}\otimes\rhoh_{\phi}$ for the sake of simplicity, since these two systems are the only ones involved in the measurement. But it should be immediately realized that we can consider general initial states of the form $\rhoh=\rhoh_{\textrm{d}}\otimes\rhoh_{\Phi}$, where $\rhoh_{\Phi}$ is the joint state of the field and all the other physical systems with which it might share entanglement that may be relevant for our experiment. For simplicity of the treatment, let us first assume that all of them are non-relativistic, in the sense that their individual dynamics can be dealt with using non-relativistic quantum mechanics and in particular they are localized. We will relax this assumption and allow for the presence of other relativistic fields at the end of the section. 

Let us denote the relevant physical systems that are not the field or the detector by $\Sigma=\{\Sigma_1,\Sigma_2,\hdots\}$. The derivation of the updated state for $\rhoh_{\Phi}$ proceeds as shown in Section~\ref{section: the updated state of the field}. The only difference that we need to take into account is that now, if $\hat{\mathcal{O}}$ is an operator acting on the Hilbert space of the whole system Hilbert space (detector, field and $\Sigma$) and $\ket{\psi_1},\ket{\psi_2}$ are detector states, then we should understand $\bra{\psi_1}\hat{\mathcal{O}}\ket{\psi_2}$ to be an operator acting on the Hilbert space $\mathcal{H}_\Phi$ of the field and the systems in $\Sigma$, such that
\begin{equation}\label{notation for sandwiched operators extended}
\bra{\Phi_1}\bra{\psi_1}\hat{\mathcal{O}}\ket{\psi_2}\ket{\Phi_2}=\bra{\psi_1,\Phi_1}\hat{\mathcal{O}}\ket{\psi_2,\Phi_2}
\end{equation}
for any $\ket{\Phi_1},\ket{\Phi_2}\in\mathcal{H}_\Phi$. Clearly, Eq.~\eqref{notation for sandwiched operators extended} is the generalization of Eq.~\eqref{notation for sandwiched operators}. It is straightforward to check that all the prescriptions given in Section~\ref{section: the update rule} still apply after this direct generalization has been made, taking into account the extra systems when keeping track of the available information. However, in this more general setup, giving the update solely in terms of $n$-point functions as in Section~\ref{section: update of n-point functions} would no longer be possible. Nevertheless, we can consider ``joint'' \textit{extended} $n$-point functions of the joint system as follows: notice, first of all, that just as any observable of the field can be expressed in terms of the field operators, any observable of the systems in a subset \mbox{$\Gamma\subseteq\Sigma$} can be expressed in terms of the rank-one operators $\proj{\gamma_l}{\gamma_m}$, as
\begin{equation}\label{observable decomposition non-relativistic system}
\hat{\mathcal{O}}_{\Gamma}=\sum_{l,m}\bra{\gamma_{l}}\hat{\mathcal{O}}_{\Gamma}\ket{\gamma_m}\proj{\gamma_l}{\gamma_{m}}
\end{equation}
for an orthonormal basis $\{\ket{\gamma_{l}}\}$ of the Hilbert space of $\Gamma$.
Thus, any operator acting on the field and $\Gamma$ can be expressed in terms of the field operators $\phih(\mathsf{x})$ and the rank-one operators $\proj{\gamma_{l}}{\gamma_m}$. We can therefore define the \textit{extended $n$-point functions} as
\begin{align}\label{extended n-point function definition non-relativistic}
\widetilde{w}_{\Gamma,n}(l,m&;\mathsf{x}_1,\hdots,\mathsf{x}_n)\\
&\coloneqq \textrm{tr}\big( \rhoh_{\Phi}\!\proj{\gamma_l}{\gamma_m}\phih(\mathsf{x}_1)\cdots\phih(\mathsf{x}_n)\big) \nonumber
\end{align}
for $n\geq 0$, where
\begin{equation}
\widetilde{w}_{\Gamma,0}(l,m)\coloneqq \textrm{tr}\big(\rhoh_{\Phi}\!\proj{\gamma_l}{\gamma_m}\big) \;.
\end{equation}
The extended $n$-point functions characterize $\rhoh_{\Phi}$. The update rule can now be given in terms of an update of the extended $n$-point functions, which can be shown to be just a modification of the update for $n$-point functions given in Section~\ref{section: update of n-point functions}. 

\paragraph{Non-selective update.}\label{paragraph: non-selective update} 

Since the non-selective update~\eqref{non-selective} is trace-preserving and it acts non-trivially only on the Hilbert space of the field, we can simply prescribe the same update of Eq.~\eqref{non-selective updated n-point function non-perturbative} for each of the extended $n$-point functions:
\begin{align}\label{non-selective updated extended n-point function}
\widetilde{w}&_{\Gamma,n}^{\textsc{NS}}(l,m;\mathsf{x}_1,\hdots,\mathsf{x}_n) \nonumber\\
&= \textrm{tr}\big(\hat{M}_{s,\psi}\rhoh_{\Phi}\hat{M}_{s,\psi}^{\dagger}\!\proj{\gamma_l}{\gamma_m} \phih(\mathsf{x}_1)\cdots\phih(\mathsf{x}_n) \big)\\
&\phantom{==}+\textrm{tr}\big(\hat{M}_{\bar{s},\psi}\rhoh_{\Phi}\hat{M}_{\bar{s},\psi}^{\dagger}\!\proj{\gamma_l}{\gamma_m} \phih(\mathsf{x}_1)\cdots\phih(\mathsf{x}_n) \big) \;. \nonumber 
\end{align}
As in Subsection~\ref{subsection: non-selective update}, this expression can be written in terms of the non-updated extended $n$-point functions. Note that, in particular, $\widetilde{w}_{\Gamma,0}^{\textsc{NS}}(l,m)=\widetilde{w}_{\Gamma,0}(l,m)$.

\paragraph{Selective update.}\label{paragraph: selective update}

Same as in Section~\ref{section: update of n-point functions}, the prescription of the update requires to keep track of where the information is accessible. This leads to a piecewise definition as in Eqs.~\eqref{selective updated n-point function def 1} and~\eqref{selective updated n-point function def 2}: let $\mathcal{P}$ be the causal future of the region in which the projective measurement on the detector is performed, 
\begin{equation}\label{selective updated extended n-point function def 1}
\widetilde{w}_{\Gamma,n}^{\textsc{S}}(l,m;\mathsf{x}_1,\hdots,\mathsf{x}_n)=\widetilde{w}_{\Gamma,n}^{\textsc{NS}}(l,m;\mathsf{x}_1,\hdots,\mathsf{x}_n)  
\end{equation}
if all $\mathsf{x}_1,\hdots,\mathsf{x}_n$ \textit{and} the systems of $\Gamma$ are outside $\mathcal{P}$, and
\begin{align}\label{selective updated extended n-point function def 2}
\widetilde{w}&_{\Gamma,n}^{\textsc{S}}(l,m;\mathsf{x}_1,\hdots,\mathsf{x}_n) \nonumber\\
&= \frac{\textrm{tr}\big(\hat{M}_{s,\psi}\rhoh_{\Phi}\hat{M}_{s,\psi}^{\dagger}\!\proj{\gamma_l}{\gamma_m} \phih(\mathsf{x}_1)\cdots\phih(\mathsf{x}_n) \big)}{\textrm{tr}_{\Phi}\big( \rhoh_{\Phi}\hat{E}_{s,\psi} \big)} 
\end{align}
otherwise. In particular,   
\begin{equation}\label{selective update rule extended order 0 def 1}
\widetilde{w}_{\Gamma,0}^{\textsc{S}}(l,m)=\widetilde{w}_{\Gamma,0}^{\textsc{NS}}(l,m)
\end{equation}\label{selective update rule extended order 0 def 2}
if the systems of $\Gamma$ are outside $\mathcal{P}$, and
\begin{equation}
\widetilde{w}_{\Gamma,0}^{\textsc{S}}(l,m)=\frac{\textrm{tr}\big(\hat{M}_{s,\psi}\rhoh_{\Phi}\hat{M}_{s,\psi}^{\dagger}\!\proj{\gamma_l}{\gamma_m}\big)}{\textrm{tr}_{\Phi}\big( \rhoh_{\Phi}\hat{E}_{s,\psi} \big)}
\end{equation}
otherwise. 

To end this section, we can address the case where the third parties sharing entanglement with the probed field are themselves relativistic fields. In that scenario, Eq.~\eqref{observable decomposition non-relativistic system} is not useful anymore, since for a basis of the Hilbert space of a field, the rank-one operators $\proj{\gamma_l}{\gamma_m}$ are not local objects and the update to the extended $n$-point function has to be defined over local regions of spacetime. %Considering that in order to properly prescribe the update of the extended $n$-functions we need its arguments to be local, the formulation above cannot be strictly applied.
Fortunately, the field itself is defined in terms of local observables. The local observables of a field $\sigma$ in $\Sigma$ can be expressed in terms of its associated field operators $\hat{\sigma}(\mathsf{x})$. Thus, for the simplest case in which the only system in $\Sigma$ is a field $\sigma$, we define the extended $n$-point function as an \textit{$(n',n)$-point function},
\begin{align}
&\widetilde{w}_{n'\!,n}(\mathsf{y}_{1},\hdots,\mathsf{y}_{n'};\mathsf{x}_1,\hdots,\mathsf{x}_{n}) \\
&\phantom{===}=\textrm{tr}\big( \rhoh_{\Phi}\,\hat{\sigma}(\mathsf{y}_1)\cdots\hat{\sigma}(\mathsf{y}_{n'}\!)\,\phih(\mathsf{x}_1)\cdots\phih_1(\mathsf{x}_{n})  \big) \;. \nonumber
\end{align}
This expression provides the extended $n$-point function that substitutes Eq.~\eqref{extended n-point function definition non-relativistic} for the case in which $\Sigma$ is one relativistic field. If there are more fields present, one can build the extended $n$-point function in an analogous fashion. Regarding the update rule, in the same spirit of the prescriptions given in Eqs.~\eqref{non-selective updated extended n-point function},\eqref{selective updated extended n-point function def 1} and \eqref{selective updated extended n-point function def 2}, 
\begin{align}\label{non-selective updated (n',n)-point function fields}
&\widetilde{w}_{n'\!,n}^{\textsc{NS}}(\mathsf{y}_1,\hdots,\mathsf{y}_{n'};\mathsf{x}_1,\hdots,\mathsf{x}_n) \nonumber\\
&=\textrm{tr}\big(\hat{M}_{s,\psi}\rhoh_{\Phi}\hat{M}_{s,\psi}^{\dagger}\hat{\sigma}(\mathsf{y}_1)\cdots\hat{\sigma}(\mathsf{y}_{n'}\!) \phih(\mathsf{x}_1)\cdots\phih(\mathsf{x}_n) \big)\\
&\;+\textrm{tr}\big(\hat{M}_{\bar{s},\psi}\rhoh_{\Phi}\hat{M}_{\bar{s},\psi}^{\dagger}\hat{\sigma}(\mathsf{y}_1)\cdots\hat{\sigma}(\mathsf{y}_{n'}\!) \phih(\mathsf{x}_1)\cdots\phih(\mathsf{x}_n) \big)  \nonumber 
\end{align}
for the non-selective case, while for the selective case,
\begin{align}\label{selective updated extended n-point function fields def 1}
\widetilde{w}_{n'\!,n}^{\textsc{S}}&(\mathsf{y}_1,\hdots,\mathsf{y}_{n'};\mathsf{x}_1,\hdots,\mathsf{x}_n) \nonumber\\
&=\widetilde{w}_{\Gamma,n}^{\textsc{NS}}(\mathsf{y}_1,\hdots,\mathsf{y}_{n'};\mathsf{x}_1,\hdots,\mathsf{x}_n)   
\end{align}
if all $\mathsf{x}_1,\hdots,\mathsf{x}_n$ \textit{and} $\mathsf{y}_1,\hdots,\mathsf{y}_{n'}$  are outside $\mathcal{P}$, and
\begin{align}\label{selective updated extended n-point function fields def 2}
&\widetilde{w}_{n'\!,n}^{\textsc{S}}(\mathsf{y}_1,\hdots,\mathsf{y}_{n'};\mathsf{x}_1,\hdots,\mathsf{x}_n) \nonumber\\
&= \frac{\textrm{tr}\big(\hat{M}_{s,\psi}\rhoh_{\Phi}\hat{M}_{s,\psi}^{\dagger}\,\hat{\sigma}(\mathsf{y}_1)\cdots\hat{\sigma}(\mathsf{y}_{n'}\!) \phih(\mathsf{x}_1)\cdots\phih(\mathsf{x}_n) \big)}{\textrm{tr}_{\Phi}\big( \rhoh_{\Phi}\hat{E}_{s,\psi} \big)} 
\end{align}
otherwise.

Finally, for the mixed case in which $\Sigma$ contains both localized non-relativistic systems and relativistic fields, we just need to use the natural combination of both formalisms, that includes rank-one operators of the form \mbox{$\proj{\gamma_l}{\gamma_m}$} for the non-relativistic systems and field operators $\hat{\sigma}(\mathsf{y})$ for the relativistic fields. 
%To end this section, observe that we can  extend this formalism to the case in which we have two or more interacting fields by considering ``joint'' {\color{red}\bf [EDU: FALTAN COSAS]} $(n_1,n_2,\hdots)$-functions,
%\begin{align}
%&\widetilde{w}_{n_1,n_2,\hdots}(\mathsf{x}_{1},\hdots,\mathsf{x}_{n_1};\mathsf{y}_1,\hdots,\mathsf{y}_{n_2};\hdots) \\
%&=\textrm{tr}\big( \rhoh_{\Phi}\,\phih_1(\mathsf{x}_1)\cdots\phih_1(\mathsf{x}_{n_1})\phih_2(\mathsf{y}_{1})\cdots\phih_2(\mathsf{y}_{n_2})  \big) \nonumber
%\end{align}
%and updating it in the straightforwardly analogous way.

\section{A practical example with detectors}\label{section: a practical example with detectors}

\begin{figure}[h]
\centering
\includegraphics[scale=1]{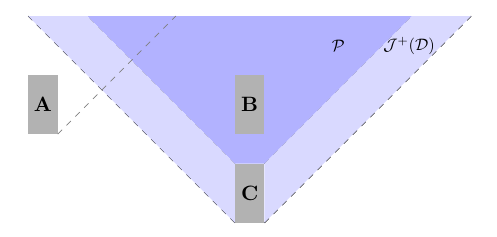}
\caption{Configuration in a slice of spacetime of the interaction regions of detectors A, B and C. In blue, the causal future of the projective measurement performed on C, $\mathcal{P}$; in pale blue, the causal future of the measurement that is not already in the future of the projective measurement, $\mathcal{J}^{+}(\mathcal{D})\setminus\mathcal{P}$.}\label{fig: example with three detectors}
\end{figure}

To further clarify how to use the formalism in a practical implementation we will consider an example involving three stationary experimenters, Alba, Blanca and Clara, each provided with a two-level Unruh-DeWitt detector. The situation, depicted in Figure~\ref{fig: example with three detectors}, is as follows\footnote{This configuration is a pretty archetypal setup in Relativistic Quantum Information in scenarios of entanglement harvesting, see e.g.~\cite{Reznik2005,Pozas2015} among many others.}:

\begin{itemize}
\item {Clara performs a measurement with her detector, by first letting it interact with the field and then performing a projective measurement on it, immediately after the interaction is switched off.}
\item {Blanca lets her detector interact with the field in the causal future of the projective measurement performed by Clara, $\mathcal{P}$.}
\item {Alba lets her detector interact with the field in a region that is spacelike separated from both Blanca's and Clara's interaction regions.}
\end{itemize}

%\begin{figure}[h]
%\begin{center}
%    \begin{tikzpicture}
%        \fill[gray!60!white] (-0.25,-0.5) rectangle (0.25,0.5);
%        \draw[dashed] (-0.25,-0.5) -- (-3.75,3);
%        \draw[dashed] (0.25,-0.5) -- (3.75,3);
%        \draw[dashed] (0.25,0.5) -- (2.75,3);
%        \draw[dashed] (-0.25,0.5) -- (-2.75,3);
%        \fill[blue!30!white] (-2.75,3) -- (-0.25,0.5) -- (0.25,0.5) -- (2.75,3); 
%        \fill[blue!15!white] (-3.75,3) -- (-0.25,-0.5) -- (-0.25,0.5) -- (-2.75,3);
%        \fill[blue!15!white] (3.75,3) -- (0.25,-0.5) -- (0.25,0.5) -- (2.75,3);
%        \node[] at (1.5,2.5) {\scalebox{0.8}{$\mathcal{P}$}};
%        \node[] at (2.7,2.5) {\scalebox{0.8}{$\mathcal{J}^{+}(\mathcal{D})$}};
%        \fill[gray!60!white] (-0.25,1) rectangle (0.25,2);
%        \fill[gray!60!white] (-3.75,1) rectangle (-3.25,2);
%        \node[] at (0,0) {\textbf{C}};
%        \node[] at (0,1.5) {\textbf{B}};
%        \node[] at (-3.5,1.5) {\textbf{A}};
%        \draw[dashed,gray] (-3.25,1) -- (-1.25,3);
%    \end{tikzpicture}
%   \caption{Configuration in a slice of spacetime of the interaction regions of detectors A, B and C. In blue, the causal future of the projective measurement performed on C, $\mathcal{P}$; in pale blue, the causal future of the measurement that is not already in the future of the projective measurement, $\mathcal{J}^{+}(\mathcal{D})\setminus\mathcal{P}$.}\label{fig: example with three detectors}
%\end{center}
%\end{figure}

We consider an initial state 
\begin{equation}
\rhoh=\rhoh_{\textsc{a}}\otimes\rhoh_{\textsc{b}}\otimes\rhoh_\textsc{c}\otimes \rhoh_\phi
\end{equation}
for the array of detectors and the field. For simplicity we have assumed that the detector that is measured starts out in a pure state, $\rhoh_\textsc{c}=\proj{\psi}{\psi}$. In the interaction picture, the interaction of the detectors with the field is given by the Hamiltonian 
\begin{equation}
\hat{H}_{I}(t)=\hat{H}_{\textsc{a}}(t)+\hat{H}_\textsc{b}(t)+\hat{H}_\textsc{c}(t) \;,
\end{equation}
where
\begin{equation}
\hat{H}_{\nu}(t)=\lambda_{\nu}\chi_{\nu}(t)\muh_{\nu}(t) \int\diff^d\bm{x}\,F_{\nu}(\bm{x})\phih(t,\bm{x})
\end{equation}
is the same Unruh-DeWitt Hamiltonian from Eq.~\eqref{UDW Hamiltonian}, for \mbox{$\nu\in\{\text{A},\textsc{B},\textsc{C}\}$}.
Now, since Clara's operations causally precede Blanca's, and since Alba is spacelike separated from both of them, the unitary operator that describes the evolution of the three detectors and the field
\begin{equation}
\hat{U}=\mathcal{T}\exp[-\ii\int_{-\infty}^{\infty}\!\!\!\!\diff t'\, \Big(\hat{H}_\textsc{a}(t')+\hat{H}_\textsc{b}(t')+\hat{H}_\textsc{c}(t')\Big)] \;
\end{equation}
can in fact be written as~\cite{Pipo2021}
\begin{equation}
\hat{U}=\hat{U}_\textsc{a}\hat{U}_\textsc{b}\hat{U}_\textsc{c}=\hat{U}_\textsc{b}\hat{U}_\textsc{c}\hat{U}_\textsc{a} \;,
\end{equation}
where
\begin{equation}
\hat{U}_\nu=\mathcal{T}\exp(-\ii\int_{-\infty}^{\infty}\!\!\!\!\diff t'\, \hat{H}_\nu(t')) \;
\end{equation}
for $\nu \in \{\textsc{A},\textsc{B},\textsc{C}\}$. %In particular, since the interaction $\hat{U}_\textsc{c}$ is already included in the $\hat{M}$ operators of the update prescription for Clara's measurement, the interaction of detectors A and B and the field ``after'' the measurement\footnote{Since Alba and Clara are spacelike separated, there is no notion of Alba's operations taking place ``after'' Clara's measurement. In this derivation it is more comfortable to proceed firstly updating the state after the measurement, and secondly considering the interaction of A (and B). Nevertheless, it will become clear when we arrive at the results that we would have obtained the same ones had we proceeded evolving the state due to Alba's interaction in the first place.} can be described by \mbox{$\hat{U}_\textsc{a}\hat{U}_\textsc{b}=\hat{U}_\textsc{b}\hat{U}_\textsc{a}$}.
In particular, we have that 
\begin{equation}\label{commutation evolutions A, B and C}
[\hat{U}_\textsc{a},\hat{U}_\textsc{b}]=[\hat{U}_\textsc{a},\hat{U}_\textsc{c}]=0 \;,
\end{equation}
and by Eq.~\eqref{M operator},
\begin{equation}\label{commutation evolution A measurement C}
[\hat{U}_\textsc{a},\hat{M}_{c,\psi}]=[\hat{U}_\textsc{a},\hat{M}_{c,\psi}^\dagger]=0 
\end{equation}
for any $\ket{c}\in\mathcal{H}_{\textsc{c}}$.

We are interested in studying how the measurement performed by Clara affects the joint partial state of Alba and Blanca, $\rhoh_{\textsc{ab}}$, as well as their individual partial states $\rhoh_{\textsc{a}}$ and $\rhoh_{\textsc{b}}$. Along the lines of previous sections, we distinguish whether the measurement performed by Clara is non-selective or selective. For the sake of clarity, in the main body of this section we will use the approach that uses a context-dependent density operator, as presented in Section~\ref{section: the update rule}, instead of the equivalent but more involved formulation based on $n$-point functions and its extensions, presented in Sections~\ref{section: update of n-point functions} and~\ref{section: generalization to the presence of entangled third parties}. Nevertheless, we have explicitly computed all the updates using the formulation of extended $n$-point functions in Appendix~\ref{appendix: a practical example using n-point functions}, showing explicitly that both methods give the same results.

\subsection{Non-selective measurement}\label{subsection: non-selective measurement}

Since it is less involved from the point of view of the update rule, let us first address the case in which Clara measures non-selectively. We will show that in the non-selective case the updated partial states of Alba and Blanca will coincide with the case where the three detectors interact with the field and we trace out the state of Clara's detector. That is, the only influence that Clara's detector has on $\hat\rho_{\textsc{ab}}$ is through its coupling to the field since no information about the measurement is assumed to be known by Alba and Blanca.

We already argued in Section~\ref{section: causal behaviour} that all observers are susceptible to being informed of the performance of the measurement without knowing its outcome. Thus, we can consider that both Alba and Blanca have access to the non-selective update of the state\footnote{Note that since Alba is spacelike separated from both Blanca and Clara, it does not matter whether we carry out first the update of Clara's measurement or the evolution due to the interaction of Alba's detector, as we saw in Section~\ref{section: causal behaviour} and becomes apparent in Eq.~\eqref{commutation evolution A measurement C}.}. For our purposes, it is simpler to consider the update of the measurement in the first place. Thus, we obtain a final joint state
\begin{align}\label{Alba-Blanca non-selective 1st derivation}
&\rhoh_{\textsc{ab}}'=\textrm{tr}_{\phi}\big( \hat{U}_{\textsc{a}}\hat{U}_{\textsc{b}} \big[ \rhoh_{\textsc{a}}\otimes\rhoh_{\textsc{b}}\otimes\hat{M}_{c,\psi}\rhoh_{\phi}\hat{M}_{c,\psi}^{\dagger} \big] \hat{U}_{\textsc{b}}^{\dagger}\hat{U}_{\textsc{a}}^{\dagger} \big) \nonumber\\
&\phantom{==}+\textrm{tr}_{\phi}\big[ \hat{U}_{\textsc{a}}\hat{U}_{\textsc{b}} \big( \rhoh_{\textsc{a}}\otimes\rhoh_{\textsc{b}}\otimes\hat{M}_{\bar{c},\psi}\rhoh_{\phi}\hat{M}_{\bar{c},\psi}^{\dagger} \big) \hat{U}_{\textsc{b}}^{\dagger}\hat{U}_{\textsc{a}}^{\dagger} \big] \\
&=\textrm{tr}_{\textsc{c},\phi}\big[ \hat{U}_{\textsc{a}}\hat{U}_{\textsc{b}}\hat{U}_{\textsc{c}}(\rhoh_{\textsc{a}}\otimes\rhoh_{\textsc{b}}\otimes\proj{\psi}{\psi}\otimes\rhoh_{\phi})\hat{U}_{\textsc{c}}^{\dagger}\hat{U}_{\textsc{b}}^{\dagger}\hat{U}_{\textsc{a}}^{\dagger} \big]  \;,\nonumber
\end{align}
where in the last step we used Eq.~\eqref{non-selective updated state as a trace}. As anticipated, this is the same result obtained for $\rhoh_\textsc{ab}$ in the case in which Clara does not perform a projective measurement on the detector at all. The partial states are
\begin{align}\label{Blanca non-selective 1st derivation}
\rhoh_{\textsc{b}}'&=\textrm{tr}_{\textsc{a}}(\rhoh'_{\textsc{ab}}) \\
&=\textrm{tr}_{\textsc{c},\phi}\big[ \hat{U}_{\textsc{b}}\hat{U}_{\textsc{c}}(\rhoh_{\textsc{b}}\otimes\proj{\psi}{\psi}\otimes\rhoh_{\phi})\hat{U}_{\textsc{c}}^{\dagger}\hat{U}_{\textsc{b}}^{\dagger} \big] \nonumber
\end{align}
and
\begin{equation}\label{Alba non-selective 1st derivation}
\rhoh_{\textsc{a}}'=\textrm{tr}_{\textsc{b}}(\rhoh'_{\textsc{ab}})=\textrm{tr}_{\phi}\big[ \hat{U}_{\textsc{a}}(\rhoh_{\textsc{a}}\otimes\rhoh_{\phi})\hat{U}_{\textsc{a}}^{\dagger} \big] \;.
\end{equation}
where in order to trace out A and B we have used Eq.~\eqref{commutation evolutions A, B and C} and the cyclic property of trace. 

The same results of Eqs.~\eqref{Alba-Blanca non-selective 1st derivation},~\eqref{Blanca non-selective 1st derivation} and~\eqref{Alba non-selective 1st derivation} are obtained by using the extended $n$-point function update formalized in Section~\ref{section: generalization to the presence of entangled third parties}, as can be explicitly seen in the calculations leading to Eqs.~\eqref{Alba-Blanca non-selective 2nd derivation},~\eqref{Blanca non-selective 2nd derivation} and~\eqref{Alba non-selective 2nd derivation} in Appendix~\ref{appendix: a practical example using n-point functions}. 

Notice in particular that $\rhoh_{\textsc{a}}'$ does not depend on the operations performed by Blanca and Clara. In fact, as we expected, this is the same result that we would have obtained had we updated the state with the interaction of Alba's detector in the first place. Note that both partial states satisfy
\begin{equation}
\rhoh_\textsc{a}'=\textrm{tr}_\textsc{b}(\rhoh_\textsc{ab}') \quad \textrm{and} \quad \rhoh_\textsc{b}'=\textrm{tr}_\textsc{a}(\rhoh_\textsc{ab}') \;.
\end{equation}
This is a consequence of the fact that for non-selective measurements, as we saw in Section~\ref{section: causal behaviour}, there is no need to make a distinction in the update for observers inside $\mathcal{P}$ and outside $\mathcal{P}$, since they may in principle have access to the same information: that a measurement whose outcome is unknown has potentially been performed. More concretely, here both Alba and Blanca are ignorant about the outcome of Clara's measurement, and therefore all three partial density operators, $\rhoh_\textsc{ab}$, $\rhoh_\textsc{a}$ and $\rhoh_\textsc{b}$ are calculated with the same amount of information about the field and its interactions.

\subsection{Selective measurement}\label{subsection: selective measurement}

The case in which Clara performs a selective measurement requires slightly more care than the non-selective one, since in this case the updated state after the measurement depends on the observer and the information that is available to them (in the language of $n$-point functions, the update is defined piecewise, unlike the non-selective case). As in the non-selective case, for the sake of formal simplicity, in this derivation we will perform the update due to Clara's measurement in the first place. We will check nevertheless that, as before and as should be required, the results are the same if we evolve the state due to Alba's interaction in the first place. 

To calculate the joint state $\rhoh_{\textsc{ab}}$, we need to take into account that the information in this state is only fully accessible by an observer that eventually has access to the information from both systems held by Alba and Blanca. In particular, such an observer has access to the outcome of Clara's measurement, since Blanca does\footnote{This line of reasoning is completely analogous to the one carried out in Section~\ref{subsubsection: two-point function} to prescribe the piecewise update of two-point functions.}. Therefore
\begin{align}\label{Alba-Blanca selective 1st derivation}
\rhoh_{\textsc{ab}}'&=\frac{\textrm{tr}_{\phi}\big[ \hat{U}_{\textsc{a}}\hat{U}_{\textsc{b}}(\rhoh_{\textsc{a}}\otimes\rhoh_{\textsc{b}}\otimes\hat{M}_{c,\psi}\rhoh_{\phi}\hat{M}_{c,\psi}^{\dagger})\hat{U}_{\textsc{b}}^{\dagger}\hat{U}_{\textsc{a}}^{\dagger}\big]}{\textrm{tr}_{\phi}\big(\rhoh_{\phi}\hat{E}_{c,\psi}\big)} \\
&=\frac{\textrm{tr}_{\phi}\big[ \hat{U}_{\textsc{a}}\hat{U}_{\textsc{b}}\hat{M}_{c,\psi}(\rhoh_{\textsc{a}}\otimes\rhoh_{\textsc{b}}\otimes\rhoh_{\phi})\hat{M}_{c,\psi}^{\dagger}\hat{U}_{\textsc{b}}^{\dagger}\hat{U}_{\textsc{a}}^{\dagger}\big]}{\textrm{tr}_{\phi}\big(\rhoh_{\phi}\hat{E}_{c,\psi}\big)} \;,\nonumber
\end{align}
where the last step is simply an abuse of notation. For calculating $\rhoh_{\textsc{b}}$, observe that since Blanca is in the causal future of the measurement performed by Clara, she has access to its outcome. Thus,
\begin{align}\label{Blanca selective 1st derivation}
\rhoh_{\textsc{b}}'&=\frac{\textrm{tr}_{\textsc{a},\phi}\big[ \hat{U}_{\textsc{a}}\hat{U}_{\textsc{b}}\hat{M}_{c,\psi}(\rhoh_{\textsc{a}}\otimes\rhoh_{\textsc{b}}\otimes\rhoh_{\phi})\hat{M}_{c,\psi}^{\dagger}\hat{U}_{\textsc{b}}^{\dagger}\hat{U}_{\textsc{a}}^{\dagger}\big]}{\textrm{tr}_{\phi}\big(\rhoh_{\phi}\hat{E}_{c,\psi}\big)} \\
&=\frac{\textrm{tr}_{\phi}\big[\hat{U}_{\textsc{b}}\hat{M}_{c,\psi}(\rhoh_{\textsc{b}}\otimes\rhoh_{\phi})\hat{M}_{c,\psi}^{\dagger}\hat{U}_{\textsc{b}}^{\dagger}\big]}{\textrm{tr}_{\phi}\big(\rhoh_{\phi}\hat{E}_{c,\psi}\big)}=\textrm{tr}_{\textsc{a}}(\rhoh_{\textsc{ab}}') \;.\nonumber 
\end{align}
Finally, if we want to obtain $\rhoh_{\textsc{a}}$, we just need to take into account that Alba does not have access to the outcome of Clara's measurement, and hence the state of the field that she deals with is the one updated non-selectively or directly the initial one (since both bring the same result, as we saw in the previous section). The result is therefore the same as in Eq.~\eqref{Alba non-selective 1st derivation} for the non-selective measurement,
\begin{equation}\label{Alba selective 1st derivation}
\rhoh_{\textsc{a}}'=\textrm{tr}_{\phi}\big[ \hat{U}_{\textsc{a}}(\rhoh_{\textsc{a}}\otimes\rhoh_{\phi})\hat{U}_{\textsc{a}}^{\dagger} \big] \;.
\end{equation}
Notice that
\begin{equation}
\rhoh_{\textsc{a}}'\neq\textrm{tr}_{\textsc{b}}\big(\rhoh_{\textsc{ab}}')\;,
\end{equation}
since $\rhoh_{\textsc{ab}}'$ was calculated for an observer that, unlike Alba, knows the outcome of Clara's measurement. In fact, if Alba eventually reaches the causal future of Clara's measurement and learns of its outcome, then the state should be updated as
\begin{align}\label{Alba selective 1st derivation after cone}
\rhoh_{\textsc{a}}''&=\frac{\textrm{tr}_{\textsc{b},\phi}\big[ \hat{U}_{\textsc{a}}\hat{U}_{\textsc{b}}\hat{M}_{c,\psi}(\rhoh_{\textsc{a}}\otimes\rhoh_{\textsc{b}}\otimes\rhoh_{\phi})\hat{M}_{c,\psi}^{\dagger}\hat{U}_{\textsc{b}}^{\dagger}\hat{U}_{\textsc{a}}^{\dagger}\big]}{\textrm{tr}_{\phi}\big(\rhoh_{\phi}\hat{E}_{c,\psi}\big)} \\
&=\frac{\textrm{tr}_{\phi}\big[\hat{U}_{\textsc{a}}\hat{M}_{c,\psi}(\rhoh_{\textsc{a}}\otimes\rhoh_{\phi})\hat{M}_{c,\psi}^{\dagger}\hat{U}_{\textsc{a}}^{\dagger}\big]}{\textrm{tr}_{\phi}\big(\rhoh_{\phi}\hat{E}_{c,\psi}\big)}=\textrm{tr}_\textsc{b}(\rhoh_{\textsc{ab}}') \;, \nonumber
\end{align}
which coincides with what happens to $\rho’_{\textsc{b}}$ in Eq.~\eqref{Blanca selective 1st derivation}.

The same results of Eqs.~\eqref{Alba-Blanca selective 1st derivation},~\eqref{Blanca selective 1st derivation},~\eqref{Alba selective 1st derivation} and~\eqref{Alba selective 1st derivation after cone} are obtained by using the extended $n$-point function update formalized in Section~\ref{section: generalization to the presence of entangled third parties}, as can be explicitly seen in the calculations leading to Eqs.~\eqref{Alba-Blanca selective 2nd derivation},~\eqref{Blanca selective 2nd derivation},~\eqref{Alba selective 2nd derivation} and~\eqref{Alba selective 2nd derivation after cone} in Appendix~\ref{appendix: a practical example using n-point functions}. 

\section{A measurement theory}\label{section: discussion}

We have proposed a measurement scheme where localized non-relativistic quantum systems that couple covariantly to the field gather information about its state. We are now in position to argue that this measurement framework has all the characteristics that one should expect from a proper measurement theory for QFT. Namely,
\begin{enumerate}
\item  \textit{It is consistent with relativistic QFT}. The measurement process consists of two steps: the interaction between the detector and the field, and the projective measurement on the detector once the interaction has been switched off in order to access the information about the field stored in it. From recently established results, we know that UDW detectors can be coupled fully covariantly to quantum fields~\cite{TalesBruno2020,TalesBruno2021}, and that the interaction with the field does not \textit{per se} allow faster-than-light signalling~\cite{Edu2015,Pipo2021}. Furthermore, when a detector is smeared, the possible signalling appears only in a restricted and controlled way if there is a third, non-pointlike detector mediating between them. Such a causality violation does not even become apparent in leading orders of perturbation theory~\cite{Pipo2021}. As for the projective measurement on the detector, in this work we have shown that the effect of performing projective measurements on detectors and updating the field state consistently is as safe from causality violations as the interaction with the field itself (Section~\ref{section: causal behaviour}). 

\item \textit{It provides an update rule}. As we have explicitly described and discussed in Sections \ref{section: the update rule},~\ref{section: update of n-point functions} and~\ref{section: generalization to the presence of entangled third parties},  we have given a consistent update rule for the field state after the measurement that respects causality---as explicitly manifested in the update of the (extended) $n$-point functions---and includes the information obtained from the outcome of the measurement in the spirit of L\"uders rule, enforcing the compatibility of sequential measurements.

\item \textit{It produces definite values for the outcome of single-shot measurements}. Since the detectors are measured through projective measurements, the outcome of a measurement is a real number that can be written down in an experimenter's notepad. 

\item \textit{It is capable of reproducing experiments}. Indeed, particle detector models have been proven to capture the features of experimental setups in quantum optics and the light-matter interaction~\cite{Edu2013,Pozas2016,Rodriguez2018,Lopp2020}, as well as the phenomenology of the measurement of other quantum fields such as, e.g., neutrinos~\cite{Bruno2020neutrino,Tales2021antiparticle}. Particle detector models are therefore directly connected with experimentally realistic setups where quantum fields are measured.
\end{enumerate}

By satisfying these four characteristics, we conclude that the measurement scheme proposed in this article constitutes a measurement theory for QFT that can still rely on the projection postulate of non-relativistic quantum mechanics to access the information in the field. This proposed scheme has the additional advantage of combining localized rank-one projective measurements (that return a definite value of a measurement outcome) with compatibility with the relativistic nature of the theory. This is something that we cannot accomplish by performing projective measurements on quantum fields, where localized projective measurements are forced to be infinite-rank~\cite{Redhead1995}.

\section{Conclusions}\label{section: conclusions}

Since Sorkin's seminal paper in 1993, it has been evident that the measurement theory of non-relativistic quantum mechanics cannot be directly imported to quantum field theory due to relativistic considerations. As Sorkin put it, ``\textit{this problem leaves the Hilbert space formulation of quantum field theory with no definite measurement theory}''~\cite{Sorkin1993}. In this paper we have proposed a way to build a measurement theory for QFT based on particle detectors that 1) has all the advantages of the measurement theory of non-relativistic quantum mechanics, in that it provides the values of single-shot experiments and there is a state update enforcing compatibility with future measurements, 2) is compatible with relativity and is safe from gross causality violations, and 3) can be easily connected to experiments. 

In order to establish the consistency of the proposed measurement scheme---consisting of 1) interaction of the detector with the probed field and 2) performing an idealized measurement on the detector and updating accordingly---we have relied on previous results establishing the covariance of the UDW detector-field coupling~\cite{TalesBruno2020,TalesBruno2021} and the compatibility of the interaction with relativity~\cite{Edu2015,Pipo2021}. In addition, in this work we have shown that the performance of the projective measurement on the detector does not introduce any causality violations, and---effectively subscribing to an epistemic interpretation of the field state---we have provided a contextual update rule for the state of the field after the measurement. This update rule has been given in full detail in terms of (extended) $n$-point functions of the field for both non-selective and selective measurements on particle detectors, and we have shown how it is implemented in a practical example. 

These results provide a formal basis for a measurement theory for QFT. Furthermore, they pave the way to fully relativistic formulations of problems where the role of measurements is central, such as the quantum Zeno effect~\cite{Misra1977,Patil2015}, the delayed choice quantum eraser experiment~\cite{Scully1982,Scully1991,Kwiat1992,Kim2000,Ma2013}, and many other similar experiments that can be performed within, e.g., the framework of the light-matter interaction.

\section{Acknowledgements}
The authors would like to thank Christopher J. Fewster, Maximilian H. Ruep and Ian Jubb for enlightening discussions. The authors also thank Maria Papageorgiou for her comments and for interesting discussions. J.P.-G. is supported by a Mike and Ophelia Lazaridis Fellowship. J.P.-G. also received the support of a fellowship from ``La Caixa'' Foundation (ID 100010434, with fellowship code LCF/BQ/AA20/11820043). L.J.G. acknowledges support through Project. No.
MICINN FIS2017-86497-C2-2-P from Spain (with extension Project. No. MICINN PID2020-118159GB-C44 under
evaluation). E.M.-M. acknowledges support through the Discovery
Grant Program of the Natural Sciences and Engineering Research Council of Canada (NSERC). E.M.-M. also
acknowledges support of his Ontario Early Researcher
award.

\appendix

\section{Causal behaviour of the selective update}\label{appendix: causal behaviour of the selective update}

In this appendix we analyze the causal structure of the selective update, confirming that it affects local operations outside the causal future of the interaction region and therefore should not be applied as a global update anywhere outside the causal support of the detector if we want the update rule to be compatible with relativity.  

Following a selective measurement, we consider the updated state of the field to be
\beq
\rhoh_{\phi}^\textrm{u}=\rhoh_{\phi}^{s,\psi}
\eeq
where $\rhoh_{\phi}^{s,\psi}$ is given by \eqref{updated state field} for a specific $\ket{s}$ coming from the result of the measurement. For this update, we will use the following estimator to evaluate where in spacetime the POVM alters the field:
\begin{align}\label{change in n-point function}
\Delta_n&(t_1,\spatial{x}_1,\cdots,t_n,\spatial{x}_{n}) \nonumber\\
&=\big(\langle \phih(t_1,\spatial{x}_1)\cdots\phih(t_{n},\spatial{x}_{n}) \rangle_{\rhoh_{\phi}^\textrm{u}} \nonumber\\
&\phantom{=}\,-\langle\phih(t_1,\spatial{x}_1)\cdots\phih(t_{n},\spatial{x}_{n}) \rangle_{\rhoh_{\phi}}\big)\langle \hat{E}_{s,\psi} \rangle_{\rhoh_{\phi}}\\
&= \langle \hat{M}_{s,\psi}^{\dagger}\phih(t_1,\spatial{x}_1)\cdots\phih(t_{n},\spatial{x}_{n})\hat{M}_{s,\psi}\rangle_{\rhoh_{\phi}} \nonumber\\
&\phantom{=}\;-\langle \hat{M}_{s,\psi}^{\dagger}\hat{M}_{s,\psi}\rangle_{\rhoh_{\phi}} \langle\phih(t_1,\spatial{x}_1)\cdots\phih(t_n,\spatial{x}_{n})\rangle_{\rhoh_{\phi}} \;. \nonumber
\end{align}
This estimator is the difference between the $n$-point function of the post-measurement and pre-measurement states of the field, but multiplied by the quantity $\textrm{tr}\big(\rhoh_{\phi}\hat{E}_{s,\psi}\big)$ to make the evaluation simpler. Note that this trace is finite and positive if the updated state is well defined. Studying where the estimator $\Delta_n$ is non-zero gives us information about the spacetime domain of the effect of the selective update.

We will first perform the analysis without making any assumptions on the pure state of the detector $\ket{\psi}$ or the initial state of the field $\rhoh_{\phi}$. In this general case we will see that already for the one-point function, $\Delta_{1}(t_1,\bm{x}_1)$ can be non-zero out of the causal future of the detector, to first order in perturbation theory. In the next subsection, we show this also happens for the updates associated with eigenstates of the detector Hamiltonian, by going to order $\lambda^2$ in the \mbox{one-point} function. We end the appendix by showing that there are certain conditions under which the update does not affect the one-point function out of the causal support of the detector. In this last case we have to use the two-point function to confirm that, as in the rest of the cases, the selective update has an effect over regions of spacetime that are spacelike separated from the measurement.

\subsection{The general case}\label{appendix subsection: the general case}

In order to analyze the change in the one-point function, we use \eqref{change in n-point function} with $n=1$ and the perturbative expansions considered at the end of Section~\ref{section: the updated state of the field}. For the first term in Eq.~\eqref{change in n-point function}, 
\begin{align}\label{expansion up to second order}
\hat{M}&_{s,\psi}^{\dagger}\phih(t_1,\bm{x}_1)\hat{M}_{s,\psi}=\hat{M}_{s,\psi}^{\dagger\;(0)}\phih(t_1,\bm{x}_1)\hat{M}_{s,\psi}^{(0)}\\
&+\lambda \left(\hat{M}_{s,\psi}^{\dagger\;(1)}\phih(t_1,\bm{x}_1)\hat{M}_{s,\psi}^{(0)}+\hat{M}_{s,\psi}^{\dagger\;(0)}\phih(t_1,\bm{x}_1)\hat{M}_{s,\psi}^{(1)} \right) \nonumber\\
&+O(\lambda^2) \;. \nonumber 
\end{align}  
Term by term, for the zeroth order,
\beq\label{expansion zeroth order}
\hat{M}_{s,\psi}^{\dagger\;(0)}\phih(t_1,\bm{x}_1)\hat{M}_{s,\psi}^{(0)}=|\!\braket{s}{\psi}\!|^{2}\,\phih(t_1,\bm{x}_1) \;.
\eeq
For the first order in $\lambda$,
\begin{align}\label{expansion first order 10}
&\hat{M}_{s,\psi}^{\dagger\;(1)}\phih(t_1,\bm{x}_1)\hat{M}_{s,\psi}^{(0)}=\ii \braket{s}{\psi} \int\diff t \,\diff^{d}\spatial{x}\,\chi(t)F(\spatial{x}) \nonumber\\
& \phantom{==========}\,\times \, \bra{s}\muh(t)\ket{\psi}^{\ast}\phih(t,\spatial{x})\phih(t_1,\bm{x}_1) 
\end{align}
and
\begin{align}\label{expansion first order 01}
&\hat{M}_{s,\psi}^{\dagger\;(0)}\phih(t_1,\bm{x}_1)\hat{M}_{s,\psi}^{(1)}=-\ii \braket{s}{\psi}^{\ast} \int\diff t \,\diff^{d}\spatial{x}\,\chi(t)F(\spatial{x}) \nonumber \\ 
&\phantom{==========}\times \, \bra{s}\muh(t)\ket{\psi}\phih(t_1,\bm{x}_1)\phih(t,\spatial{x})
\end{align}
which add up to
\begin{align}\label{expansion first order}
&\left(\hat{M}_{s,\psi}^{\dagger}\phih(t_1,\bm{x}_1)\hat{M}_{s,\psi}\right)^{(1)} \nonumber\\
&=2 \int\diff t \,\diff^{d}\spatial{x} \,\chi(t) F(\spatial{x}) \Im(\braket{\psi}{s}\bra{s}\muh(t)\ket{\psi}) \nonumber\\
&\phantom{=====}\;\times \phih(t,\spatial{x})\phih(t_1,\bm{x}_1)\\
&\phantom{=}\;-\ii \int\diff t \,\diff^{d}\spatial{x}\,\chi(t)F(\spatial{x}) \braket{\psi}{s}\bra{s}\muh(t)\ket{\psi} \nonumber\\
&\phantom{===}\;\phantom{=}\phantom{=}\;\times [\phih(t_1,\bm{x}_1),\phih(t,\spatial{x})] \;. \nonumber
\end{align}
On the other hand, for the second term in \eqref{change in n-point function} we have
\begin{align}\label{expansion up to second order denominator}
\hat{M}&_{s,\psi}^{\dagger}\hat{M}_{s,\psi}=\hat{M}_{s,\psi}^{\dagger\;(0)}\hat{M}_{s,\psi}^{(0)}\\
&+\lambda(\hat{M}_{s,\psi}^{\dagger\;(1)}\hat{M}_{s,\psi}^{(0)}+\hat{M}_{s,\psi}^{\dagger\;(0)}\hat{M}_{s,\psi}^{(1)})+O(\lambda^2) \nonumber
\end{align}
with
\beq\label{expansion zeroth order denominator}
\left(\hat{M}_{s,\psi}^{\dagger}\hat{M}_{s,\psi}\right)^{(0)}=\hat{M}_{s,\psi}^{\dagger\;(0)}\hat{M}_{s,\psi}^{(0)}=|\!\braket{s}{\psi}\!|^{2}
\eeq
and
\begin{align}\label{expansion first order denominator}
\left(\hat{M}_{s,\psi}^{\dagger}\hat{M}_{s,\psi}\right)^{(1)}&=2\int\diff t \, \diff^{d}\spatial{x} \,\chi(t) F(\spatial{x}) \\
& \times\, \Im(\braket{\psi}{s}\bra{s}\muh(t)\ket{\psi})\phih(t,\spatial{x}) \;. \nonumber
\end{align}
We are now set to examine $\Delta_1(t_1,\bm{x}_1)$. At zeroth order it is trivial that there is no difference. Up to first order in $\lambda$,
\begin{align}\label{eq. Delta1(1)}
&\Delta_1(t_1,\bm{x}_1)=\lambda\Delta_1(t_1,\bm{x}_1)^{(1)}+O(\lambda^2)\\
&=2\lambda\int \diff t \, \diff^{d}\spatial{x} \,\chi(t) F(\spatial{x}) \Im(\braket{\psi}{s}\bra{s}\muh(t)\ket{\psi}) \nonumber\\
&\qquad \qquad \qquad \qquad \qquad \quad \times\langle \phih(t,\spatial{x})\phih(t_1,\bm{x}_1) \rangle_{\rhoh_{\phi}} \nonumber\\
&\phantom{=}\;-\ii\lambda \int \diff t \, \diff^{d}\spatial{x} \,\chi(t) F(\spatial{x}) \braket{\psi}{s}\bra{s}\muh(t)\ket{\psi} \nonumber\\
& \qquad \qquad \qquad \qquad \qquad \times \langle [ \phih(t,\spatial{x}),\phih(t_1,\bm{x}_1)] \rangle_{\rhoh_{\phi}} \nonumber\\
&\phantom{=}\;-2\lambda \int \diff t \, \diff^{d}\spatial{x} \,\chi(t)F(\spatial{x}) \Im(\braket{\psi}{s}\bra{s}\muh(t)\ket{\psi}) \nonumber\\
&\qquad \qquad \qquad \quad \times \langle \phih(t,\spatial{x}) \rangle_{\rhoh_{\phi}} \langle \phih(t_1,\bm{x}_1) \rangle_{\rhoh_{\phi}} + O(\lambda^2) \nonumber\\
&= 2\lambda \int \diff t \, \diff^{d}\spatial{x} \,\chi(t) F(\spatial{x}) \Im(\braket{\psi}{s}\bra{s}\muh(t)\ket{\psi}) \nonumber\\
&\qquad \qquad \qquad \qquad \qquad \times \textrm{Cov}_{\rhoh_{\phi}}\big[\phih(t,\spatial{x}),\phih(t_1,\bm{x}_1)\big] \nonumber \\
&\phantom{=}\;-\ii\lambda \int \diff t \, \diff^{d}\spatial{x} \, \chi(t) F(\spatial{x}) \braket{\psi}{s}\bra{s}\muh(t)\ket{\psi} \nonumber\\
&\qquad \qquad \qquad \quad \times\langle [ \phih(t,\spatial{x}),\phih(t_1,\bm{x}_1)] \rangle_{\rhoh_{\phi}} + O(\lambda^2) \nonumber
\end{align}
where $\textrm{Cov}_{\rhoh}[A,B]=\langle AB \rangle_{\rhoh} - \langle A \rangle_{\rhoh} \langle B \rangle_{\rhoh}$. The term in the last line, depending on the commutator, is definitely \textit{not} contributing to $\Delta_{1}$ when $(t_1,\bm{x}_1)$ is out of the causal future of the interaction region. However, the term in the penultimate line \textit{will} contribute in general to $\Delta_{1}$ not being zero in that same case. Indeed, \mbox{$\braket{\psi}{s}\bra{s}\muh(t)\ket{\psi}$} is not real in general, and the correlations of the field \mbox{$\textrm{Cov}_{\rhoh_{\phi}}\big[\phih(t,\spatial{x}),\phih(t_1,\bm{x}_1)\big]$} will not vanish in general if $(t,\spatial{x})$ and $(t_1,\bm{x}_1)$ are spacelike separated. The reason why $\Delta_1$ does not vanish everywhere outside the causal support of the detector is that once the detector starts interacting with the field, it gets entangled with it (in a way that respects causality~\cite{Edu2015,Pipo2021}). As the state of the field will in general show spacelike correlations~\cite{ReehSchlieder,Summers1985,Summers1987,Redhead1995,Simidzija2018}, the projection operator destroys some of these correlations. The entanglement between the detector and the field generated by their interaction thus hinders the possibility of applying the selective update outside the causal future of the detector in a way consistent with the relativistic framework of QFT. Not even in the---singular but generally less problematic in causality-related issues---case in which the detector is considered to be pointlike and the interaction sudden, that is, with $\chi(t)=\delta(t)$ and $F(\spatial{x})=\delta(\spatial{x})$, is the update safe from being non-causal. Indeed, choosing
\beq
\ket{\psi}=\frac{1}{\sqrt{2}}(\ii\ket{g}+\ket{e}) \quad \ket{s}=\ket{e} \quad \rhoh_{\phi}=\ket{0}\!\bra{0} 
\eeq
we have $\langle \phih(t,\spatial{x}) \rangle=0$ for every $(t,\spatial{x}) \in \mathcal{M}$ and therefore
\begin{align}
\Delta_{1}(t_1,\bm{x}_1)=&\lambda \langle \phih(0,0)\phih(t_1,\bm{x}_1) \rangle_{\rhoh_{\phi}} \\
& + \frac{\lambda}{2} \langle [ \phih(0,0),\phih(t_1,\bm{x}_1) ] \rangle_{\rhoh_{\phi}}+O(\lambda^2) \;. \nonumber
\end{align}
For spacelike $(t_1,\bm{x}_1)$, this gives
\beq
\Delta_{1}(t_1,\bm{x}_1)=\lambda \langle \phih(0,0) \phih(t_1,\bm{x}_1) \rangle_{\rhoh_{\phi}} + O(\lambda^2)
\eeq
which in general does not vanish. 

\subsection{The ground and excited states}\label{appendix subsection: the ground and excited states}

By examining \eqref{eq. Delta1(1)} one observes that $\Delta_{1}^{(1)}$ cancels out of the causal future of the interaction region if both the initial state $\ket{\psi}$ and the state associated with the projection $\ket{s}$ are eigenstates of the free Hamiltonian of the detector, that is, if \mbox{$\ket{\psi},\ket{s} \in \{\ket{g},\ket{e}\}$}. These are important states, and one could wonder if for these states the selective update could be safe from showing the non-causal features we saw in the previous subsection. The answer is no, as can be checked by simply analyzing the next order of $\Delta_1$ in perturbation theory. Proceeding as before,
\begin{align}\label{eq. expansion second order}
&\left( \hat{M}_{s,\psi}^{\dagger}\phih(t_1,\bm{x}_1)\hat{M}_{s,\psi} \right)^{(2)}=\hat{M}_{s,\psi}^{\dagger \; (2)}\phih(t_1,\bm{x}_1)\hat{M}_{s,\psi}^{(0)} \nonumber\\
&+\hat{M}_{s,\psi}^{\dagger\;(0)}\phih(t_1,\bm{x}_1)\hat{M}_{s,\psi}^{(2)}+\hat{M}_{s,\psi}^{\dagger\;(1)}\phih(t_1,\bm{x}_1)\hat{M}_{s,\psi}^{(1)} \\
&=-\int \diff t \,\diff t' \,\diff^{d}\spatial{x} \,\diff^{d}\spatial{x}' \chi(t)\chi(t')F(\spatial{x})F(\spatial{x}') \cdot \mathcal{C} \nonumber
\end{align}
where
\begin{align}
&\mathcal{C}=\theta(t-t') \\
&\phantom{\;}\times \big(\!\braket{s}{\psi}\bra{\psi}\muh(t')\muh(t)\ket{s}\phih(t',\spatial{x}')\phih(t,\spatial{x})\phih(t_1,\bm{x}_1) \nonumber\\
&\phantom{\;}+\braket{\psi}{s}\bra{s}\muh(t)\muh(t')\ket{\psi}\phih(t_1,\bm{x}_1)\phih(t,\spatial{x})\phih(t',\spatial{x}') \big) \nonumber\\
&\phantom{\;}-\bra{\psi}\muh(t)\ket{s}\bra{s}\muh(t')\ket{\psi}\phih(t,\spatial{x})\phih(t_1,\bm{x}_1)\phih(t',\spatial{x}') \;. \nonumber
\end{align} 
Now,
\beq
\theta(t-t')+\theta(t'-t)=1
\eeq
almost everywhere, as the diagonal set $\{t=t'\} \subset \R{2}$ in which the equality does not hold has zero Lebesgue measure. Therefore, for a smooth switching function $\chi$ or, in general, one switching not involving delta functions, we can write
\begin{align}
&\mathcal{C}=\theta(t'-t) \nonumber\\
&\phantom{=}\;\times\phih(t,\spatial{x}) \big(\braket{s}{\psi}\bra{\psi}\muh(t)\muh(t')\ket{s}\phih(t',\spatial{x}')\phih(t_1,\bm{x}_1) \nonumber \\
&\phantom{=}\;-\bra{\psi}\muh(t)\ket{s}\bra{s}\muh(t')\ket{\psi}\phih(t_1,\bm{x}_1)\phih(t',\spatial{x}') \big) \nonumber\\
&\phantom{=}\;+\theta(t-t') \\
&\phantom{=}\;\times(\braket{\psi}{s}\bra{s}\muh(t)\muh(t')\ket{\psi}\phih(t_1,\bm{x}_1)\phih(t,\spatial{x}) \nonumber\\
&\phantom{=}\;-\bra{\psi}\muh(t)\ket{s}\bra{s}\muh(t')\ket{\psi}\phih(t,\spatial{x})\phih(t_1,\bm{x}_1))\phih(t',\spatial{x}') \nonumber
\end{align}
and in particular
\begin{align}
&\braket{s}{\psi}\bra{\psi}\muh(t)\muh(t')\ket{s}\phih(t',\spatial{x}')\phih(t_1,\bm{x}_1) \nonumber\\
&-\bra{\psi}\muh(t)\ket{s}\bra{s}\muh(t')\ket{\psi}\phih(t_1,\bm{x}_1)\phih(t',\spatial{x}') \nonumber\\
&=\braket{s}{\psi}\bra{\psi}\muh(t)\muh(t') \ket{s} [\phih(t',\spatial{x}'),\phih(t_1,\bm{x}_1)]\\
&\phantom{=}\;+(\braket{s}{\psi}\bra{\psi}\muh(t)\muh(t') \ket{s} \nonumber\\
&\phantom{=}\;-\bra{\psi}\muh(t)\ket{s}\bra{s}\muh(t')\ket{\psi})\phih(t_1,\bm{x}_1)\phih(t',\spatial{x}') \nonumber
\end{align}
and
\begin{align}
&\braket{\psi}{s}\bra{s}\muh(t)\muh(t')\ket{\psi}\phih(t_1,\bm{x}_1)\phih(t,\spatial{x}) \nonumber\\
&-\bra{\psi}\muh(t)\ket{s}\bra{s}\muh(t')\ket{\psi}\phih(t,\spatial{x})\phih(t_1,\bm{x}_1) \nonumber\\
&=\braket{\psi}{s}\bra{s}\muh(t)\muh(t') \ket{\psi} [\phih(t_1,\bm{x}_1),\phih(t,\spatial{x})]\\
&\phantom{=}\;+(\braket{\psi}{s}\bra{s}\muh(t)\muh(t') \ket{\psi} \nonumber\\
&\phantom{=}\;-\bra{\psi}\muh(t)\ket{s}\bra{s}\muh(t')\ket{\psi})\phih(t,\spatial{x})\phih(t_1,\bm{x}_1) \;. \nonumber
\end{align}
Now, since the factors accompanying $\mathcal{C}$ in the integral of Eq.~\eqref{eq. expansion second order} are symmetric in $t$ and $t'$, we can safely exchange both time parameters in the $\theta(t'-t)$ term, hence rewriting
\begin{align}\label{expr. C final}
&\mathcal{C}=\theta(t-t')\Big[ \braket{s}{\psi} \bra{\psi} \muh(t')\muh(t) \ket{s} \phih(t',\spatial{x}')\\
&\phantom{=}\;\times [\phih(t,\spatial{x}),\phih(t_1,\bm{x}_1)]+ \big(\braket{s}{\psi}\bra{\psi}\muh(t')\muh(t)\ket{s} \nonumber\\
&\phantom{=}\;-\bra{\psi}\muh(t')\ket{s}\bra{s}\muh(t)\ket{\psi} \big)\phih(t',\spatial{x}')\phih(t_1,\bm{x}_1)\phih(t,\spatial{x}) \nonumber\\
&\phantom{=}\;+\braket{\psi}{s}\bra{s}\muh(t)\muh(t')\ket{\psi}[\phih(t_1,\bm{x}_1),\phih(t,\spatial{x})]\phih(t',\spatial{x}') \nonumber\\
&\phantom{=}\;+\big(\braket{\psi}{s}\bra{s}\muh(t)\muh(t')\ket{\psi} \nonumber\\
&\phantom{=}\;-\bra{\psi}\muh(t)\ket{s}\bra{s}\muh(t')\ket{\psi} \big)\phih(t,\spatial{x})\phih(t_1,\bm{x}_1)\phih(t',\spatial{x}') \Big] \;. \nonumber
\end{align}
We still have to compute
\begin{align}
\left( \hat{M}_{s,\psi}^{\dagger}\hat{M}_{s,\psi} \right)^{(2)}=&\,\,\hat{M}_{s,\psi}^{\dagger\;(2)}\hat{M}_{s,\psi}^{(0)}+\hat{M}_{s,\psi}^{\dagger\;(0)}\hat{M}_{s,\psi}^{(2)}\\
&+\hat{M}_{s,\psi}^{\dagger\;(1)}\hat{M}_{s,\psi}^{(1)} \;. \nonumber 
\end{align}
We proceed exactly as above, the difference being that we do not have the field operator $\phih(t_1,\bm{x}_1)$ in between anymore. In the same spirit, we get
\begin{align}
&\left(\hat{M}_{s,\psi}^{\dagger}\hat{M}_{s,\psi}\right)^{(2)}=- \int \diff t \,\diff t' \,\diff^{d}\spatial{x} \,\diff^{d}\spatial{x}' \chi(t)\chi(t')\\
& \phantom{===============}\times \, F(\spatial{x})F(\spatial{x}') \cdot \mathcal{I} \nonumber
\end{align}
where
\begin{align}\label{expr. G final}
\mathcal{I}&=\theta(t-t')\Big[(\big(\braket{s}{\psi}\bra{\psi}\muh(t')\muh(t)\ket{s}\\
&\phantom{=}\;-\bra{\psi}\muh(t')\ket{s}\bra{s}\muh(t)\ket{\psi}\big)\phih(t',\spatial{x}')\phih(t,\spatial{x}) \nonumber\\
&\phantom{=}\;+\big(\braket{\psi}{s}\bra{s}\muh(t)\muh(t')\ket{\psi} \nonumber\\
&\phantom{=}\;-\bra{\psi}\muh(t)\ket{s}\bra{s}\muh(t')\ket{\psi}\big)\phih(t,\spatial{x})\phih(t',\spatial{x}') \Big]  \nonumber
\end{align}

\noindent
We realize that the terms that are not included in the commutators in \eqref{expr. C final} are conjugate to each other, and that the same happens with the terms in \eqref{expr. G final}. Putting everything together, we get
\begin{align}
&\Delta_{1}^{(2)}=- \int \diff t \, \diff t' \, \diff^{d}\spatial{x}\, \diff^{d}\spatial{x}'\,\chi(t)\chi(t')\\
&\phantom{=======} \times \, F(\spatial{x})F(\spatial{x}') \theta(t-t') \,\cdot \,\mathcal{R} \nonumber
\end{align}
where
\begin{align}
&\mathcal{R}= \braket{s}{\psi}\bra{\psi}\muh(t')\muh(t)\ket{s}\langle \phih(t',\spatial{x}')[\phih(t,\spatial{x}),\phih(t_1,\bm{x}_1)] \rangle_{\rhoh_{\phi}} \nonumber \\
&\phantom{=}\;+\braket{\psi}{s}\bra{s}\muh(t)\muh(t')\ket{\psi}\langle [\phih(t_1,\bm{x}_1),\phih(t,\spatial{x})]\phih(t',\spatial{x}')\rangle_{\rhoh_{\phi}} \nonumber\\
&\phantom{=}\;+2\Re(\mathcal{S})
\end{align}
and
\begin{align}\label{expr. S Delta1(2)}
\mathcal{S}&=\big(\braket{\psi}{s}\bra{s}\muh(t)\muh(t')\ket{\psi}\\
&\phantom{=}\;-\bra{\psi}\muh(t)\ket{s}\bra{s}\muh(t')\ket{\psi}\big) \nonumber\\
&\phantom{=}\;\times\,\big(\langle \phih(t,\spatial{x})\phih(t_1,\bm{x}_1)\phih(t',\spatial{x}')\rangle_{\rhoh_{\phi}} \nonumber\\
&\phantom{=}\;-\langle \phih(t,\spatial{x})\phih(t',\spatial{x}') \rangle_{\rhoh_{\phi}} \langle \phih(t_1,\bm{x}_1) \rangle_{\rhoh_{\phi}}\big) \;. \nonumber  
\end{align}
Now, the first two terms of $\mathcal{R}$ are proportional to commutators, so that if $(t_1,\bm{x}_1)$ is spacelike separated from the interaction region, they become zero. However, $\mathcal{S}$ is not zero in general in that case, nor purely imaginary. In fact, this is the term that allows us to confirm that even for the ground and excited states the selective update should not be applied outside the causal future of the detector. For example, consider the case in which the field is initially in a coherent state $\rhoh_{\phi}=\ket{\varphi}\!\bra{\varphi}$. For these states, the correlation functions of one, two and three points are
\begin{align}
&\langle \phih(\sptime{x}) \rangle_{\rhoh_{\phi}}= \varphi(\sptime{x})\;,\\
&\langle \phih(\sptime{x})\phih(\sptime{y}) \rangle_{\rhoh_{\phi}}=\varphi(\sptime{x})\varphi(\sptime{y}) + w_{\textrm{vac}}(\sptime{x},\sptime{y})\;,\\
&\langle \phih(\sptime{x})\phih(\sptime{y})\phih(\sptime{z}) \rangle_{\rhoh_{\phi}}=\varphi(\sptime{x})\varphi(\sptime{y})\varphi(\sptime{z}) \\
&+\varphi(\sptime{x})w_{\textrm{vac}}(\sptime{y},\sptime{z}) +\varphi(\sptime{y})w_{\textrm{vac}}(\sptime{x},\sptime{z})+\varphi(\sptime{z})w_{\textrm{vac}}(\sptime{x},\sptime{y}) \;. \nonumber 
\end{align}
where $\sptime{x},\sptime{y},\sptime{z} \in \mathcal{M}$, $\varphi(\sptime{x})$ is the field amplitude of the coherent state at $\sptime{x}$ and 
\beq
w_{\textrm{vac}}(\sptime{x},\sptime{y})=\bra{0}\phih(\sptime{x})\phih(\sptime{y})\ket{0}
\eeq
is the two-point Wightman function. If we additionally consider $\ket{\psi}=\ket{g}$ and $\ket{s}=\ket{e}$, then 
\begin{align}
&\braket{\psi}{s}\bra{s}\mu(t)\mu(t')\ket{\psi}-\bra{\psi}\mu(t)\ket{s}\bra{s}\mu(t')\ket{\psi}\\
&\; =-e^{-\ii\Omega (t-t')} \nonumber
\end{align}
and
\begin{align}
&\langle \phi(t,\spatial{x})\phih(t_1,\bm{x}_1)\phi(t',\spatial{x}')\rangle_{\rhoh_{\phi}} \nonumber\\
&\qquad \quad\! -\langle \phi(t,\spatial{x})\phi(t',\spatial{x}') \rangle \langle \phih(t_1,\bm{x}_1) \rangle_{\rhoh_{\phi}}\\
&\qquad \qquad \qquad\!\! =\varphi(t,\spatial{x})w_{\textrm{vac}}(t_1,\spatial{x}_1,t',\spatial{x}') \nonumber\\
&\qquad \qquad \qquad \qquad \quad\!\!\!+\varphi(t',\spatial{x}')w_{\textrm{vac}}(t,\spatial{x},t_1,\spatial{x}_1) \;. \nonumber
\end{align}
The product of both terms is not always purely imaginary, as the sum of it and its conjugate is non-zero in general. We conclude that, under these conditions, \mbox{$\Delta_{1}^{(2)}(t_1,\bm{x}_1)$} does not cancel out for every $(t_1,\bm{x}_1)$ spacelike separated from the interaction region. 

\subsection{Gaussian states}\label{appendix subsection: Gaussian states}

Observation of \eqref{expr. S Delta1(2)} reveals that still, if the initial state of the field is the vacuum, or a thermal state, the selective update does not affect the one-point function outside the causal support of the detector. What these states share that makes $\mathcal{S}$ in \eqref{expr. S Delta1(2)} cancel out are its vanishing one-point and three-point functions. In particular, we can prove the following: 

\textit{In the previous setting, if $\ket{s}$ and $\ket{\psi}$ are in $\{\ket{g},\ket{e}\}$ and $\rhoh_{\phi}$ is a Gaussian state with $\langle \phih(\sptime{x}) \rangle_{\rhoh_{\phi}}=0$ for every $\sptime{x} \in \mathcal{M}$, then the selective POVM update does not affect the one-point function outside the causal future of the interaction region.}

To prove this claim, we proceed simply by examination of the general term
\begin{align}
\Delta_{1}^{(p,q)} \coloneqq& \langle \hat{M}^{\dagger\;(p)}\phih(t_1,\bm{x}_1)\hat{M}_{s,\psi}^{(q)} \rangle_{\rhoh_{\phi}} \nonumber\\
&- \langle \hat{M}_{s,\psi}^{\dagger\;(p)}\hat{M}_{s,\psi}^{(q)} \rangle \langle \phih(t_1,\bm{x}_1) \rangle_{\rhoh_{\phi}} \;,
\end{align}
where the second term vanishes because of the assumption $\langle \phih(\sptime{x}) \rangle_{\rhoh_{\phi}}=0$. The exact same kind of calculation carried out before yields
\begin{align}\label{general term}
\Delta_{1}^{(p,q)}&=\ii^{p+3q} \int \diff \mathfrak{t}_1 \hdots \diff \mathfrak{t}_p \,\diff \mathfrak{t}_1'\hdots\diff \mathfrak{t}_q'\,\diff^{d}\spatial{z}_1
\hdots \diff^{d}\spatial{z}_p\, \nonumber\\
&\phantom{=}\;\times \diff^{d}\spatial{z}_1'\hdots\diff^{d}\spatial{z}_q'\,\theta(\mathfrak{t}_1-\mathfrak{t}_2)\cdots\theta(\mathfrak{t}_{p-1}-\mathfrak{t}_p) \nonumber\\
&\phantom{=}\;\times\;\theta(\mathfrak{t}_1'-\mathfrak{t}_2')\cdots\theta(\mathfrak{t}_{q-1}'-\mathfrak{t}_q) \chi(\mathfrak{t}_1)\cdots\chi(\mathfrak{t}_q') \nonumber\\
&\phantom{=}\;\times F(\spatial{z}_1)\cdots F(\spatial{z}_q')\bra{\psi}\muh(\mathfrak{t}_p)\cdots\muh(\mathfrak{t}_1)\ket{s} \nonumber\\
&\phantom{=}\;\times \bra{s}\muh(\mathfrak{t}_1')\cdots\muh(\mathfrak{t}_q')\ket{\psi} \langle \phih(\mathfrak{t}_p,\spatial{z}_p)\cdots\phih(\mathfrak{t}_1,\spatial{z}_1) \nonumber\\
&\phantom{=}\;\times \phih(t_1,\bm{x}_1)\phih(\mathfrak{t}_1',\spatial{z}_1')\cdots\phih(\mathfrak{t}_q,\spatial{z}_q) \rangle_{\rhoh_{\phi}} 
\end{align}
In order to analyze this expression, we first calculate the general form of the operator $\muh(\mathfrak{t}_{1})\hdots\muh(\mathfrak{t}_N)$: let us define
\beq
T\equiv \sum_{n=1}^{N}(-1)^{n-1}\mathfrak{t}_n \;,
\eeq
then if $N$ is odd, in the ordered basis $\{\ket{g},\ket{e}\}$,
\beq
\muh(\mathfrak{t}_1)\hdots\muh(\mathfrak{t}_N)=\left(
\begin{array}{cc}
0 & e^{-\ii\Omega T}\\
e^{\ii\Omega T} & 0 \\
\end{array}\right)
\eeq
while if $N$ is even,
\beq
\muh(\mathfrak{t}_{1})\hdots\muh(\mathfrak{t}_{N})= \left(
\begin{array}{cc}
e^{-\ii\Omega T} & 0 \\
0 & e^{\ii\Omega T} \\
\end{array}\right) \;.
\eeq
Thus, if $\ket{s}$ and $\ket{\psi}$ are in $\{\ket{g},\ket{e}\}$, the term $\bra{\psi}\muh(\mathfrak{t}_p)\cdots\muh(\mathfrak{t}_1)\ket{s}\bra{s}\muh(\mathfrak{t}_1')\cdots\muh(\mathfrak{t}_q')\ket{\psi}$ in \eqref{general term} only survives if
\begin{enumerate}
    \item[(1)] $\ket{s}\neq\ket{\psi}$ and both $p$ and $q$ are odd, or
    \item[(2)] $\ket{s}=\ket{\psi}$ and both $p$ and $q$ are even.
\end{enumerate}
Now, the last factor of the integral in \eqref{general term} is the \mbox{$(p+q+1)$-point} correlation function
\beq
\langle \phi(\mathfrak{t}_p,\spatial{z}_p)\cdots\phi(\mathfrak{t}_1,\spatial{z}_1) \phih(t_1,\bm{x}_1)\phi(\mathfrak{t}_1',\spatial{z}_1')\cdots\phi(\mathfrak{t}_q,\spatial{z}_q) \rangle_{\rhoh_{\phi}} \;.
\eeq
But for both cases (1) and (2), the parity of $p$ and $q$ is the same, so that $p+q+1$ is odd, and therefore the correlation function above is zero, since the state $\rhoh_{\phi}$ is Gaussian \cite{Haag1996,Hollands2015}. We conclude that under the stated hypotheses,
\beq
\Delta_{1}^{(p,q)}(t_1,\bm{x}_1)=0
\eeq
for every $(p,q)$ and therefore
\beq
\Delta_{1}(t_1,\bm{x}_1)=0
\eeq
for \textit{every} point $(t_1,\bm{x}_1)$.

Observe that rather than showing that the conditions of the claim guarantee that the selective update does not affect the one-point function outside the causal support of the interaction region, we have proven that it does not affect it \textit{at all}. As a consequence, to show that the selective update alters the state of the field outside the causal future of the detector under the hypotheses of the claim's statement, we need to consider the two-point function of the updated state. 

First, we have by \eqref{change in n-point function} for $n=2$ that
\begin{align}
\Delta_2(\mathsf{x}_1,\mathsf{x}_2)&=\langle \hat{M}_{s,\psi}^{\dagger}\phih(\mathsf{x}_1)\phih(\mathsf{x}_2)\hat{M}_{s,\psi} \rangle_{\rhoh_{\phi}}\\
& \phantom{==}-\langle \hat{M}_{s,\psi}^{\dagger}\hat{M}_{s,\psi}\rangle_{\rhoh_{\phi}}\langle \phih(\mathsf{x}_1)\phih(\mathsf{x}_2)\rangle_{\rhoh_{\phi}} \;. \nonumber 
\end{align}
Following the same calculations as in Subsection~\ref{appendix subsection: the general case} of this appendix, but with two field operators instead of one, it is straightforward to see that, up to second order in $\lambda$,
\begin{align}
&\Delta_{2}(t_1,\bm{x}_1,t_2,\bm{x}_2) \nonumber\\
&= 2\lambda \int \diff t \, \diff^{d}\spatial{x} \,\chi(t) F(\spatial{x}) \Im(\braket{\psi}{s}\bra{s}\muh(t)\ket{\psi}) \nonumber\\
&\qquad \qquad \times \textrm{Cov}_{\rhoh_{\phi}}\big[\phih(t,\spatial{x}),\phih(t_1,\bm{x}_1)\phih(t_2,\bm{x}_2)\big] \\
&\phantom{=}\;-\ii\lambda \int \diff t \, \diff^{d}\spatial{x} \, \chi(t) F(\spatial{x}) \braket{\psi}{s}\bra{s}\muh(t)\ket{\psi} \nonumber\\
&\qquad \qquad \qquad \times\langle [ \phih(t,\spatial{x}),\phih(t_1,\bm{x}_1)\phih(t_2,\bm{x}_2)] \rangle_{\rhoh_{\phi}} \nonumber\\
&\phantom{=}\;-\lambda^2 \int \diff t \, \diff t' \, \diff^{d}\spatial{x}\, \diff^{d}\spatial{x}'\,\chi(t)\chi(t') \nonumber\\
& \qquad \qquad \qquad \quad \times \, F(\spatial{x})F(\spatial{x}') \theta(t-t') \,\cdot \,\mathcal{R}_2 \nonumber\\
&\phantom{=}\;+O(\lambda^3) \nonumber
\end{align}
where
\begin{align}
&\mathcal{R}_2= \braket{s}{\psi}\bra{\psi}\muh(t')\muh(t)\ket{s} \\
&\qquad \quad \times\langle \phih(t',\spatial{x}')[\phih(t,\spatial{x}),\phih(t_1,\bm{x}_1)\phih(t_2,\bm{x}_2] \rangle_{\rhoh_{\phi}} \nonumber\\
&\phantom{==}\;+\braket{\psi}{s}\bra{s}\muh(t)\muh(t')\ket{\psi} \nonumber\\
&\qquad \quad \times\langle [\phih(t_1,\bm{x}_1)\phih(t_2,\bm{x}_2,\phih(t,\spatial{x})]\phih(t',\spatial{x}')\rangle_{\rhoh_{\phi}} \nonumber\\
&\phantom{==}\;+2\Re(\mathcal{S}_2) \nonumber
\end{align}
with
\begin{align}\label{expr. S2}
\mathcal{S}_2&=\big(\braket{\psi}{s}\bra{s}\muh(t)\muh(t')\ket{\psi}\\
&\phantom{=}\qquad\qquad\;-\bra{\psi}\muh(t)\ket{s}\bra{s}\muh(t')\ket{\psi}\big) \nonumber\\
&\;\times\,\big(\langle \phih(t,\spatial{x})\phih(t_1,\bm{x}_1)\phih(t_2,\bm{x}_2)\phih(t',\spatial{x}')\rangle_{\rhoh_{\phi}} \nonumber\\
&\;\;-\langle \phih(t,\spatial{x})\phih(t',\spatial{x}') \rangle_{\rhoh_{\phi}} \langle \phih(t_1,\bm{x}_1)\phih(t_2,\bm{x}_2) \rangle_{\rhoh_{\phi}}\big) \;. \nonumber  
\end{align} 
When we consider \mbox{$\ket{s},\ket{\psi}\in\{\ket{g},\ket{e}\}$}, the first order in $\lambda$ becomes zero. Moreover, when $(t_1,\bm{x}_1)$ and $(t_2,\bm{x}_2)$ are spacelike separated from the detector, the first two terms of $\mathcal{R}_2$, that depend on commutators, are zero, and only the real part of $\mathcal{S}_2$ remains. When $\ket{s}$ and $\ket{\psi}$ are eigenstates of the detector's Hamiltonian, then it can be checked that the first factor in $\mathcal{S}_2$ is \mbox{$\pm e^{\pm\ii\Omega(t-t')}$}, where the signs depend on whether we take the ground or the excited state for each of $\ket{s}$ and $\ket{\psi}$. For the second factor, because the field state is Gaussian, it holds that
\begin{align}
&w_{4}(\mathsf{x},\mathsf{x}_1,\mathsf{x}_2,\mathsf{x}')-w_2(\mathsf{x},\mathsf{x}')w_2(\mathsf{x}_1,\mathsf{x}_2)\\
&\phantom{=\,}=w_2(\mathsf{x},\mathsf{x}_1)w_2(\mathsf{x}_2,\mathsf{x})+w_2(\mathsf{x},\mathsf{x}_2)w_2(\mathsf{x}_1,\mathsf{x}') \;. \nonumber
\end{align}
This makes apparent that if we exchange $t$ and $t'$, the modified $\mathcal{S}_2$ is the complex conjugate of the original. Taking advantage of the fact that only the real part of $\mathcal{S}_2$ contributes, and proceeding as in Subsection~\ref{appendix subsection: the ground and excited states} to get rid of the Heaviside step functions $\theta$, we conclude that under the conditions of the claim above,
\begin{align}\label{general delta 2 lemma}
&\Delta_2(t_1,\bm{x}_1,t_2,\bm{x}_2)=\pm\lambda^2\int\diff t\,\diff t'\,\diff^d \bm{x}\,\diff^d \bm{x}' \nonumber\\
&\quad\times\chi(t)\chi(t') F(\bm{x})F(\bm{x}')\,e^{\pm\ii\Omega(t-t')}\\
&\quad\times \big(w_2(t,\bm{x},t_1,\bm{x}_1)w_2(t_2,\bm{x}_2,t',\bm{x}') \nonumber\\
&\qquad+w_2(t,\bm{x},t_2,\bm{x}_2)w_2(t_1,\bm{x}_1,t',\bm{x}')\big)+O(\lambda^3)\;, \nonumber
\end{align}
which as a distribution is non-zero in general. Consider for example the case in which the initial state of the field is the vacuum. In that case the last factor of the integrand involving the two-point functions is
\begin{align}
&\lambda^2\int\frac{\diff^d\bm{k}}{2(2\pi)^d\omega_{\bm{k}}}\int\frac{\diff^d\bm{k}'}{2(2\pi)^d\omega_{\bm{k}'}}\Big( e^{-\ii[\omega_{\bm{k}}(t-t_1)-\bm{k}\cdot(\bm{x}-\bm{x}_1)]} \nonumber\\[2mm]
&\times e^{-\ii[\omega_{\bm{k}'}(t_2-t')-\bm{k}'\cdot(\bm{x}_2-\bm{x}')]}+e^{-\ii[\omega_{\bm{k}}(t-t_2)-\bm{k}\cdot(\bm{x}-\bm{x}_2)]} \nonumber\\[1mm]
& \times e^{-\ii[\omega_{\bm{k}'}(t_1-t')-\bm{k}'\cdot(\bm{x}_1-\bm{x}')]} \Big) \;. 
\end{align}
Particularizing for $\ket{\psi}=\ket{g}$ and $\ket{s}=\ket{e}$ and using \eqref{general delta 2 lemma}, we get that if $(t_1,\bm{x}_1)$ and $(t_2,\bm{x}_2)$ are spacelike separated from the detector,
\begin{align}
&\Delta_2(t_1,\bm{x}_1,t_2,\bm{x}_2)=\frac{\lambda^2}{4(2\pi)^{d-1}}\int\frac{\diff^d\bm{k}}{\omega_{\bm{k}}}\int\frac{\diff^d\bm{k}'}{\omega_{\bm{k}'}} \nonumber\\[2mm]
&\qquad \quad\times \widetilde{\chi}(\omega_{\bm{k}}+\Omega)\widetilde{\chi}(\omega_{\bm{k}'}+\Omega)^* \widetilde{F}(\bm{k})^*\widetilde{F}(\bm{k}')\\[2mm]
&\qquad \quad \times \Big( e^{\ii(\omega_{\bm{k}}t_1-\omega_{\bm{k}'}t_2-\bm{k}\cdot\bm{x}_1+\bm{k}'\cdot\bm{x}_2)} \nonumber\\
&\qquad \qquad + e^{-\ii(\omega_{\bm{k}}t_2-\omega_{\bm{k}'}t_1-\bm{k}\cdot\bm{x}_2+\bm{k}'\cdot\bm{x}_1)}\Big)+O(\lambda^3)\;. \nonumber
\end{align}
This expression does not cancel out in general, as can be checked considering for example Gaussian smearings and switchings.
%Prescribing specific switchings and smearings (for example, Gaussians), the distribution associated to this $\Delta_2$ can be numerically evaluated for functions supported in regions spacelike separated from the detector, checking that it does not cancel out in general. 

With these calculations we have discarded the last of the cases that remained open to the possibility of applying the selective update globally in a way compatible with causality. We thus conclude that the selective update cannot be applied outside the causal future of the interaction region if we want the update to be consistent with the relativistic nature of QFT.

\section{Update rules for {\itshape n\,}-point functions}\label{appendix: update rules for n-point functions}  

In this appendix we give some of the details behind the perturbative results of sections~\ref{subsection: non-selective update} and~\ref{subsection: selective update}, where we formulated the update rule for the $n$-point functions explicitly in terms of the initial $n$-point functions to first order in $\lambda$. We will reuse some of the calculations already performed in Appendix~\ref{appendix: causal behaviour of the selective update}.

\subsection{Non-selective case}\label{appendix subsection: non-selective case}

After a non-selective measurement, we consider the update $w_{n}^{\textsc{NS}}$ given in~\eqref{non-selective updated n-point function non-perturbative} for the $n$-point function. By~\eqref{expansion zeroth order}, we have that 
\begin{align}
\big(\hat{M}_{l,\psi}^{\dagger}\phih(t_1,\bm{x}_1)&\hdots\phih(t_n,\bm{x}_n)\hat{M}_{l,\psi} \big)^{(0)}\\
&=\braket{\psi}{l}\braket{l}{\psi}\phih(t_1,\bm{x}_1)\hdots\phih(t_n,\bm{x}_n) \nonumber
\end{align}
for $l=s,\bar{s}$. Since $\proj{s}{s}+\proj{\bar{s}}{\bar{s}}=\mathds{1}_{\textrm{d}}$ and $\ket{\psi}$ is normalized, the zeroth order of $w_{n}^{\textsc{NS}}$ is
\begin{align}
&w_{n}^{\textsc{NS}}(t_1,\bm{x}_1,\hdots,t_n,\bm{x}_n)^{(0)} \nonumber\\
&\phantom{======}=\langle \phih(t_1,\bm{x}_1)\cdots\phih(t_n,\bm{x}_n) \rangle_{\rhoh_{\phi}}\\
&\phantom{============}=w_n(t_1,\bm{x}_1,\hdots,t_n,\bm{x}_n) \;. \nonumber
\end{align}
For the first order, by~\eqref{expansion first order 10} we have
\begin{align}
&\hat{M}_{l,\psi}^{\dagger\;(1)}\phih(t_1,\bm{x}_1)\cdots\phih(t_n,\bm{x}_n)\hat{M}_{l,\psi}^{(0)} \nonumber\\
&=\ii \braket{\psi}{l}^{\ast} \int\diff t \,\diff^{d}\spatial{x}\,\chi(t)F(\spatial{x})\bra{l}\muh(t)\ket{\psi}^{\ast}\\
&\phantom{====}\;\times\phih(t,\spatial{x})\phih(t_1,\bm{x}_1)\cdots\phih(t_n,\bm{x}_n) \nonumber
\end{align}
and by~\eqref{expansion first order 01}
\begin{align}
&\hat{M}_{l,\psi}^{\dagger\;(0)}\phih(t_1,\bm{x}_1)\cdots\phih(t_n,\bm{x}_n)\hat{M}_{l,\psi}^{(1)} \nonumber\\
&=-\ii \braket{\psi}{l} \int\diff t \,\diff^{d}\spatial{x}\,\chi(t)F(\spatial{x})\bra{l}\muh(t)\ket{\psi}\\
&\phantom{====}\;\times\phih(t_1,\bm{x}_1)\cdots\phih(t_n,\bm{x}_n)\phih(t,\spatial{x}) \;, \nonumber
\end{align}
for $l=s,\bar{s}$. Taking the expectation values and taking into account again that $\{\ket{s},\ket{\bar{s}}\}$ form an orthonormal basis of the Hilbert space of the detector,
\begin{align}
&w_{n}^{\textsc{NS}}(t_1,\bm{x}_1,\hdots,t_n,\bm{x}_n)^{(1)} \nonumber\\
&\phantom{=}=\ii\int\diff t\,\diff^d\bm{x}\,\chi(t)F(\bm{x})\bra{\psi}\hat{\mu}(t)\ket{\psi}\\
&\phantom{===\;}\times\big(\langle \phih(t,\bm{x})\phih(t_1,\bm{x}_1)\cdots\phih(t_n,\bm{x}_n)\rangle_{\rhoh_{\phi}} \nonumber\\[1mm]
&\phantom{======\;}-\langle\phih(t_1,\bm{x}_1)\cdots\phih(t_n,\bm{x}_n)\phih(t,\bm{x}) \rangle_{\rhoh_{\phi}}\big) \;.\nonumber
\end{align}
This equation, along with the zeroth order and the definition of $n$-point functions, yields~\eqref{non-selective updated n-point function}. The particularization to $n=2$ leads immediately to~\eqref{non-selective updated two-point function}. For $n=1$, we can use the property
\begin{equation}
    w_{2}(t,\bm{x},t_1,\bm{x}_1)^*=w_{2}(t_1,\bm{x}_1,t,\bm{x}) \;.
\end{equation}
Thence,
\begin{align}
\ii w_2(t,\bm{x},t_1,\bm{x}_1)-\ii w_2(&t_1,\bm{x}_1,t,\bm{x})\\
&=\Im\big[w_2(t_1,\bm{x}_1,t,\bm{x})\big] \;, \nonumber
\end{align}
which yields~\eqref{non-selective updated one-point function}. 

In a completely analogous way, we can perform the somewhat more tedious calculations leading to the expression for $w_{n}^{\textsc{NS}}$ to second order in $\lambda$. Using the expansion in~\eqref{eq. expansion second order} and Eqs.~\eqref{M operator order 0}, \eqref{M operator order 1} and \eqref{M operator order 2},
\begin{align}
&\hat{M}_{l,\psi}^{\dagger\;(2)}\phih(t_1,\bm{x}_1)\cdots\phih(t_n,\bm{x}_n)\hat{M}_{l,\psi}^{(0)} \nonumber\\
&=-\braket{\psi}{l}^*\int\diff t\,\diff t'\,\diff^d\bm{x}\,\diff^d\bm{x}'\,\theta(t-t')\chi(t)\chi(t') \\
&\phantom{===\;}\times F(\bm{x}) F(\bm{x}') \bra{s}\hat{\mu}(t)\hat{\mu}(t')\ket{\psi}^*\nonumber\\
&\phantom{===\;}\times \phih(t',\bm{x}')\phih(t,\bm{x}) \phih(t_1,\bm{x}_1)\cdots\phih(t_n,\bm{x}_n) \;, \nonumber\\
&\hat{M}_{l,\psi}^{\dagger\;(0)}\phih(t_1,\bm{x}_1)\cdots\phih(t_n,\bm{x}_n)\hat{M}_{l,\psi}^{(2)} \nonumber\\
&=-\braket{\psi}{l}\int\diff t\,\diff t'\,\diff^d\bm{x}\,\diff^d\bm{x}'\,\theta(t-t')\chi(t)\chi(t') \\
&\phantom{===\;}\times F(\bm{x}) F(\bm{x}') \bra{s}\hat{\mu}(t)\hat{\mu}(t')\ket{\psi}\nonumber\\
&\phantom{===\;}\times \phih(t_1,\bm{x}_1)\cdots\phih(t_n,\bm{x}_n) \phih(t,\bm{x})\phih(t',\bm{x}') \;, \nonumber\\
&\hat{M}_{l,\psi}^{\dagger\;(1)}\phih(t_1,\bm{x}_1)\cdots\phih(t_n,\bm{x}_n)\hat{M}_{l,\psi}^{(1)} \nonumber\\
&=\braket{\psi}{l}\int\diff t\,\diff t'\,\diff^d\bm{x}\,\diff^d\bm{x}'\,\chi(t)\chi(t')F(\bm{x}) F(\bm{x}') \\
&\phantom{===\;}\times\bra{\psi}\hat{\mu}(t)\ket{s} \bra{s}\hat{\mu}(t')\ket{\psi}\nonumber\\
&\phantom{===\;}\times \phih(t,\bm{x})\phih(t_1,\bm{x}_1)\cdots\phih(t_n,\bm{x}_n) \phih(t',\bm{x}') \;, \nonumber
\end{align}
for $l=s,\bar{s}$. Taking again the expectation values, we arrive at the second order contribution to $w_{n}^{\textsc{NS}}$,
\begin{align}
&w_{n}^{\textsc{NS}}(t_1,\bm{x}_1,\hdots,t_n,\bm{x}_n)^{(2)} \nonumber\\
&=-\int\diff t\,\diff t'\,\diff^d\bm{x}\,\diff^d\bm{x}'\,\chi(t)\chi(t')F(\bm{x})F(\bm{x}')\\
&\times\Big[ \theta(t-t') \big( \bra{\psi}\hat{\mu}(t')\hat{\mu}(t)\ket{\psi} \nonumber\\
& \phantom{=====}\times w_{n+2}(t',\bm{x}',t,\bm{x},t_1,\bm{x}_1,\hdots,t_n,\bm{x}_n) \nonumber\\
&\phantom{\times \Big( \theta(t-t')\!}+\bra{\psi}\hat{\mu}(t')\hat{\mu}(t)\ket{\psi} \nonumber\\
& \phantom{=====}\times w_{n+2}(t_1,\bm{x}_1,\hdots,t_n,\bm{x}_n,t,\bm{x},t',\bm{x}') \big) \nonumber\\
&\phantom{\times\Big[}-\bra{\psi}\hat{\mu}(t)\hat{\mu}(t')\ket{\psi}\nonumber\\
&\phantom{=====}\times w_{n+2}(t,\bm{x},t_1,\bm{x}_1,\hdots,t_n,\bm{x}_n,t',\bm{x}') \Big] \;. \nonumber
\end{align}
All together,
\begin{align}\label{non-selective updated n-function up to second order in lambda}
w_{n}^{\textsc{NS}}&(t_1,\bm{x}_1,\hdots,t_n,\bm{x}_n)=w_{n}(t_1,\bm{x}_1,\hdots,t_n,\bm{x}_n)\\
&+\ii\lambda\int\diff t\,\diff^{d}{\bm{x}}\,\chi(t)F(\bm{x})\bra{\psi}\hat{\mu}(t)\ket{\psi} \nonumber\\
&\phantom{====}\times \big( w_{n+1}(t,\bm{x},t_1,\bm{x}_1,\hdots,t_n,\bm{x}_n) \nonumber\\[2mm]
&\phantom{=======}-w_{n+1}(t_1,\bm{x}_1,\hdots,t_n,\bm{x}_n,t,\bm{x}) \big) \nonumber\\
&-\lambda^2\int\diff t\,\diff t'\,\diff^d\bm{x}\,\diff^d\bm{x}'\,\chi(t)\chi(t')F(\bm{x})F(\bm{x}') \nonumber\\
&\phantom{==\;}\times\Big[ \theta(t-t') \big( \bra{\psi}\hat{\mu}(t')\hat{\mu}(t)\ket{\psi} \nonumber\\
& \phantom{=====}\times w_{n+2}(t',\bm{x}',t,\bm{x},t_1,\bm{x}_1,\hdots,t_n,\bm{x}_n) \nonumber\\
&\phantom{==\;\times\Big[}+\bra{\psi}\hat{\mu}(t')\hat{\mu}(t)\ket{\psi} \nonumber\\
& \phantom{=====}\times w_{n+2}(t_1,\bm{x}_1,\hdots,t_n,\bm{x}_n,t,\bm{x},t',\bm{x}') \big) \nonumber\\
&\phantom{==\;}-\bra{\psi}\hat{\mu}(t)\hat{\mu}(t')\ket{\psi}\nonumber\\
&\phantom{=====}\times w_{n+2}(t,\bm{x},t_1,\bm{x}_1,\hdots,t_n,\bm{x}_n,t',\bm{x}') \Big] \; \nonumber\\
&+O(\lambda^3) \;. \nonumber
\end{align}

\subsection{Selective case}\label{appendix subsection: selective case}

For the selective update $w_n^{\textsc{S}}$, we consider the update corresponding to the case in which not all the points in the argument are outside the causal future of the region in which the projective measurement on the detector is performed, $\mathcal{P}$ (otherwise the update is just the non-selective one). Recalling \eqref{selective updated n-point function def 2}, we need to consider two expansions. First, by~\eqref{expansion up to second order},
\begin{align}
&\langle \hat{M}_{s,\psi}^{\dagger}\phih(t_1,\bm{x}_1)\cdots\phih(t_n,\bm{x}_n)\hat{M}_{s,\psi}\rangle_{\rhoh_{\phi}} \label{numerator n-point functions expansion to first order in lambda}\\
&\phantom{\;}=\langle\hat{M}_{s,\psi}^{\dagger\;(0)}\phih(t_1,\bm{x}_1)\cdots\phih(t_n,\bm{x}_n)\hat{M}_{s,\psi}^{(0)}\rangle_{\rhoh_{\phi}} \nonumber\\
&\phantom{\;=\;}+\lambda \Big( \langle\hat{M}_{s,\psi}^{\dagger\;(1)}\phih(t_1,\bm{x}_1)\cdots\phih(t_n,\bm{x}_n)\hat{M}_{s,\psi}^{(0)}\rangle_{\rhoh_{\phi}} \nonumber\\
&\phantom{=====}+\langle\hat{M}_{s,\psi}^{\dagger\;(0)}\phih(t_1,\bm{x}_1)\cdots\phih(t_n,\bm{x}_n)\hat{M}_{s,\psi}^{(1)}\rangle_{\rhoh_{\phi}}\Big) \nonumber\\
&\phantom{\;=\;}+O(\lambda^2) \;. \nonumber
\end{align}
And second, by~\eqref{expansion up to second order denominator},
\begin{align}
\langle \hat{E}_{s,\phi}\rangle_{\rhoh_{\phi}}&=\langle\hat{M}_{s,\psi}^{\dagger}\hat{M}_{s,\psi}\rangle_{\rhoh_{\phi}} \label{denominator n-point functions expansion to first order in lambda}\\
&=\langle\hat{M}_{s,\psi}^{\dagger\;(0)}\hat{M}_{s,\psi}^{(0)}\rangle_{\rhoh_{\phi}}+\lambda\Big(\langle\hat{M}_{s,\psi}^{\dagger\;(1)}\hat{M}_{s,\psi}^{(0)}\rangle_{\rhoh_{\phi}} \nonumber\\
&\phantom{=}\;\;+\langle\hat{M}_{s,\psi}^{\dagger\;(0)}\hat{M}_{s,\psi}^{(1)}\rangle_{\rhoh_{\phi}}\Big)+O(\lambda^2) \;. \nonumber
\end{align}
If the zeroth order term is not zero,
\begin{equation}
\langle\hat{M}_{s,\psi}^{\dagger\;(0)}\hat{M}_{s,\psi}^{(0)}\rangle_{\rhoh_{\phi}}=|\!\braket{s}{\psi}\!|^2 \neq 0 \;,
\end{equation}
we can give an expansion for the inverse
\begin{align}\label{denominator to order 1 in lambda}
&\big(\langle \hat{E}_{s,\psi}\rangle_{\rhoh_{\phi}}\big)^{-1} \nonumber\\
&=\big(\langle \hat{M}_{s,\psi}^{\dagger\;(0)}\hat{M}_{s,\psi}^{(0)}\rangle_{\rhoh_{\phi}}\big)^{-1}-\lambda\big(\langle \hat{M}_{s,\psi}^{\dagger\;(0)}\hat{M}_{s,\psi}^{(0)}\rangle_{\rhoh_{\phi}}\big)^{-2}\\
&\phantom{===}\times\Big(\langle\hat{M}_{s,\psi}^{\dagger\;(1)}\hat{M}_{s,\psi}^{(0)}\rangle_{\rhoh_{\phi}}+\langle\hat{M}_{s,\psi}^{\dagger\;(0)}\hat{M}_{s,\psi}^{(1)}\rangle_{\rhoh_{\phi}}\Big) \nonumber\\
&\phantom{=====}+O(\lambda^2) \;. \nonumber
\end{align}
Thus, by \eqref{expansion zeroth order} and \eqref{expansion zeroth order denominator}, if $\braket{s}{\psi}\neq 0$, the zeroth order of $w_{n}^{\textsc{S}}$ is
\begin{align}
&w_{n}^{\textsc{S}}(t_1,\bm{x}_1,\hdots,t_n,\bm{x}_n)^{(0)}=\big( \langle\hat{M}_{s,\psi}^{\dagger\;(0)}\hat{M}_{s,\psi}^{(0)}\rangle_{\rhoh_{\phi}} \big)^{-1} \nonumber\\
&\phantom{===\;}\times\langle\hat{M}_{s,\psi}^{\dagger\;(0)}\phih(t_{1,\bm{x}_1})\cdots\phih(t_n,\bm{x}_n)\hat{M}_{s,\psi}^{(0)}\rangle_{\rhoh_{\phi}} \label{selective update zeroth order non-orthogonal states}\\
&=w_{n}(t_1,\bm{x}_1,\hdots,t_n,\bm{x}_n) \;.\nonumber
\end{align}
Also, for the first order,
\begin{align}
&w_{n}^{\textsc{S}}(t_1,\bm{x}_1,\hdots,t_n,\bm{x}_n)^{(1)}=\big( \langle\hat{M}_{s,\psi}^{\dagger\;(0)}\hat{M}_{s,\psi}^{(0)}\rangle_{\rhoh_{\phi}} \big)^{-1} \nonumber\\
&\phantom{==}\times\Big( \langle\hat{M}_{s,\psi}^{\dagger\;(1)}\phih(t_1,\bm{x}_1)\cdots\phih(t_n,\bm{x}_n)\hat{M}_{s,\psi}^{(0)}\rangle_{\rhoh_{\phi}} \nonumber\\
&\phantom{===}+\langle\hat{M}_{s,\psi}^{\dagger\;(0)}\phih(t_1,\bm{x}_1)\cdots\phih(t_n,\bm{x}_n)\hat{M}_{s,\psi}^{(1)}\rangle_{\rhoh_{\phi}}\Big) \label{selective update first order non-orthogonal states}\\
&-\big(\langle \hat{M}_{s,\psi}^{\dagger\;(0)}\hat{M}_{s,\psi}^{(0)}\rangle_{\rhoh_{\phi}}\big)^{-2} \nonumber \\
&\phantom{==}\times\Big(\langle\hat{M}_{s,\psi}^{\dagger\;(1)}\hat{M}_{s,\psi}^{(0)}\rangle_{\rhoh_{\phi}}+\langle\hat{M}_{s,\psi}^{\dagger\;(0)}\hat{M}_{s,\psi}^{(1)}\rangle_{\rhoh_{\phi}}\Big) \nonumber\\
&\phantom{==}\times\langle\hat{M}_{s,\psi}^{\dagger\;(0)}\phih(t_1,\bm{x}_1)\cdots\phih(t_n,\bm{x}_n)\hat{M}_{s,\psi}^{(0)}\rangle_{\rhoh_{\phi}} \;.\nonumber
\end{align}
By \eqref{expansion first order} and \eqref{expansion first order denominator},
\begin{align}
&w_{n}^{\textsc{S}}(t_1,\bm{x}_1,\hdots,t_n,\bm{x}_n)^{(1)} \nonumber\\
&=\frac{1}{|\!\braket{s}{\psi}\!|^{2}}\int\diff t\,\diff^d\bm{x}\,\chi(t)F(\bm{x})\Big( \ii\braket{s}{\psi}\bra{\psi}\hat{\mu}(t)\ket{s} \nonumber\\
&\phantom{=}\times \langle \phih(t,\bm{x})\phih(t_1,\bm{x}_1)\cdots\phih(t_n,\bm{x}_n) \rangle_{\rhoh_{\phi}} -\ii\braket{\psi}{s} \nonumber\\
&\phantom{=}\times\bra{s}\hat{\mu}(t)\ket{\psi}\langle \phih(t_1,\bm{x}_1)\cdots\phih(t_n,\bm{x}_n)\phih(t,\bm{x}) \rangle_{\rhoh_{\psi}}\\
&\phantom{=}-2\Im(\braket{\psi}{s}\bra{s}\hat{\mu}(t)\ket{\psi}) \nonumber\\
&\phantom{=}\times\langle \phih(t_1,\bm{x}_1)\cdots\phih(t_n,\bm{x}_n)\rangle_{\rhoh_{\phi}}\langle\phih(t,\bm{x})\rangle_{\rhoh_{\phi}} \Big) \;. \nonumber
\end{align}
This equation, along with the one for the zeroth order, yields~\eqref{selective update n-point function perturbative}. Particularization for $n=2$ gives~\eqref{selective update two-point function perturbative} immediately. For the one-point function, it only remains to use 
\begin{equation}
    w_{2}(t,\bm{x},t_1,\bm{x}_1)^*=w_{2}(t_1,\bm{x}_1,t,\bm{x}) \;
\end{equation}
as for the non-selective case, to get the final expression in \eqref{selective update one-point function perturbative}.

We can get involved in more cumbersome calculations in order to arrive at the second order terms of the expression for $w_n^{\textsc{S}}$ when $\braket{s}{\psi}\neq 0$. First, by~\eqref{eq. expansion second order},
\begin{align}
\langle\hat{M}_{s,\psi}^{\dagger}&\phih(t_1,\bm{x}_1)\cdots\phih(t_n,\bm{x}_n)\hat{M}_{s,\psi}\rangle_{\rhoh_{\phi}}^{(2)} \nonumber\\
&=\langle\hat{M}_{s,\psi}^{\dagger\;(2)}\phih(t_1,\bm{x}_1)\cdots\phih(t_n,\bm{x}_n)\hat{M}_{s,\psi}^{(0)}\rangle_{\rhoh_{\phi}} \\
&\phantom{=\;}+\langle\hat{M}_{s,\psi}^{\dagger\;(0)}\phih(t_1,\bm{x}_1)\cdots\phih(t_n,\bm{x}_n)\hat{M}_{s,\psi}^{(2)}\rangle_{\rhoh_{\phi}} \nonumber\\
&\phantom{=\;}+\langle\hat{M}_{s,\psi}^{\dagger\;(1)}\phih(t_1,\bm{x}_1)\cdots\phih(t_n,\bm{x}_n)\hat{M}_{s,\psi}^{(1)}\rangle_{\rhoh_{\phi}} \nonumber \;.
\end{align}
Removing the field operators in this last equation we get the expansion to second order of $\langle\hat{E}_{s,\psi}\rangle_{\rhoh_{\phi}}$, and thus for its inverse the second order contribution is
\begin{align}
&(\langle\hat{E}_{s,\psi}\rangle_{\rhoh_{\phi}}^{-1})^{(2)}=\langle\hat{M}_{s,\psi}^{\dagger\;(0)}\hat{M}_{s,\psi}^{(0)}\rangle_{\rhoh_{\phi}}^{-3} \\
&\phantom{==}\times \Big[ \big( \langle\hat{M}_{s,\psi}^{\dagger\;(1)}\hat{M}_{s,\psi}^{(0)}\rangle_{\rhoh_{\phi}}+\langle\hat{M}_{s,\psi}^{\dagger\;(0)}\hat{M}_{s,\psi}^{(1)}\rangle_{\rhoh_{\phi}} \big)^2 \nonumber\\
&\phantom{==\times\Big[}-\langle\hat{M}_{s,\psi}^{\dagger\;(0)}\hat{M}_{s,\psi}^{(0)}\rangle_{\rhoh_{\phi}}\big( \langle\hat{M}_{s,\psi}^{\dagger\;(2)}\hat{M}_{s,\psi}^{(0)}\rangle_{\rhoh_{\phi}} \nonumber\\
&\phantom{==\times\Big[}+\langle\hat{M}_{s,\psi}^{\dagger\;(0)}\hat{M}_{s,\psi}^{(2)}\rangle_{\rhoh_{\phi}}+\langle\hat{M}_{s,\psi}^{\dagger\;(1)}\hat{M}_{s,\psi}^{(1)}\rangle_{\rhoh_{\phi}}\big) \Big] \;.\nonumber 
\end{align}
Taking also into account the first order contributions in Eqs.~\eqref{numerator n-point functions expansion to first order in lambda} and~\eqref{denominator to order 1 in lambda}, and proceeding along the lines of the calculations for $w_n^{\textsc{S}\,(0)}$ and $w_n^{\textsc{S}\,(1)}$ we get that
\begin{align}\label{second order of selective update n-point functions non-orthogonal}
&w_{n}^{\textsc{S}}(t_1,\bm{x}_1,\hdots,t_n,\bm{x}_n)^{(2)} \\
&=-\frac{1}{|\!\braket{s}{\psi}\!|^2}\int\diff t\,\diff t'\,\diff^d\bm{x}\,\diff^d\bm{x}'\,\chi(t)\chi(t') \nonumber\\
&\times F(\bm{x})F(\bm{x}') \bigg( \theta(t-t')\mathcal{J} +\mathcal{K} -\frac{\mathcal{L}}{|\!\braket{s}{\psi}\!|^2}\bigg) \;, \nonumber
\end{align}
where
\begin{align}
&\mathcal{J}=\braket{s}{\psi}\bra{\psi}\hat{\mu}(t')\mu(t)\ket{s} \\
&\phantom{===}\times \big(w_{n+2}(t',\bm{x}',t,\bm{x},t_1,\bm{x}_1,\hdots,t_n,\bm{x}_n) \nonumber\\
&\phantom{=====}-w_2(t',\bm{x}',t,\bm{x})w_n(t_1,\bm{x}_1,\hdots,t_n,\bm{x}_n) \big) \nonumber\\
&\phantom{\mathcal{J}=}+\braket{\psi}{s}\bra{s}\hat{\mu}(t)\hat{\mu}(t')\ket{\psi} \nonumber\\ 
&\phantom{===}\times \big( w_{n+2}(t_1,\bm{x}_1,\hdots,t_n,\bm{x}_n,t,\bm{x},t',\bm{x}') \nonumber\\
&\phantom{=====}-w_n(t_1,\bm{x}_1,\hdots,t_n,\bm{x}_n)w_2(t,\bm{x},t',\bm{x}') \big) \;, \nonumber
\end{align}
also
\begin{align}
&\mathcal{K}=\bra{\psi}\hat{\mu}(t)\ket{s}\bra{s}\hat{\mu}(t')\ket{\psi} \\
&\phantom{=\,}\times \big( w_{n+2}(t,\bm{x},t_1,\bm{x}_1,\hdots,t_n,\bm{x}_n,t',\bm{x}') \nonumber\\
&\phantom{==}-w_1(t,\bm{x})w_{n+1}(t_1,\bm{x}_1,\hdots,t_n,\bm{x}_n,t',\bm{x}') \nonumber\\
&\phantom{==}-w_{n+1}(t,\bm{x},t_1,\bm{x}_1,\hdots,t_n,\bm{x}_n)w_1(t',\bm{x}') \nonumber\\
&\phantom{==}-w_2(t,\bm{x},t',\bm{x}')w_n(t_1,\bm{x}_1,\hdots,t_n,\bm{x}_n) \nonumber\\
&\phantom{==}+2w_1(t,\bm{x})w_1(t',\bm{x}')w_n(t_1,\bm{x}_1,\hdots,t_n,\bm{x}_n) \big) \nonumber   
\end{align}
and
\begin{align}
&\mathcal{L}=2\textrm{Re}\big(\!\braket{s}{\psi}^2\bra{\psi}\hat{\mu}(t)\ket{s}\bra{\psi}\hat{\mu}(t')\ket{s}\!\big) \\
&\phantom{===}\times w_1(t,\bm{x})w_1(t',\bm{x}')w_n(t_1,\bm{x}_1,\hdots,t_n,\bm{x}_n) \nonumber\\
&\phantom{\mathcal{L}=}+\braket{s}{\psi}^2\bra{\psi}\hat{\mu}(t)\ket{s}\bra{\psi}\hat{\mu}(t')\ket{s} \nonumber\\
&\phantom{===}\times w_1(t,\bm{x})w_{n+1}(t',\bm{x}',t_1,\bm{x}_1,\hdots,t_n,\bm{x}_n) \nonumber \\
&\phantom{\mathcal{L}=}+\braket{\psi}{s}^2\bra{s}\hat{\mu}(t)\ket{\psi}\bra{s}\hat{\mu}(t')\ket{\psi} \nonumber\\
&\phantom{===}\times w_1(t,\bm{x})w_{n+1}(t_1,\bm{x}_1,\hdots,t_n,\bm{x}_n,t',\bm{x}') \;. \nonumber
\end{align}

We finish this appendix by addressing the selective update when $\braket{s}{\psi}=0$. In this case, $\hat{M}_{s,\psi}^{(0)}=0$, and therefore, by~\eqref{numerator n-point functions expansion to first order in lambda} and~\eqref{denominator n-point functions expansion to first order in lambda}, the zeroth order and first order contributions of both the numerator and the denominator in~\eqref{selective updated n-point function def 2} cancel out. Hence,  
\begin{align}
&\langle \hat{M}_{s,\psi}^{\dagger}\phih(t_1,\bm{x}_1)\cdots\phih(t_n,\bm{x}_n)\hat{M}_{s,\psi}\rangle_{\rhoh_{\phi}} \\
&\phantom{\;}=\lambda^2\langle\hat{M}_{s,\psi}^{\dagger\;(1)}\phih(t_1,\bm{x}_1)\cdots\phih(t_n,\bm{x}_n)\hat{M}_{s,\psi}^{(1)}\rangle_{\rhoh_{\phi}} \nonumber\\
&\phantom{\;=\;}+\lambda^3 \Big( \langle\hat{M}_{s,\psi}^{\dagger\;(2)}\phih(t_1,\bm{x}_1)\cdots\phih(t_n,\bm{x}_n)\hat{M}_{s,\psi}^{(1)}\rangle_{\rhoh_{\phi}} \nonumber\\
&\phantom{=====}+\langle\hat{M}_{s,\psi}^{\dagger\;(1)}\phih(t_1,\bm{x}_1)\cdots\phih(t_n,\bm{x}_n)\hat{M}_{s,\psi}^{(2)}\rangle_{\rhoh_{\phi}}\Big) \nonumber\\
&\phantom{\;=\;}+O(\lambda^4) \;, \nonumber
\end{align}
and in particular,
\begin{align}
\langle \hat{E}_{s,\phi}\rangle_{\rhoh_{\phi}}&=\langle\hat{M}_{s,\psi}^{\dagger}\hat{M}_{s,\psi}\rangle_{\rhoh_{\phi}} \\
&=\lambda^2\langle\hat{M}_{s,\psi}^{\dagger\;(1)}\hat{M}_{s,\psi}^{(1)}\rangle_{\rhoh_{\phi}}+\lambda^3\Big(\langle\hat{M}_{s,\psi}^{\dagger\;(2)}\hat{M}_{s,\psi}^{(1)}\rangle_{\rhoh_{\phi}} \nonumber\\
&\phantom{=}\;\;+\langle\hat{M}_{s,\psi}^{\dagger\;(1)}\hat{M}_{s,\psi}^{(2)}\rangle_{\rhoh_{\phi}}\Big)+O(\lambda^4) \;. \nonumber
\end{align}
Since the update depends on the quotient of both expressions, we can drop the factor $\lambda^2$ and proceed as we did formerly with the case \mbox{$\braket{s}{\psi}\neq0$}. For the zeroth order, as in~\eqref{selective update zeroth order non-orthogonal states},
\begin{align}
&w_{n}^{\textsc{S}}(t_1,\bm{x}_1,\hdots,t_n,\bm{x}_n)^{(0)}=\big( \langle\hat{M}_{s,\psi}^{\dagger\;(1)}\hat{M}_{s,\psi}^{(1)}\rangle_{\rhoh_{\phi}} \big)^{-1} \label{selective update zeroth order orthogonal states}\\
&\phantom{===\;}\times\langle\hat{M}_{s,\psi}^{\dagger\;(1)}\phih(t_{1,\bm{x}_1})\cdots\phih(t_n,\bm{x}_n)\hat{M}_{s,\psi}^{(1)}\rangle_{\rhoh_{\phi}} \;. \nonumber
\end{align}
And for the first order, as in~\eqref{selective update first order non-orthogonal states},
\begin{align}
&w_{n}^{\textsc{S}}(t_1,\bm{x}_1,\hdots,t_n,\bm{x}_n)^{(1)}=\big( \langle\hat{M}_{s,\psi}^{\dagger\;(1)}\hat{M}_{s,\psi}^{(1)}\rangle_{\rhoh_{\phi}} \big)^{-1} \nonumber\\
&\phantom{==}\times\Big( \langle\hat{M}_{s,\psi}^{\dagger\;(2)}\phih(t_1,\bm{x}_1)\cdots\phih(t_n,\bm{x}_n)\hat{M}_{s,\psi}^{(1)}\rangle_{\rhoh_{\phi}} \nonumber\\
&\phantom{===}+\langle\hat{M}_{s,\psi}^{\dagger\;(1)}\phih(t_1,\bm{x}_1)\cdots\phih(t_n,\bm{x}_n)\hat{M}_{s,\psi}^{(2)}\rangle_{\rhoh_{\phi}}\Big) \label{selective update first order orthogonal states}\\
&-\big(\langle \hat{M}_{s,\psi}^{\dagger\;(1)}\hat{M}_{s,\psi}^{(1)}\rangle_{\rhoh_{\phi}}\big)^{-2} \nonumber \\
&\phantom{==}\times\Big(\langle\hat{M}_{s,\psi}^{\dagger\;(2)}\hat{M}_{s,\psi}^{(1)}\rangle_{\rhoh_{\phi}}+\langle\hat{M}_{s,\psi}^{\dagger\;(1)}\hat{M}_{s,\psi}^{(2)}\rangle_{\rhoh_{\phi}}\Big) \nonumber\\
&\phantom{==}\times\langle\hat{M}_{s,\psi}^{\dagger\;(1)}\phih(t_1,\bm{x}_1)\cdots\phih(t_n,\bm{x}_n)\hat{M}_{s,\psi}^{(1)}\rangle_{\rhoh_{\phi}} \;.\nonumber
\end{align}
For the sake of clarity, let us denote
\begin{align}
&\mathcal{F}_{n}=\int\diff t\,\diff t'\,\diff^d\bm{x}\,\diff^d\bm{x}'\,\chi(t)\chi(t')F(\bm{x})F(\bm{x}') \\
&\phantom{=====} \times \bra{\psi}\hat{\mu}(t)\ket{s}\bra{s}\hat{\mu}(t')\ket{\psi} \nonumber\\[2mm]
&\phantom{=====}\times w_{n+2}(t,\bm{x},t_1,\bm{x}_1,\hdots,t_n,\bm{x}_n,t',\bm{x}') \nonumber
\end{align}
and
\begin{align}
&\mathcal{G}_{n}=\ii\int\diff t\,\diff t'\,\diff t''\,\diff^d\bm{x}\,\diff^d\bm{x}'\,\diff^d\bm{x}''\,\chi(t)\chi(t')\chi(t'') \\
& \times F(\bm{x})F(\bm{x}') F(\bm{x}'') \,\theta(t-t') \nonumber\\
&\times \Big(\bra{s}\hat{\mu}(t'')\ket{\psi}\bra{\psi}\hat{\mu}(t')\hat{\mu}(t)\ket{s} \nonumber\\
&\phantom{==}\times w_{n+3}(t',\bm{x}',t,\bm{x},t_1,\bm{x}_1,\hdots,t_n,\bm{x}_n,t'',\bm{x}'') \nonumber\\[1.5mm]
&-\bra{\psi}\hat{\mu}(t'')\ket{s}\bra{s}\hat{\mu}(t)\hat{\mu}(t')\ket{\psi} \nonumber\\
&\phantom{=\,}\times w_{n+3}(t'',\bm{x}'',t_1,\bm{x}_1,\hdots,t_n,\bm{x}_n,t,\bm{x},t',\bm{x}') \Big)\;. \nonumber
\end{align}
Thus, by~\eqref{selective update zeroth order orthogonal states} and~\eqref{selective update first order orthogonal states}, when $\braket{s}{\psi}=0$ we have
\begin{align}\label{selective update n-point function orthogonal}
    &w_{n}^{\textsc{S}}(t_1,\bm{x}_1,\hdots,t_n,\bm{x}_n)\\
    &\phantom{==}=\frac{\mathcal{F}_{n}}{\mathcal{F}_{0}} +\frac{\lambda \mathcal{G}_{n}}{\mathcal{F}_{0}}-\frac{\lambda \mathcal{G}_{0}\mathcal{F}_{n}}{\mathcal{F}_{0}^2} + O(\lambda^2) \;.\nonumber
\end{align}

\section{A practical example using {\itshape n\,}-point functions}\label{appendix: a practical example using n-point functions}

In this appendix we study the practical example considered in Section~\ref{section: a practical example with detectors} using the approach based on $n$-point functions to implement the update rule. In particular, we will calculate the joint partial state $\rhoh_{\textsc{ab}}$, and the partial states $\rhoh_{\textsc{a}}$ and $\rhoh_{\textsc{b}}$, using $n$-point functions and its extensions as presented in Sections~\ref{section: update of n-point functions} and~\ref{section: generalization to the presence of entangled third parties}. As we will show, the results are the same as those obtained in Section~\ref{section: a practical example with detectors} using a context-dependent density operator. 

\subsection{Non-selective case}

We consider the case in which Clara performs a non-selective measurement in the first place. For an initial state of the field $\rhoh_\phi$, the initial $n$-point functions are
\begin{equation}\label{Eq. 71}
w_n(\mathsf{x}_1,\hdots,\mathsf{x}_n)=\textrm{tr}_{\phi}\big( \rhoh_{\phi}\phih(\mathsf{x}_1)\cdots\phih(\mathsf{x}_n) \big) \;.
\end{equation}
After the measurement, we update $w_n$ to $w_{n}^{\textsc{NS}}$ following the prescription in Eq.~\eqref{non-selective updated n-point function non-perturbative},
\begin{align}
w&_{n}^{\textsc{NS}}(\mathsf{x}_1,\hdots,\mathsf{x}_n) \nonumber\\
&=\textrm{tr}_{\phi}\big( \hat{M}_{c,\psi}\rhoh_{\phi}\hat{M}_{c,\psi}^{\dagger}\phih(\mathsf{x}_1)\cdots\phih(\mathsf{x}_n) \big) \\
&\phantom{==}+\textrm{tr}_{\phi}\big( \hat{M}_{\bar{c},\psi}\rhoh_{\phi}\hat{M}_{\bar{c},\psi}^{\dagger}\phih(\mathsf{x}_1)\cdots\phih(\mathsf{x}_n) \big) \nonumber\\
&\phantom{====}=\textrm{tr}_{\textsc{c},\phi}\big[ \hat{U}_{\textsc{c}}(\,\proj{\psi}{\psi}\otimes\rhoh_{\phi})\hat{U}_{\textsc{c}}^{\dagger}\phih(\mathsf{x}_1)\cdots\phih(\mathsf{x}_n) \big] \;, \nonumber
\end{align}
where $\hat M_{c,\psi}$ is the $\hat M$ operator as defined in~\eqref{M operator}, for $\ket{c}$ and $\ket{\psi}$ states of the Hilbert space of Clara's detector. Now, we need to include the knowledge about the initial states of detectors A and B in the $n$-point functions using the extended formalism described in Section~\ref{section: generalization to the presence of entangled third parties},
\begin{align}
&\widetilde{w}_{\Gamma,n}(k,l;\mathsf{x}_1,\hdots,\mathsf{x}_n) \nonumber\\
&\phantom{===}=\textrm{tr}\big[ \hat{U}_{\textsc{c}}(\rhoh_{\textsc{a}}\otimes\rhoh_{\textsc{b}}\otimes\proj{\psi}{\psi}\otimes\rhoh_{\phi})\hat{U}_{\textsc{c}}^{\dagger} \\
&\phantom{======}\times \proj{k}{l}\phih(\mathsf{x}_1)\cdots\phih(\mathsf{x}_n) \big] \;, \nonumber
\end{align}
where $\Gamma\subseteq\{A,B\}$ and $\ket{k},\ket{l}$ are elements of an orthonormal basis of the Hilbert space of $\Gamma$. We now update $\widetilde{w}_{\Gamma,n}$ taking into account the time evolution of the detectors A and B coupled to the field as
\begin{align}\label{extended n-point function final practical example}
&\widetilde{w}_{\Gamma,n}'(k,l;\mathsf{x}_1,\hdots,\mathsf{x}_n)  \nonumber\\
&\phantom{===}=\textrm{tr}\big[\hat{U}_{\textsc{a}}\hat{U}_{\textsc{b}}\hat{U}_{\textsc{c}}(\rhoh_{\textsc{a}}\otimes\rhoh_{\textsc{b}}\otimes\proj{\psi}{\psi}\otimes\rhoh_{\phi}) \\
&\phantom{========}\times\hat{U}_{\textsc{c}}^{\dagger}\hat{U}_{\textsc{b}}^{\dagger}\hat{U}_{\textsc{a}}^{\dagger}\proj{k}{l}\phih(\mathsf{x}_1)\cdots\phih(\mathsf{x}_n) \big] \;.\nonumber
\end{align}
Once we have obtained the extended $n$-point function~\eqref{extended n-point function final practical example}, we are in position to calculate the different partial states we are interested in. In particular,
\begin{align}
&\bra{l_\textsc{a},l_\textsc{b}}\rhoh_{\textsc{ab}}'\ket{k_\textsc{a},k_\textsc{b}}=\widetilde{w}_{\{\textsc{a,b}\},0}'((k_\textsc{a},k_\textsc{b}),(l_\textsc{a},l_\textsc{b})) \nonumber\\
&=\textrm{tr}\big[ \hat{U}_{\textsc{a}}\hat{U}_{\textsc{b}}\hat{U}_{\textsc{c}} (\rhoh_{\textsc{a}}\otimes\rhoh_{\textsc{b}}\otimes\proj{\psi}{\psi}\otimes\rhoh_{\phi}) \nonumber\\
&\phantom{==========}\times\hat{U}_{\textsc{c}}^{\dagger}\hat{U}_{\textsc{b}}^{\dagger}\hat{U}_{\textsc{a}}^{\dagger} \proj{k_\textsc{a},k_\textsc{b}}{l_\textsc{a},l_\textsc{b}} \big]  \\
&=\bra{l_\textsc{a},l_\textsc{b}}\textrm{tr}_{\textsc{c},\phi}\big[\hat{U}_{\textsc{a}}\hat{U}_{\textsc{b}}\hat{U}_{\textsc{c}} \nonumber \\
&\phantom{\;==}\times (\rhoh_{\textsc{a}}\otimes\rhoh_{\textsc{b}}\otimes\proj{\psi}{\psi}\otimes\rhoh_{\phi})\hat{U}_{\textsc{c}}^{\dagger}\hat{U}_{\textsc{b}}^{\dagger}\hat{U}_{\textsc{a}}^{\dagger} \big]\ket{k_\textsc{a},k_\textsc{b}} \;, \nonumber
\end{align}
where $\ket{k_\nu},\ket{l_\nu}\in\{\ket{g_\nu},\ket{e_\nu}\}$ for $\nu\in\{\textsc{A},\textsc{B}\}$. Therefore,
\begin{align}\label{Alba-Blanca non-selective 2nd derivation}
&\rhoh_{\textsc{ab}}'=\textrm{tr}_{\textsc{c},\phi}\big[ \hat{U}_{\textsc{a}}\hat{U}_{\textsc{b}}\hat{U}_{\textsc{c}} \\
&\phantom{====}\times(\rhoh_{\textsc{a}}\otimes\rhoh_{\textsc{b}}\otimes\proj{\psi}{\psi}\otimes\rhoh_{\phi})\hat{U}_{\textsc{c}}^{\dagger}\hat{U}_{\textsc{b}}^{\dagger}\hat{U}_{\textsc{a}}^{\dagger} \big] \nonumber
\end{align}
as we obtained in Section~\ref{section: a practical example with detectors}. For the partial state $\rhoh_\textsc{b}'$,
\begin{align}
&\bra{l_\textsc{b}}\rhoh_{\textsc{b}}'\ket{k_\textsc{b}}=\widetilde{w}_{\textsc{b},0}'(k_\textsc{b},l_\textsc{b})\\
&\phantom{==}=\textrm{tr}\big[ \hat{U}_{\textsc{a}}\hat{U}_{\textsc{b}}\hat{U}_{\textsc{c}} (\rhoh_{\textsc{a}}\otimes\rhoh_{\textsc{b}}\otimes\proj{\psi}{\psi}\otimes\rhoh_{\phi}) \nonumber\\
&\phantom{=============}\times\hat{U}_{\textsc{c}}^{\dagger}\hat{U}_{\textsc{b}}^{\dagger}\hat{U}_{\textsc{a}}^{\dagger} \proj{k_\textsc{b}}{l_\textsc{b}} \big] \nonumber
\end{align}
so that, proceeding as for $\rhoh_\textsc{ab}'$ above,
\begin{align}\label{Blanca non-selective 2nd derivation}
&\rhoh_{\textsc{b}}'=\textrm{tr}_{\textsc{a,c},\phi}\big[\hat{U}_{\textsc{a}}\hat{U}_{\textsc{b}}\hat{U}_{\textsc{c}} \nonumber\\
&\phantom{====}\times(\rhoh_{\textsc{a}}\otimes\rhoh_{\textsc{b}}\otimes\proj{\psi}{\psi}\otimes\rhoh_{\phi})\hat{U}_{\textsc{c}}^{\dagger}\hat{U}_{\textsc{b}}^{\dagger}\hat{U}_{\textsc{a}}^{\dagger} \big]  \\
&\phantom{\rhoh_{\textsc{b}}^f}=\textrm{tr}_{\textsc{c},\phi}\big[ \hat{U}_{\textsc{b}}\hat{U}_{\textsc{c}}(\rhoh_{\textsc{b}}\otimes\proj{\psi}{\psi}\otimes\rhoh_{\phi})\hat{U}_{\textsc{c}}^{\dagger}\hat{U}_{\textsc{b}}^{\dagger} \big] \;.\nonumber%\\
%&=\textrm{tr}_\textsc{a}(\rhoh_{\textsc{ab}}^f) \;, \nonumber
\end{align}
Analogously,
\begin{align}
&\bra{l_\textsc{a}}\rhoh_{\textsc{a}}'\ket{k_\textsc{a}}=\widetilde{w}_{\textsc{a},0}'(k_\textsc{a},l_\textsc{a})\\
&\phantom{==}=\textrm{tr}\big[ \hat{U}_{\textsc{a}}\hat{U}_{\textsc{b}}\hat{U}_{\textsc{c}} (\rhoh_{\textsc{a}}\otimes\rhoh_{\textsc{b}}\otimes\proj{\psi}{\psi}\otimes\rhoh_{\phi}) \nonumber\\
&\phantom{=============}\times\hat{U}_{\textsc{c}}^{\dagger}\hat{U}_{\textsc{b}}^{\dagger}\hat{U}_{\textsc{a}}^{\dagger} \proj{k_\textsc{b}}{l_\textsc{b}} \big] \;, \nonumber
\end{align}
thus getting
\begin{align}\label{Alba non-selective 2nd derivation}
&\rhoh_{\textsc{a}}'=\textrm{tr}_{\textsc{b,c},\phi}\big[ \hat{U}_{\textsc{a}}\hat{U}_{\textsc{b}}\hat{U}_{\textsc{c}} \nonumber\\
&\phantom{====}\times(\rhoh_{\textsc{a}}\otimes\rhoh_{\textsc{b}}\otimes\proj{\psi}{\psi}\otimes\rhoh_{\phi})\hat{U}_{\textsc{c}}^{\dagger}\hat{U}_{\textsc{b}}^{\dagger}\hat{U}_{\textsc{a}}^{\dagger} \big] \\
&\phantom{\rhoh_{\textsc{a}}^f}=\textrm{tr}_{\phi}\big[ \hat{U}_{\textsc{a}}(\rhoh_{\textsc{a}}\otimes\rhoh_{\phi})\hat{U}_{\textsc{a}}^{\dagger} \big] \;.\nonumber%\\
%&=\textrm{tr}_\textsc{b}(\rhoh_{\textsc{ab}}^f) \;, \nonumber
\end{align}
We see that the results obtained for the density operators associated with the partial states $\rhoh_{\textsc{ab}}'$, $\rhoh_\textsc{a}$ and $\rhoh_\textsc{b}$ using the $n$-point function formalism for implementing the update rule are the same as those obtained with the context-dependent density operator formalism.

\subsection{Selective case}

Let us now analyze the case in which the measurement performed by Clara is selective. In this case, the approach based on $n$-point functions has the advantage that the analysis of where the information is accessible is already contained in the piecewise definition of the selective update, so the calculations are more systematic. As a downside, it is more cumbersome than the density operator approach. Starting from the same initial $n$-point functions of Eq.~\eqref{Eq. 71}, after the measurement we update $w_n$ to $w_n^{\textsc{S}}$ following the prescription of Eqs.~\eqref{selective updated n-point function def 1} and~\eqref{selective updated n-point function def 2}:
\begin{align}
&w_{n}^{\textsc{S}}(\mathsf{x}_1,\hdots,\mathsf{x}_n) \nonumber\\
&\phantom{===}=\textrm{tr}_{\textsc{c},\phi}\big[ \hat{U}_{\textsc{c}}(\,\proj{\psi}{\psi}\otimes\rhoh_{\phi})\hat{U}_{\textsc{c}}^{\dagger}\phih(\mathsf{x}_1)\cdots\phih(\mathsf{x}_n) \big]
\end{align}
if all $\mathsf{x}_1,\hdots,\mathsf{x}_n$ are outside $\mathcal{P}$, and
\begin{equation}
w_{n}^{\textsc{S}}(\mathsf{x}_1,\hdots,\mathsf{x}_n)= \frac{\textrm{tr}\big(\hat{M}_{c,\psi}\rhoh_{\phi} \hat{M}_{c,\psi}^{\dagger}\phih(\mathsf{x}_1)\cdots\phih(\mathsf{x}_n) \big)}{\textrm{tr}_{\phi}\big( \rhoh_{\phi}\hat{E}_{c,\psi} \big)} 
\end{equation}
otherwise. Now, to include detectors A and B in the picture we use the extended formalism we introduced in Section~\ref{section: generalization to the presence of entangled third parties}. Since B is in the causal future of the measurement, we have
\begin{align}
&\widetilde{w}_{\{\textsc{a,b}\},n}^\textsc{S}((k_\textsc{a},k_\textsc{b}),(l_\textsc{a},l_\textsc{b});\mathsf{x}_1,\hdots,\mathsf{x}_n) \nonumber\\
&=\textrm{tr}_{\textsc{a,b},\phi}\big[ \hat{M}_{c,\psi}(\rhoh_{\textsc{a}}\otimes\rhoh_{\textsc{b}}\otimes\rhoh_\phi)\hat{M}_{c,\psi}^{\dagger} \\
&\phantom{=}\times\proj{k_\textsc{a},k_\textsc{b}}{l_\textsc{a},l_\textsc{b}}\phih(\mathsf{x}_1)\cdots\phih(\mathsf{x}_n) \big] \textrm{tr}_{\phi}\big(\rhoh_{\phi}\hat{E}_{c,\psi}\big)^{-1} \nonumber
\end{align}
and
\begin{align}
&\widetilde{w}_{\textsc{b},n}^\textsc{S}((k_\textsc{b},l_\textsc{b});\mathsf{x}_1,\hdots,\mathsf{x}_n) \nonumber\\
&\phantom{=}=\textrm{tr}_{\textsc{a,b},\phi}\big[ \hat{M}_{c,\psi}(\rhoh_{\textsc{a}}\otimes\rhoh_{\textsc{b}}\otimes\rhoh_\phi)\hat{M}_{c,\psi}^{\dagger} \nonumber\\
&\phantom{==}\times\proj{k_\textsc{b}}{l_\textsc{b}}\phih(\mathsf{x}_1)\cdots\phih(\mathsf{x}_n) \big] \textrm{tr}_{\phi}\big(\rhoh_{\phi}\hat{E}_{c,\psi}\big)^{-1} \\
&\phantom{=}=\textrm{tr}_{\textsc{b},\phi}\big[ \hat{M}_{c,\psi}(\rhoh_{\textsc{b}}\otimes\rhoh_\phi)\hat{M}_{c,\psi}^{\dagger} \nonumber\\
&\phantom{==}\times\proj{k_\textsc{b}}{l_\textsc{b}}\phih(\mathsf{x}_1)\cdots\phih(\mathsf{x}_n) \big]\textrm{tr}_{\phi}\big(\rhoh_{\phi}\hat{E}_{c,\psi}\big)^{-1} \;. \nonumber
\end{align}
Now, because A is not in the causal future of the measurement,
\begin{align}
&\widetilde{w}_{\textsc{a},n}^{\textsc{S}}((k_\textsc{a},l_\textsc{a});\mathsf{x}_1,\hdots,\mathsf{x}_n) \nonumber\\
&\phantom{==}=\textrm{tr}\big[ \hat{U}_{\textsc{c}}(\rhoh_{\textsc{a}}\otimes\rhoh_{\textsc{b}}\otimes\proj{\psi}{\psi}\otimes\rhoh_{\phi})\hat{U}_{\textsc{c}}^{\dagger} \\
&\phantom{=========}\times\proj{k_\textsc{a}}{l_\textsc{a}}\phih(\mathsf{x}_1)\cdots\phih(\mathsf{x}_n) \big] \; \nonumber
\end{align}
if $\mathsf{x}_1,\hdots,\mathsf{x}_n \notin \mathcal{P}$, and
\begin{align}
&\widetilde{w}_{\textsc{a},n}^\textsc{S}((k_\textsc{a},l_\textsc{a});\mathsf{x}_1,\hdots,\mathsf{x}_n) \\
&\phantom{=}=\textrm{tr}_{\textsc{a},\phi}\big[ \hat{M}_{c,\psi}(\rhoh_{\textsc{a}}\otimes\rhoh_\phi)\hat{M}_{c,\psi}^{\dagger} \nonumber\\
&\phantom{==}\times\proj{k_\textsc{a}}{l_\textsc{a}}\phih(\mathsf{x}_1)\cdots\phih(\mathsf{x}_n) \big]\textrm{tr}_{\phi}\big(\rhoh_{\phi}\hat{E}_{c,\psi}\big)^{-1}  \nonumber
\end{align}
otherwise.

After taking into account the interaction of the field with A and B, we update the extended $n$-point functions accordingly: for the $n$-point functions involving both detectors A and B,
\begin{align}
&\widetilde{w}_{\{\textsc{a,b}\},n}'((k_\textsc{a},k_\textsc{b}),(l_\textsc{a},l_\textsc{b});\mathsf{x}_1,\hdots,\mathsf{x}_n) \nonumber\\
&\phantom{==}=\textrm{tr}_{\textsc{a,b},\phi}\big[\hat{U}_{\textsc{a}}\hat{U}_{\textsc{b}} \hat{M}_{c,\psi}(\rhoh_{\textsc{a}}\otimes\rhoh_{\textsc{b}}\otimes\rhoh_\phi)\hat{M}_{c,\psi}^{\dagger} \\
&\phantom{====}\times\hat{U}_{\textsc{b}}^\dagger\hat{U}_{\textsc{a}}^\dagger\proj{k_\textsc{a},k_\textsc{b}}{l_\textsc{a},l_\textsc{b}}\phih(\mathsf{x}_1)\cdots\phih(\mathsf{x}_n) \big] \nonumber\\
&\phantom{======}\times\textrm{tr}_{\phi}\big(\rhoh_{\phi}\hat{E}_{c,\psi}\big)^{-1} \;, \nonumber
\end{align}
and for the ones involving only B,
\begin{align}
&\widetilde{w}_{\textsc{b},n}'((k_\textsc{b},l_\textsc{b});\mathsf{x}_1,\hdots,\mathsf{x}_n) \nonumber\\
&\phantom{==}=\textrm{tr}_{\textsc{a,b},\phi}\big[\hat{U}_{\textsc{a}}\hat{U}_{\textsc{b}} \hat{M}_{c,\psi}(\rhoh_{\textsc{a}}\otimes\rhoh_{\textsc{b}}\otimes\rhoh_\phi)\hat{M}_{c,\psi}^{\dagger} \\
&\phantom{====}\times\hat{U}_{\textsc{b}}^\dagger\hat{U}_{\textsc{a}}^\dagger\proj{k_\textsc{b}}{l_\textsc{b}}\phih(\mathsf{x}_1)\cdots\phih(\mathsf{x}_n) \big] \nonumber\\
&\phantom{======}\times\textrm{tr}_{\phi}\big(\rhoh_{\phi}\hat{E}_{c,\psi}\big)^{-1} \;. \nonumber
\end{align}
For the extended $n$-point functions involving A, however, we get
\begin{align}
&\widetilde{w}_{\textsc{a},n}'((k_\textsc{a},l_\textsc{a});\mathsf{x}_1,\hdots,\mathsf{x}_n) \nonumber\\
&\phantom{==}=\textrm{tr}\big[ \hat{U}_{\textsc{a}}\hat{U}_{\textsc{b}}\hat{U}_{\textsc{c}}(\rhoh_{\textsc{a}}\otimes\rhoh_{\textsc{b}}\otimes\proj{\psi}{\psi}\otimes\rhoh_{\phi}) \\
&\phantom{=======}\times\hat{U}_{\textsc{c}}^\dagger\hat{U}_{\textsc{b}}^\dagger\hat{U}_{\textsc{a}}^{\dagger}\proj{k_\textsc{a}}{l_\textsc{a}}\phih(\mathsf{x}_1)\cdots\phih(\mathsf{x}_n) \big] \; \nonumber
\end{align}
if $\mathsf{x}_1,\hdots,\mathsf{x}_n$ and Alba are outside $\mathcal{P}$, and 
\begin{align}
&\widetilde{w}_{\textsc{a},n}'((k_\textsc{a},l_\textsc{a});\mathsf{x}_1,\hdots,\mathsf{x}_n) \nonumber\\
&\phantom{==}=\textrm{tr}_{\textsc{a,b},\phi}\big[\hat{U}_{\textsc{a}}\hat{U}_{\textsc{b}} \hat{M}_{c,\psi}(\rhoh_{\textsc{a}}\otimes\rhoh_{\textsc{b}}\otimes\rhoh_\phi)\hat{M}_{c,\psi}^{\dagger} \\
&\phantom{====}\times\hat{U}_{\textsc{b}}^\dagger\hat{U}_{\textsc{a}}^\dagger\proj{k_\textsc{a}}{l_\textsc{a}}\phih(\mathsf{x}_1)\cdots\phih(\mathsf{x}_n) \big] \nonumber\\
&\phantom{======}\times\textrm{tr}_{\phi}\big(\rhoh_{\phi}\hat{E}_{c,\psi}\big)^{-1} \nonumber
\end{align}
otherwise. This is a complete account of the relevant extended $n$-point functions for the practical example we are dealing with, so that now we can calculate the partial states. In particular,
\begin{align}
&\bra{l_\textsc{a},l_\textsc{b}}\rhoh_{\textsc{ab}}'\ket{k_\textsc{a},k_\textsc{b}}=\widetilde{w}_{\{\textsc{a,b}\},0}'((k_\textsc{a},k_\textsc{b}),(l_\textsc{a},l_\textsc{b})) \nonumber\\
&=\textrm{tr}_{\textsc{a,b},\phi}\big[\hat{U}_{\textsc{a}}\hat{U}_{\textsc{b}} \hat{M}_{c,\psi}(\rhoh_{\textsc{a}}\otimes\rhoh_{\textsc{b}}\otimes\rhoh_\phi)\hat{M}_{c,\psi}^{\dagger} \\
&\phantom{====}\times\hat{U}_{\textsc{b}}^\dagger\hat{U}_{\textsc{a}}^\dagger\proj{k_\textsc{a},k_\textsc{b}}{l_\textsc{a},l_\textsc{b}}\big]\textrm{tr}_{\phi}\big(\rhoh_{\phi}\hat{E}_{c,\psi}\big)^{-1} \nonumber
\end{align}
giving
\begin{equation}\label{Alba-Blanca selective 2nd derivation}
\rhoh_{\textsc{ab}}'=\frac{\textrm{tr}_{\phi}\big[ \hat{U}_{\textsc{a}}\hat{U}_{\textsc{b}}\hat{M}_{c,\psi}(\rhoh_{\textsc{a}}\otimes\rhoh_{\textsc{b}}\otimes\rhoh_{\phi})\hat{M}_{c,\psi}^{\dagger}\hat{U}_{\textsc{b}}^{\dagger}\hat{U}_{\textsc{a}}^{\dagger}\big]}{\textrm{tr}_{\phi}\big(\rhoh_{\phi}\hat{E}_{c,\psi}\big)} \;.
\end{equation}
Analogously,
\begin{align}
&\bra{l_\textsc{b}}\rhoh_{\textsc{b}}'\ket{k_\textsc{b}}=\widetilde{w}_{\textsc{b},0}'(k_\textsc{b},l_\textsc{b}) \nonumber\\
&=\textrm{tr}_{\textsc{a,b},\phi}\big[\hat{U}_{\textsc{a}}\hat{U}_{\textsc{b}} \hat{M}_{c,\psi}(\rhoh_{\textsc{a}}\otimes\rhoh_{\textsc{b}}\otimes\rhoh_\phi)\hat{M}_{c,\psi}^{\dagger} \\
&\phantom{====}\times\hat{U}_{\textsc{b}}^\dagger\hat{U}_{\textsc{a}}^\dagger\proj{k_\textsc{b}}{l_\textsc{b}}\big]\textrm{tr}_{\phi}\big(\rhoh_{\phi}\hat{E}_{c,\psi}\big)^{-1} \nonumber %\\
%&=\frac{\bra{l_\textsc{b}}\textrm{tr}_{\phi}\big[\hat{U}_{\textsc{b}} \hat{M}_{s,\psi}(\rhoh_{\textsc{b}}\otimes\rhoh_\phi)\hat{M}_{s,\psi}^{\dagger}\hat{U}_{\textsc{b}}^\dagger\big]\ket{k_\textsc{b}}}{\textrm{tr}_{\phi}\big(\rhoh_{\phi}\hat{E}_{s,\psi}\big)} \nonumber
\end{align}
gives
\begin{equation}\label{Blanca selective 2nd derivation}
\rhoh_{\textsc{b}}'=\frac{\textrm{tr}_{\phi}\big[\hat{U}_{\textsc{b}} \hat{M}_{c,\psi}(\rhoh_{\textsc{b}}\otimes\rhoh_\phi)\hat{M}_{c,\psi}^{\dagger}\hat{U}_{\textsc{b}}^\dagger\big]}{\textrm{tr}_{\phi}\big(\rhoh_{\phi}\hat{E}_{c,\psi}\big)} \;.
\end{equation}
On the other hand, since A stays spacelike separated from the causal future of Clara's measurement $\mathcal{P}$,
\begin{align}
&\bra{l_\textsc{a}}\rhoh_{\textsc{a}}'\ket{k_\textsc{a}}=\widetilde{w}_{\textsc{a},0}'(k_\textsc{a},l_\textsc{a})\\
&\phantom{==}=\bra{l_\textsc{a}}\textrm{tr}_{\textsc{b,c},\phi}\big[\hat{U}_{\textsc{a}}\hat{U}_{\textsc{b}} \hat{U}_{\textsc{c}} \nonumber\\
&\phantom{====}\times(\rhoh_{\textsc{a}}\otimes\rhoh_{\textsc{b}}\otimes\proj{\psi}{\psi}\otimes\rhoh_{\phi})\hat{U}_{\textsc{c}}^{\dagger}\hat{U}_{\textsc{b}}^\dagger\hat{U}_{\textsc{a}}^\dagger \big]\ket{k_\textsc{a}} \; \nonumber \\
&\phantom{==}=\bra{l_\textsc{a}} \textrm{tr}_\phi\big[ \hat{U}_{\textsc{a}}(\rhoh_{\textsc{a}}\otimes\rhoh_\phi)\hat{U}_{\textsc{a}}^\dagger \big]\ket{k_\textsc{a}} \nonumber
\end{align}
yielding
\begin{equation}\label{Alba selective 2nd derivation}
\rhoh_{\textsc{a}}'=\textrm{tr}_{\phi}\big[ \hat{U}_{\textsc{a}}(\rhoh_{\textsc{a}}\otimes\rhoh_\phi)\hat{U}_{\textsc{a}}^\dagger \big] \neq \textrm{tr}_\textsc{b}(\rhoh_{\textsc{ab}}') \;,
\end{equation}
as expected. Finally, if  Alba eventually reaches $\mathcal{P}$, then following the prescribed update rule for the case when A is inside $\mathcal{P}$,
\begin{align}
&\bra{l_\textsc{a}}\rhoh_{\textsc{a}}''\ket{k_\textsc{a}}=\widetilde{w}_{\textsc{a},0}'(k_\textsc{a},l_\textsc{a}) \nonumber\\
&=\frac{\bra{l_\textsc{a}}\textrm{tr}_{\textsc{b},\phi}\big[ \hat{U}_{\textsc{a}}\hat{U}_{\textsc{b}}\hat{M}_{c,\psi}(\rhoh_{\textsc{a}}\otimes\rhoh_{\textsc{b}}\otimes\rhoh_{\phi})\hat{M}_{c,\psi}^{\dagger}\hat{U}_{\textsc{b}}^{\dagger}\hat{U}_{\textsc{a}}^{\dagger}\big]\ket{k_\textsc{a}}}{\textrm{tr}_{\phi}\big(\rhoh_{\phi}\hat{E}_{c,\psi}\big)} \nonumber\\
&=\frac{\bra{l_\textsc{a}}\textrm{tr}_{\phi}\big[\hat{U}_{\textsc{a}}\hat{M}_{c,\psi}(\rhoh_{\textsc{a}}\otimes\rhoh_{\phi})\hat{M}_{c,\psi}^{\dagger}\hat{U}_{\textsc{a}}^{\dagger}\big]\ket{k_\textsc{a}}}{\textrm{tr}_{\phi}\big(\rhoh_{\phi}\hat{E}_{c,\psi}\big)}
\end{align}
yielding
\begin{equation}\label{Alba selective 2nd derivation after cone}
\rhoh_{\textsc{a}}''=\frac{\textrm{tr}_{\phi}\big(\hat{U}_{\textsc{a}}\hat{M}_{c,\psi}[\rhoh_{\textsc{a}}\otimes\rhoh_{\phi}]\hat{M}_{c,\psi}^{\dagger}\hat{U}_{\textsc{a}}^{\dagger}\big)}{\textrm{tr}_{\phi}\big(\rhoh_{\phi}\hat{E}_{c,\psi}\big)}=\textrm{tr}_{\textsc{b}}\big(\rhoh_{\textsc{ab}'}\big) \;. 
\end{equation}
These results are again identical to those obtained in Section~\ref{section: a practical example with detectors}, giving some insight to the equivalence of both formalisms, while showing a practical example of how to perform the calculations in each of them.

\bibliography{references}
\end{document}